\documentclass[pdflatex,sn-nature]{sn-jnl}

\usepackage{graphicx}
\usepackage{amsmath,amssymb,amsfonts,braket}
\usepackage{bm}
\usepackage{booktabs}

\usepackage{csquotes}
\MakeOuterQuote{"}
\usepackage{siunitx}
\renewcommand{\v}[1]{\bm{\mathrm{#1}}}
\newcommand{\m}[1]{\bm{\mathsf{#1}}}

\begin{document}

\title[]{Orbital angular momentum in crystalline solids: non-equilibrium theory}

\author[1,2]{\fnm{Sangeeta} \sur{Sharma}}
\author[3]{\fnm{Peter} \sur{Elliott}}
\author*[1]{\fnm{Samuel} \sur{Shallcross}}\email{phsss75@gmail.com}

\affil[1]{\orgname{Max-Born-Institute for Non-Linear Optics},
\orgaddress{\street{Max-Born Strasse 2A}, \postcode{12489}, \city{Berlin}, \country{Germany}}}

\affil[2]{\orgdiv{Institute for Theoretical Solid-State Physics},
\orgname{Freie Universit\"at Berlin},
\orgaddress{\street{Arnimallee 14}, \postcode{14195}, \city{Berlin}, \country{Germany}}}

\affil[3]{\orgdiv{Scientific Computing Department},
\orgname{Science and Technology Facilities Council UK Research and Innovation (STFC-UKRI)},
\orgaddress{\street{Rutherford Appleton Laboratory, Harwell Campus}, \city{Didcot}, \postcode{OX11 0QX}, \country{United Kingdom}}}

\abstract{We develop a general operator theory of orbital angular momentum (OAM) in periodic solids that applies equally to equilibrium and coherently driven states. Employing the crystal momentum representation of Adams and Blount, we construct a periodic-solid OAM operator exactly corresponding to the symmetrized 
$\hat{\mathbf r}\times\hat{\mathbf v}$ observable, and show that it separates into local circulation (LC) and itinerant circulation (IC) sectors that are Hermitian and $U(1)$, and more generally $U(N)$, gauge covariant. In equilibrium, the LC and IC sectors reduce to the established modern theory of orbital magnetism; out of equilibrium, the same operator structure exposes new contributions generated by laser-written momentum-space textures and interband coherence. Applied to gapped graphene, we uncover a rich phenomenology of ultrafast non-equilibrium orbital angular momentum, including a quasi-stationary light-induced OAM. This latter finding, underpinned by a close connection between band-manifold phase texture and nonequilibrium OAM, points towards  new possibilities  for light control over OAM in solids as well as a potentially rich role for topological band structures in designed OAM materials.}

\maketitle

\section{Introduction}\label{sec:introduction}

Orbital angular momentum plays a crucial role in modern condensed matter physics. Experimentally, unmistakable signatures of OAM currents~\cite{Choi2023OHE,Seifert2023BallisticOAM} promise remarkable new routes to the control of magnetic order~\cite{GoLee2020OrbitalTorque,Ding2020OrbitalSpin,
Kim2021OrbitalInjection,Lee2021OrbitalTorque,
GoJoLeeKlaeuiMokrousov2021}, while canonical effects such as the Rashba surface-state spin polarization have been reinterpreted as ``shadows'' cast by an underlying OAM physics~\cite{Park2011RashbaOAM,Park2012ChiralOAM}. While the modern theory of OAM in periodic solids is now well established and extremely successful at equilibrium{~\cite{XiaoShiNiu2005,
ThonhauserCeresoliVanderbiltResta2005,
CeresoliThonhauserVanderbiltResta2006,
ShiVignaleXiaoNiu2007,SouzaVanderbilt2008}}, a correspondingly general theory has not, however, emerged in the non-equilibrium case.

Two reasons for this can be adduced. On the one hand, the equilibrium theory is framed in its commonly employed form in terms of band-resolved, diagonal expectation values, a structure inherently unsuited to strongly non-equilibrium physics in which superpositions over band indices play the central role. On the other hand, orbital angular momentum in a periodic solid involves, already at equilibrium, delicate questions rooted in the fact that OAM requires the position operator through the product $\mathbf r\times\mathbf v$, while $\mathbf r$ itself is not a well-defined periodic operator in the Bloch representation{~\cite{ThonhauserCeresoliVanderbiltResta2005,
CeresoliThonhauserVanderbiltResta2006,SouzaVanderbilt2008}}.

A number of recent works have sought to extend orbital angular momentum and orbital-magnetization theory beyond their conventional equilibrium setting. These include current-induced orbital magnetization and orbital Edelstein responses~\cite{
ZhongMooreSouza2016,
YodaYokoyamaMurakami2018,
HaraBahramyMurakami2020,
JohanssonGobelHenkBibesMertig2021,
BaekHanCheonLee2024},
orbital-current, torque, and pumping phenomena~\cite{
GoLeeOppeneerBlugelMokrousov2024,
AtenciaArovasCulcer2024,
LiuCullenArovasCulcer2025,
HanKoOhLeeLeeKim2025},
and quantum-geometric orbital magnetoelectric response~\cite{
CullenArovasRaimondiCulcer2025}. There have also been recent efforts towards formulating genuinely non-equilibrium dynamical descriptions of OAM in periodic solids~\cite{
XiaoRenXiong2021,
ToppTormaKennesMitra2022,
DongLinGuDuan2024,
YaoGoMokrousovMurakami2026,
MrudulOppeneer2026}.
The obstacles for the latter goal are, however, formidable{~\cite{BurgosAtenciaAgarwalCulcer2024}}. 

A naive off-diagonal continuation of the equilibrium Bloch orbital moment $\mathbf m_{\v k}$, followed for example by evaluation of $\operatorname{Tr}[\rho\mathbf m]$ for a non-equilibrium $\rho$, fails the test of gauge covariance: as shown by Cysne \emph{et al.} additional Berry-connection terms are required to construct a gauge-covariant matrix-valued extension of the equilibrium orbital moment $\v m_{\v k}$~\cite{CysneSouzaRappoport2026}. Gauge covariance alone, however, does not uniquely prescribe how this equilibrium quantity should be extended into the band-off-diagonal sector. An alternative route has been developed by Mokrousov and co-workers, {who, starting from the full modern theory of equilibrium orbital magnetism, employ} gauge-covariant Wannier constructions to formulate orbital operators and response frameworks~\cite{LeeBaekSastgesMokrousovLeeGo2026,
SastgesBaekLeeLeeMokrousovGo2026}. {These approaches represent substantial progress; however, they} do not yet meet what appear to be essential criteria of a complete non-equilibrium theory: a single periodic-solid orbital angular momentum operator that (i) treats equilibrium and arbitrary nonequilibrium states on exactly the same footing, with both equilibrium OAM and dynamical $\mathbf L(t)$ obtained solely by changing the quantum state entering the expectation value, and (ii) can be shown, for arbitrary states, to be equivalent to the {symmetrized orbital angular momentum operator} $\left(\hat{\mathbf r}\times\hat{\mathbf v}-\hat{\mathbf v}\times\hat{\mathbf r}\right)/2$ represented in a form appropriate to a periodic solid.

Here we resolve this situation by introducing an operator formulation of OAM in periodic solids that, when evaluated on equilibrium states, reduces to the familiar modern-theory "local circulation" and "itinerant circulation" expressions{~\cite{ThonhauserCeresoliVanderbiltResta2005,
CeresoliThonhauserVanderbiltResta2006,
ShiVignaleXiaoNiu2007,SouzaVanderbilt2008}}, while for general nonequilibrium states reveals qualitatively new structures associated with momentum-space phase texture, interband coherence, and mixed local itinerant contributions. In this way, the familiar equilibrium theory appears as a limit of a more general operator theory, rather than serving as the starting point from which nonequilibrium expressions are inferred. Using gapped graphene as an example, we show that this approach gives rise to rich ultrafast nonequilibrium phenomena: large transient OAM and, more strikingly, a large quasi-stationary post-pulse OAM, controllable by pulse parameters, and supported by a light-written momentum-space phase texture.

\section{Theory}\label{sec:theory}

Here we detail an "operator approach" to orbital angular momentum in a {periodic} solid. In Sec.~2.1 we derive a four-operator representation of the unit cell orbital angular momentum operator comprising (i) a pure interband "local circulation" (LC) operator, (ii) a pure intraband "itinerant circulation" (IC) operator, {(iii) two mixed sector operators}. These are shown to be gauge covariant and Hermitian, with the LC and IC operators reducing to the well-known forms of the modern theory of orbital magnetism in periodic solids upon taking the expectation value of an equilibrium state (Secs.~2.2 and 2.3). The mixed sector operators are shown to yield {a} vanishing contribution to the expectation value of orbital angular momentum in this limit.

The distinction between ``home-cell'' and bulk orbital angular momentum is then developed directly at the operator level through the behaviour of these sectors under an active translation within a Born--von K\'arm\'an (BvK) supercell, Sec.~2.3. The boundary-crossing cell exposes a second IC contribution which, when redistributed on a per-cell basis, produces a factor-of-two difference between the home-cell IC and its contribution to the total bulk orbital angular momentum. 
As BvK periodic boundary conditions represent an arbitrary division of a periodic solid into "bulk" and "boundary", this construction is analogous to that of the modern theory of 
orbital magnetism~\cite{CeresoliThonhauserVanderbiltResta2006}, in which a boundary contribution supplies the second half of the equilibrium itinerant circulation term; here however it is promoted to the operator level construction suitable for going beyond equilibrium.

Finally, we turn to the new features that the operator theory reveals in the far-from-equilibrium regime of laser-driven periodic matter, Sec.~2.4. A crucial distinction between the LC and IC sectors is that the LC sectors act pointwise on the BZ torus, whereas the IC sector contains the covariant derivative in momentum space. This provides a direct link between momentum-space phase structures -- such as those imprinted by a laser pulse -- and a post-pulse quasi-stationary OAM.

\subsection{Orbital angular momentum in periodic solids: an operator approach}
\label{4op}

Orbital angular momentum, whose appropriately symmetrized operator reads

\begin{equation}
\hat{\mathbf L}
=
\frac{1}{2}
\left(
\hat{\mathbf r}\times\hat{\mathbf v}
-
\hat{\mathbf v}\times\hat{\mathbf r}
\right)
\label{eq:oam}
\end{equation}
represents an elementary quantum-mechanical observable that is highly non-trivial in the context of a periodic solid. Here, $\hat{\mathbf r}$ is the position operator and

\begin{equation}
\hat{\mathbf v} = -i[\hat{\mathbf r},\hat H]
\label{eq:velocity-op}
\end{equation}
the velocity operator, and we have used atomic units $m_e=e=\hbar=1$ so that the orbital magnetic moment is related to Eq.~\eqref{eq:oam} by $\boldsymbol{\mu}_L=-\mu_B\mathbf L$.

The origin of the difficulty of $\hat{{\mathbf L}}$ in a periodic solid arises from the incompatibility of $\hat{\v r}$ with translational symmetry, and in fact matrix elements of $\hat{\v r}$ with Bloch eigenstates $\langle\phi_{n\mathbf{k}}|\hat{\mathbf r}|\phi_{m\mathbf{k}'}\rangle$ possess no well-defined value~\cite{Blount1962}. This difficulty can be circumvented via a localized Wannier representation, the approach of the modern theories of polarization and orbital magnetism. However a second route exists, the "crystal momentum representation" (CMR), that retains the Bloch function basis while, at the same time, avoiding 
entirely matrix elements involving Bloch eigenstates. In this approach the action of an operator $\hat{O}$ is shifted from a state  

\begin{equation}
|\Psi\rangle
=
\frac{1}{V_{\mathrm{BZ}}}
\int_{\mathrm{BZ}} d\mathbf{k}
\sum_n
|\phi_{n\mathbf{k}}\rangle c_{n\mathbf{k}},
\label{eq:state-expansion}
\end{equation}
i.e. $\hat{O}\ket{\Psi}$, to acting on the "\emph{coefficient section}" $c_{\v k}$, i.e. $O c_{\v k}$~\cite{Blount1962}. Here the term coefficient section denotes the vector of expansion coefficients at each crystal momentum. Crucially, inherent in this approach is that the transfer of operator action from the quantum state to the coefficient section {\it preserves the operator algebra}~\cite{Adams1953CrystalMomentum,Blount1962}. Thus, any sequence of operators $\hat O_1\cdots\hat O_N$ acting on $\ket{\Psi}$ is mapped to the corresponding sequence $O_1\cdots O_N$ acting on $c_{\mathbf k}$,

\begin{equation}
\hat O_1\cdots\hat O_N\ket{\Psi}
\longleftrightarrow
O_1\cdots O_N c_{\mathbf k}
\label{eq:blount-composition}
\end{equation}
where the ordering of operators is retained, and where differential operators are understood to act forward on all quantities to their right. Consequently, operator products and commutators may be evaluated directly within the CMR, and expectation values of such operators obtained via

\begin{equation}
\langle\hat O\rangle
= \frac{1}{V_{\mathrm{BZ}}}
\int_{\mathrm{BZ}} d\mathbf k\,
c_{\mathbf k}^{\dagger}O\,c_{\mathbf k}
\label{eq:expectation-coeff}
\end{equation}
thus avoiding all direct use of mathematically ill defined Bloch function matrix elements.

A programme to develop a theory of angular momentum in periodic solids that satisfies the two criteria set out in the introduction is thus clear: identify the CMR operators corresponding to $\hat{\v r}$ and $\hat{\v v}$, and insert these into Eq.~\ref{eq:oam}. Both the CMR forms of $\v r$ and $\v v$ are, in fact, widely known in band theory. For clarity of exposition, as well as because subtle features of these operators play an important role in analysing the resulting $\v L$ operator, we first briefly review these two CMR operators; complete details are presented in Supplemental Information Sec.~S1. 

\emph{The position operator}: Within the crystal momentum representation the position operator takes the form

\begin{equation}
\mathbf r = i\v D_{\v k} + \mathbf d(\mathbf k)
\label{eq:blount-position}
\end{equation}
where the covariant derivative

\begin{equation}
i\mathbf D_{\mathbf k}
= i\nabla_{\mathbf k} +
{\boldsymbol{\mathcal A}(\mathbf k)}
\label{eq:covariant-derivative}
\end{equation}
combines a momentum space derivative with the band diagonal Berry connection matrix

\begin{equation}
[{\boldsymbol{\mathcal A}(\mathbf k)}]_{nm}
= \delta_{nm}\langle u_{n\mathbf k}| i\nabla_{\mathbf k}
|u_{n\mathbf k}\rangle
\label{eq:berry-diagonal}
\end{equation}
to yield a gauge covariant operator,
while the dipole matrix $\v d$ retains off-diagonal Berry connection terms 

\begin{equation}
[\mathbf d(\mathbf k)]_{nm}
= (1-\delta_{nm})
\langle u_{n\mathbf k}|
i\nabla_{\mathbf k}
|u_{m\mathbf k}\rangle
\label{eq:berry-offdiagonal}
\end{equation}
Before proceeding further we wish to draw the reader's attention to two shorthand notations that will be employed throughout this work: (i) all objects will be written as matrices or vectors in band space, as is implied in Eq.~\ref{eq:blount-position}-\ref{eq:covariant-derivative}, and (ii) we will subsequently suppress explicit $\mathbf k$-dependence of band-space quantities for which no ambiguity can arise. Thus $E$, $\mathbf d$, and $\boldsymbol{\mathcal A}$ will denote $E(\mathbf k)$, $\mathbf d(\mathbf k)$, and $\boldsymbol{\mathcal A}(\mathbf k)$, respectively.

This CMR brings out key features of the \emph{nature} of the position operator in a periodic solid, that will prove useful in analysing the orbital angular momentum operator. Firstly, the decomposition of $\v r$ into covariant derivative and dipole matrix is, as is clear from Eqs.~\ref{eq:covariant-derivative}-\ref{eq:berry-offdiagonal}, a decomposition into intra- and inter-band terms. Secondly, the behaviour of these two terms under active translation is strikingly different: employing the CMR translation operator $T_{\v a} = e^{-i\v k.\v a}$ we find

\begin{eqnarray}
T_{\v a}^{-1} i 
\v D_{\v k} T_{\v a} & = & i\v D_{\v k} + \v a \label{eq:T-covar}\\
T_{\v a}^{-1} \v d T_{\v a} & = & \v d \label{eq:T-d}
\end{eqnarray}
and thus while the covariant derivative $i\v D_{\v k}$ transforms as a coordinate, the dipole matrix elements $\v d$ transform as a coordinate difference.

\emph{The velocity operator}: The CMR operator corresponding to the velocity operator will, 
as $\v v = \dot{\v r} = -i[\v r,E]$, inherit the two-term intra- and inter-band structure of the position operator. Upon insertion of Eq.~\ref{eq:blount-position}, we immediately find

\begin{eqnarray}
\v v & = & i\dot{\v D}_{\v k} + \dot{\v d} \\
 & = & \nabla_{\mathbf k}E - i[\mathbf d,E]
\end{eqnarray}
with intra-band $i\dot{\v D}_{\v k} = -i[i\v D_{\v k},E] = \m\nabla_{\v k} E$, and inter-band $\dot{\v d} = -i[\v d,E]$. Finally, we note that under the action of $T_{\v a}$ the velocity operator transforms as

\begin{equation}
T_{\v a}^{-1} \v v T_{\v a} = \v v
\end{equation}
expressing the fact that the constant vector $\v a$ in Eq.~\ref{eq:T-covar} vanishes upon taking the time derivative.

\emph{The orbital angular momentum operator}: Having introduced and described the position and velocity operators in the CMR, we now combine them to form the  orbital angular momentum operator. This is most conveniently expressed by introducing the Hermitian antisymmetric bilinear


\begin{table}[!t]
\centering
\renewcommand{\arraystretch}{1.4}
\setlength{\tabcolsep}{40pt}
\resizebox{0.8\textwidth}{!}{%
\begin{tabular}{@{}ccc@{}}
\toprule
&
$\displaystyle \dot{\mathbf d}=-i[\mathbf d,E]$
&
$\displaystyle \dot{\mathbf R}=\nabla_{\mathbf k}E$
\\
\midrule
$\displaystyle \mathbf d=\mathbf r-\mathbf R$
&
$\displaystyle
\mathbf L_{\rm LC}
=
\mathcal L\!\left(\mathbf d,\dot{\mathbf d}\right)$
&
$\displaystyle
\mathbf L_{\rm LC}^{\rm mix}
=
\mathcal L\!\left(\mathbf d,\dot{\mathbf R}\right)$
\\
$\displaystyle \mathbf R=i\mathbf D_{\mathbf k}$
&
$\displaystyle
\mathbf L_{\rm IC}^{\rm mix}
=
\mathcal L\!\left(\mathbf R,\dot{\mathbf d}\right)$
&
$\displaystyle
\mathbf L_{\rm IC}
=
\mathcal L\!\left(\mathbf R,\dot{\mathbf R}\right)$
\\
\bottomrule
\end{tabular}%
}
\vspace{0.3cm}
\caption{\textbf{Orbital angular momentum decomposition in a periodic solid.}
Defining the Hermitian antisymmetric bilinear $\mathcal L(\mathbf X,\mathbf Y)\equiv(\mathbf X\times\mathbf Y-\mathbf Y\times\mathbf X)/2$, orbital angular momentum is $\mathbf L=\mathcal L(\mathbf r,\mathbf v)$, and insertion of the decompositions $\mathbf r=\mathbf R+\mathbf d$ and $\mathbf v=\dot{\mathbf R}+\dot{\mathbf d}$ yields the four operators shown in the table. The LC and IC sectors differ in the choice of lever arm, $\mathbf d$ or $\mathbf R$, while the two columns distinguish relative interband motion from intraband  motion. The mixed terms are the crossed pairings of these two position and velocity sectors. As shown in Secs.~2.2 and 2.3, the diagonal terms reduce in the equilibrium limit to the familiar LC and IC forms of the modern theory of orbital magnetism.}
\label{tab:oam-four-sector}
\end{table}


\begin{equation}
\mathcal L(\mathbf X,\mathbf Y)\equiv\frac12(\mathbf X\times\mathbf Y-\mathbf Y\times\mathbf X)
\end{equation}
with which we may write $\v L$ as

\begin{eqnarray}
\v L & = & \mathcal L(\v r, \v v) \\
 & = & \mathcal L(i\v D_{\v k} + \v d, i\dot{\v D}_{\v k} + \dot{\v d})
 \label{eq:Lbilinear}
\end{eqnarray}

Employing the preceding analysis of the constituent position and velocity operators, the essential structure of Eq.~\ref{eq:Lbilinear} can immediately be discerned. Of the four orbital angular momentum terms implied, two, namely

\begin{equation}
\mathbf L_{\mathrm{LC}}^{\mathrm{tot}} = \mathcal L(\v d, i\dot{\v D}_{\v k} + \dot{\v d}),
\label{LC}
\end{equation}
will be \emph{invariant under translation} $T_{\v a}$ (as both constituents $\v d$ and $\v v = i\dot{\v D}_{\v k} + \dot{\v d}$ are invariant) and so represent an orbital angular momentum \emph{about a centre}. These terms thus correspond to "local circulation" (LC) orbital angular momentum operators, with centre operator $i\v D_{\v k}$ and "lever arm" operator $\v d = \v r-i\v D_{\v k}$.

On the other hand 

\begin{equation}
\mathbf L_{\mathrm{IC}}^{\mathrm{tot}} = \mathcal L(i\v D_{\v k}, i\dot{\v D}_{\v k} + \dot{\v d})
\label{IC}
\end{equation}
under translation obeys $\mathbf L_{\mathrm{IC}}^{\mathrm{tot}} \to \mathbf L_{\mathrm{IC}}^{\mathrm{tot}} + \v a \times \v v$. Here there is no defined centre, and this represents therefore an angular momentum associated with the spatially distributed motion of centres, rather than internal circulation measured relative to those centres. These terms thus represent "itinerant circulation" (IC) orbital angular momentum operators.

Expanding the velocity $\v v = i\dot{\v D}_{\v k} + \dot{\v d}$ in Eqs.~\ref{LC} and \ref{IC} yields two operators in each of the LC and IC sectors. In the LC sector this comprises a \emph{pure inter-band} operator $\mathbf L_{\mathrm{LC}} = \mathcal L(\v d,\dot{\v d})$ and one mixing intra- and inter-band transitions, $\mathbf L_{\mathrm{LC}}^{\mathrm{mix}} = \mathcal L(\v d, i\dot{\v D}_{\v k})$. In contrast for the IC sector we have a \emph{pure intra-band} operator $\mathbf L_{\mathrm{IC}} = \mathcal L(i\v D_{\v k}, i\dot{\v D}_{\v k})$, and one mixing intra- and inter-band transitions, $\mathbf L_{\mathrm{IC}}^{\mathrm{mix}} = \mathcal L(i\v D_{\v k},\dot{\v d})$. It is the pure intra- and inter-band operators that, when acting on equilibrium states, recover the familiar modern theory of orbital magnetism results, as we show in the next section, while the operators coupling intra- and inter-band excitation yield identically zero contribution to OAM in the equilibrium limit.

This $2\times2$ intra- and inter-band sector structure of these four OAM operators is presented in Table~\ref{tab:oam-four-sector}. Writing out each operator explicitly we have

\begin{eqnarray}
\mathbf L_{\mathrm{LC}}
&=&
\frac{i}{2} \left([\mathbf d,E]\times\mathbf d
-
\mathbf d\times[\mathbf d,E] \right)
\label{O1} \\
\mathbf L_{\mathrm{LC}}^{\mathrm{mix}}
&=&
\frac{1}{2} \left( \mathbf d\times\nabla_{\mathbf k}E
-
\nabla_{\mathbf k}E\times\mathbf d \right)
\label{O2} \\
\mathbf L_{\mathrm{IC}}
&=&
-\nabla_{\mathbf k}E\times i\mathbf D_{\mathbf k}
\label{O3} \\
\mathbf L_{\mathrm{IC}}^{\mathrm{mix}}
&=&
-\frac{i}{2} \left(
i\mathbf D_{\mathbf k}\times[\mathbf d,E]
-
[\mathbf d,E]\times i\mathbf D_{\mathbf k}
\right)
\label{O4}
\end{eqnarray}

We now discuss some basic properties of these operators, and generalizations that we reserve for  Supplemental Information.

\emph{Itinerant circulation lever arm}: In equilibrium, where $\langle\mathbf v\rangle=0$, the expectation value of the pure IC operator is invariant under translation as $\left<\mathbf L_{\mathrm{IC}}^{\mathrm{tot}}\right> \to \left<\mathbf L_{\mathrm{IC}}^{\mathrm{tot}}\right> + \v a \times \left<\v v\right>$. The intrinsic IC orbital momentum is therefore that given by the IC operator itself. Out of equilibrium, however, a net current $\langle\mathbf v\rangle \ne 0$ can occur and in this case the intrinsic IC must be obtained by subtracting the lever arm orbital angular momentum

\begin{equation}
\left\langle
\mathbf L_{\mathrm{IC}}^{\mathrm{tot}}
\right\rangle_{\mathrm{int}} =
\left\langle \mathbf L_{\mathrm{IC}}^{\mathrm{tot}}
\right\rangle
-
\left\langle \mathbf r_{\mathrm{intra}} \right\rangle
\times
\left\langle \mathbf v \right\rangle
\label{eq:IC-intrinsic-centered}
\end{equation}

\emph{Gauge covariance}: Each of the four operators Eqs.~\ref{O1}-\ref{O4} is manifestly $U(1)$ gauge covariant, inheriting this property from the gauge covariance of $i\mathbf D_{\mathbf k}$ and $\mathbf d$, together with the gauge invariance of the diagonal energy matrix $E$. Interestingly, this decomposition can be generalized to $U(N)$ gauge covariance in which the LC/IC structure is then a spectral-geometric decomposition of position and velocity; we detail this in Supplemental Information Sec.~S2. 

\emph{Topological obstruction}: For topologically non-trivial bands, the absence of a globally smooth Bloch frame does not modify the local CMR position operator: on each smooth patch one still has $\mathbf r=i\mathbf D_{\mathbf k}+\mathbf d$, with the obstruction entering only through the global patching of these local representations. The same remains true for position-containing composite operators such as $r_i v_j$ and $v_j r_i$: provided their common operator domain is respected, their representation extends patchwise and generates no additional local term. Consequently the orbital angular momentum operator retains the same local form in a Chern band. For details we refer the reader to Supplemental Information Sec.~S3.

\emph{Hermiticity}: For the pure LC, mixed-LC, and mixed-IC operators, Hermiticity is structurally apparent, while for the pure IC operator it follows immediately from

\begin{equation}
i\mathbf D_{\mathbf k}\times\nabla_{\mathbf k}E
=
-\nabla_{\mathbf k}E\times i\mathbf D_{\mathbf k}
\label{eq:IC-hermiticity}
\end{equation}
where we have used the curl-gradient identity,
${
\nabla_{\mathbf k}\times\nabla_{\mathbf k}E=0}$.

Being both Hermitian and $U(1)$ gauge covariant each of the four operators thus constitutes a well-defined orbital angular momentum operator.

\emph{Filled-band phase freedom}: In a completely filled band multiplication of each occupied one-particle state by an arbitrary $\mathbf k$-dependent phase leaves the occupied Slater determinant unchanged apart from an overall phase. In the coefficient-section representation this corresponds to $c_{\mathbf k} \longrightarrow c'_{\mathbf k}
= e^{i\eta(\mathbf k)}c_{\mathbf k}$
and any physical observable of the filled band must therefore be invariant under this transformation. For the local circulation operators this follows immediately, as they act pointwise in $\mathbf k$ and so the phases $e^{i\eta({\v k})}$ cancel directly in the corresponding matrix elements. For the pure IC operator, however, this transformation generates an additional total derivative term
$\mathbf L_{\mathrm{IC}}
\longrightarrow
\mathbf L_{\mathrm{IC}}
+
\nabla_{\mathbf k}E\times\nabla_{\mathbf k}\eta(\mathbf k)$
and thus ${\mathbf L}_{\mathrm{IC}}$ is not invariant pointwise in $\mathbf k$. For a filled band in equilibrium, however, provided the one-form is globally defined on the Brillouin-zone torus, the BZ torus integral of the total derivative vanishes, and hence the integrated IC observable is invariant under filled-band phase freedom. This argument can be extended to equilibrium metals, Supplemental Information Sec.~S4.


\subsection{Equilibrium limit}

{An important limit that must be obeyed by the operator formalism described in the previous section is that, in equilibrium, it reproduce the modern theory of orbital magnetism.}
This limit evidently resides in the band-diagonal expectation values of the four operators, and we therefore see immediately that the "mixed" operators, involving both intraband and interband sectors, contribute identically zero OAM in equilibrium as these operators have zero elements on the diagonal. This can be seen "by inspection" as (i) $\v d$ is {off-diagonal} in band space, and so {$\mathbf L_{\mathrm{LC}}^{\mathrm{mix}}$} must also have zeros on the diagonal, and (ii) $[\v d,E]$ is also zero on the diagonal, from which it then follows that {$\mathbf L_{\mathrm{IC}}^{\mathrm{mix}}$} also possesses zeros on the diagonal.

We now therefore consider the equilibrium limits of the pure LC and IC operators. For simplicity of exposition, we present here the case of an isolated, topologically trivial band insulator at $T=0$.

For the LC operator, the diagonal elements may be written as

\begin{eqnarray}
[{\mathbf L_{\rm LC}}]_{nn}
&=&
\frac{i}{2}
\sum_m
(E_{m\mathbf k}-E_{n\mathbf k})
\mathbf d_{nm}\times\mathbf d_{mn}
+\mathrm{h.c.}
\\
&=&
\frac{i}{2}
\sum_m
\left(
\langle u_{n\mathbf k}|i\nabla_{\mathbf k}u_{m\mathbf k}\rangle
E_{m\mathbf k}
-
E_{n\mathbf k}
\langle u_{n\mathbf k}|i\nabla_{\mathbf k}u_{m\mathbf k}\rangle
\right)
\times
\langle u_{m\mathbf k}|i\nabla_{\mathbf k}u_{n\mathbf k}\rangle
+\mathrm{h.c.}\nonumber
\\
&=&
\frac{i}{2}
\left[
\langle\nabla_{\mathbf k}u_{n\mathbf k}|
\left(
\sum_m
E_{m\mathbf k}
|u_{m\mathbf k}\rangle
\langle u_{m\mathbf k}|
\right)
\times
|\nabla_{\mathbf k}u_{n\mathbf k}\rangle
\right.
\nonumber\\
&&\left.
-
\langle\nabla_{\mathbf k}u_{n\mathbf k}|
E_{n\mathbf k}
\left(
\sum_m
|u_{m\mathbf k}\rangle
\langle u_{m\mathbf k}|
\right)
\times
|\nabla_{\mathbf k}u_{n\mathbf k}\rangle
\right]
+\mathrm{h.c.}
\end{eqnarray}
Employing the identity operator and the spectral representation of the Hamiltonian in the Hilbert space of periodic functions $|u_{n\mathbf k}\rangle$,

\begin{equation}
H_{\mathbf k}
=
\sum_m
E_{m\mathbf k}
|u_{m\mathbf k}\rangle
\langle u_{m\mathbf k}|,
\end{equation}
we find immediate reduction to the compact form

\begin{equation}
[{\mathbf L_{\rm LC}}]_{nn}
=
i\,
\langle\nabla_{\mathbf k}u_{n\mathbf k}|
(H_{\mathbf k}-E_{n\mathbf k})
\times
|\nabla_{\mathbf k}u_{n\mathbf k}\rangle .
\end{equation}
For an equilibrium band $n$, the corresponding coefficient section is a band-space vector with a single nonzero component,

\begin{equation}
\mathbf c_{\mathbf k}^{(n)}
=
\sqrt{f_{n\mathbf k}}\,\mathbf e_n
\label{eigu}
\end{equation}
where $\mathbf e_n$ denotes the unit vector in band space with components $(\mathbf e_n)_m=\delta_{mn}$, and $f_{n\mathbf k}$ provides bookkeeping of the occupied states relative to the Fermi level.

\begin{equation}
f_{n\mathbf k}
=
\begin{cases}
1, & E_{n\mathbf k}<\mu\\[2pt]
0, & E_{n\mathbf k}>\mu
\end{cases}
\end{equation}
The equilibrium expectation value is therefore obtained by evaluating the coefficient-section expectation value for each occupied band and summing over bands,

\begin{equation}
\langle {\mathbf L_{\rm LC}}\rangle
=
\frac{1}{V_{\rm BZ}}
\sum_n
\int_{\rm BZ}d\mathbf k\,
\big[\mathbf c_{\mathbf k}^{(n)}\big]^\dagger
{\mathbf L_{\rm LC}}
\mathbf c_{\mathbf k}^{(n)},
\end{equation}
and inserting the preceding expression for $\mathbf c_{\mathbf k}^{(n)}$, Eq.~\ref{eigu}, then directly yields the familiar equilibrium local circulation orbital angular momentum
\cite{XiaoShiNiu2005,
ThonhauserCeresoliVanderbiltResta2005,
CeresoliThonhauserVanderbiltResta2006,ShiVignaleXiaoNiu2007},

\begin{equation}
\langle {\mathbf L_{\rm LC}}\rangle
=
-\frac{1}{V_{\rm BZ}}
\sum_n
\int_{\rm BZ}d\mathbf k\,
f_{n\mathbf k}\,
\operatorname{Im}
\left[
\langle\nabla_{\mathbf k}u_{n\mathbf k}|
(H_{\mathbf k}-E_{n\mathbf k})
\times
|\nabla_{\mathbf k}u_{n\mathbf k}\rangle
\right].
\label{eq:L-LC-equilibrium}
\end{equation}
For the IC operator the explicit momentum-space derivative vanishes identically,

\begin{equation}
\nabla_{\mathbf k}\mathbf c_{\mathbf k}^{(n)}=0,
\end{equation}
and so the covariant derivative contributes only through the diagonal Berry connection. The expectation value of the IC operator may then immediately be written as

\begin{equation}
\langle {\mathbf L_{\rm IC}}\rangle
=
-\frac{1}{V_{\rm BZ}}
\int_{\rm BZ}d\mathbf k
\sum_n
f_{n\mathbf k}\,
\nabla_{\mathbf k}E_{n\mathbf k}
\times
\boldsymbol{\mathcal A}_n .
\end{equation}
To make further progress we follow the standard route
\cite{XiaoShiNiu2005,ThonhauserCeresoliVanderbiltResta2005,
CeresoliThonhauserVanderbiltResta2006}
and introduce the function

\begin{equation}
G_{n\mathbf k}
=
-\int_{E_{n\mathbf k}}^{\mu}
d\epsilon\,f(\epsilon),
\end{equation}
which upon differentiation gives

\begin{equation}
\nabla_{\mathbf k}G_{n\mathbf k}
=
f_{n\mathbf k}\,
\nabla_{\mathbf k}E_{n\mathbf k}.
\end{equation}
The IC expectation value may therefore be rewritten as

\begin{equation}
\langle {\mathbf L_{\rm IC}}\rangle
=
-\frac{1}{V_{\rm BZ}}
\int_{\rm BZ}d\mathbf k
\sum_n
\nabla_{\mathbf k}G_{n\mathbf k}
\times
\boldsymbol{\mathcal A}_n .
\end{equation}
At zero temperature, using $G_{n\mathbf k}=f_{n\mathbf k}(E_{n\mathbf k}-\mu)$ and

\begin{equation}
\nabla_{\mathbf k}\times
\left(
G_{n\mathbf k}\boldsymbol{\mathcal A}_n
\right)
=
\nabla_{\mathbf k}G_{n\mathbf k}
\times
\boldsymbol{\mathcal A}_n
+
G_{n\mathbf k}
\nabla_{\mathbf k}\times
\boldsymbol{\mathcal A}_n ,
\end{equation}
we find, using the fact that for a topologically trivial band admitting a globally smooth periodic gauge the Brillouin-zone integral of a total derivative vanishes,

\begin{equation}
\langle {\mathbf L_{\rm IC}}\rangle
=
\frac{1}{V_{\rm BZ}}
\int_{\rm BZ}d\mathbf k
\sum_n
f_{n\mathbf k}
(E_{n\mathbf k}-\mu)
\boldsymbol{\Omega}_{n\mathbf k},
\label{eq:L-IC-equilibrium}
\end{equation}
where the Berry curvature is defined by

\begin{equation}
\boldsymbol{\Omega}_{n\mathbf k}
=
\nabla_{\mathbf k}\times\boldsymbol{\mathcal A}_n
\end{equation}
This has precisely the Berry-curvature structure of the familiar equilibrium itinerant circulation term, but differs by a factor of 2 from the expectation value of the itinerant circulation \cite{ThonhauserCeresoliVanderbiltResta2005,
CeresoliThonhauserVanderbiltResta2006}. We now resolve this apparent factor-of-two discrepancy.

\subsection{Cell and bulk orbital angular momentum}

We first consider the expectation value of the total {$\mathbf L$}. For an insulating set of fully occupied topologically trivial bands, combining Eqs.~\eqref{eq:L-LC-equilibrium} and \eqref{eq:L-IC-equilibrium} yields

\begin{equation}
\langle {\mathbf L}\rangle
=
-\frac{1}{V_{\rm BZ}}
\sum_n
\int_{\rm BZ}d\mathbf k\,
f_{n\mathbf k}\,
\operatorname{Im}
\left[
\langle\nabla_{\mathbf k}u_{n\mathbf k}|
H_{\mathbf k}
\times
|\nabla_{\mathbf k}u_{n\mathbf k}\rangle
\right]
\end{equation}
{In obtaining this form, the term proportional to $\mu$ vanishes because the Brillouin-zone integral of $\boldsymbol{\Omega}_{n\mathbf k}$ is zero for the fully occupied topologically trivial bands considered here.}

For a topologically trivial isolated full band, these coefficient sections correspond to the same occupied subspace as the corresponding home-cell Wannier function,

\begin{equation}
|W_n\rangle
=
\frac{1}{V_{\rm BZ}}
\int_{\rm BZ}d\mathbf k\,
|\phi_{n\mathbf k}\rangle
\end{equation}
and indeed the preceding equation is the well-known equilibrium result for this case.

However, this is not equal to the total bulk expectation value of {$\mathbf L$}. To obtain this we first recall that under active translation, while LC transforms ``as a density'', i.e. {$\mathbf L_{\rm LC}(\mathbf r)\rightarrow \mathbf L_{\rm LC}(\mathbf r-\mathbf a_i)$}, the itinerant orbital angular momentum transforms with a ``lever arm'' term,

\begin{equation}
{\mathbf L_{\rm IC}^{\rm tot}}
\xrightarrow{T_{\mathbf a_i}}
{\mathbf L_{\rm IC}^{\rm tot}}
+
\mathbf a_i\times\mathbf v
\end{equation}
where $\v a_i$ represents a lattice vector.
As we are describing an infinite bulk, we must take recourse to Born--von K\'arm\'an (BvK) periodic boundary conditions. To this end we consider a BvK cell consisting of $2N+1$ unit cells along the lattice direction $\mathbf a_i$, labelled by $j=-N,\ldots,N$. While the BvK periodicity introduces no additional structure for {$\mathbf L_{\rm LC}$}, for {$\mathbf L_{\rm IC}$} we find a dramatic difference between the "bulk" and "boundary" terms of the BvK cell:

\begin{equation}
\left(\mathbf L_{\rm IC}^{\rm tot}\right)_j
\xrightarrow{T_{\mathbf a_i}}
\begin{cases}
\left(\mathbf L_{\rm IC}^{\rm tot}\right)_j
+
\mathbf a_i\times\mathbf v,
&
-N\leq j\leq N-1
\\[4pt]
\left(\mathbf L_{\rm IC}^{\rm tot}\right)_j
-
2N\mathbf a_i\times\mathbf v,
&
j=N
\end{cases}
\end{equation}

{Strikingly, while $2N$ of the unit cells transform according to the usual lever-arm law, the boundary cell does not, and instead produces a lever-arm change scaled by $2N$, exactly equal in magnitude and opposite in sign to the combined change of all the other cells. This boundary term is the BvK representation of the same position-dependent part of the angular-momentum operator whose unit-cell-position dependence in the Blount representation is carried entirely by $i\mathbf D_{\mathbf k}$. Redistributing the boundary contribution over the $2N$ non-boundary cells gives one additional copy of the same per-cell IC functional. The bulk per-cell IC is therefore twice the home-cell IC. At equilibrium this identification can be checked directly against the Wannier formulation: the home-cell Wannier moment contains our $\mathbf L_{\rm LC}+\mathbf L_{\rm IC}$, since the $(H-E)$ and $E$ pieces sum to $H$, while the boundary contribution of the modern theory supplies the second $E$, or IC, term~\cite{CeresoliThonhauserVanderbiltResta2006}.}

{This is the origin of the factor of 2 in the IC term, and is closely analogous to the argument employed in the Wannier derivation of equilibrium bulk orbital angular momentum. In the operator formalism here it appears through BvK boundary conditions together with the active-translation lever-arm rule of IC, which exposes a second ``boundary'' orbital angular momentum contribution. As the BvK cell is \emph{entirely arbitrary}, and $N$ may take any value, there is no rigid conceptual demarcation between ``bulk'' and ``boundary'' terms in IC; the distinction depends on how the BvK boundary is constructed.}

In terms of the four orbital angular momentum operators, the corresponding bulk operator is therefore

\begin{equation}
\mathbf L_{\rm bulk}
=
\mathbf L_{\rm LC}
+
\mathbf L_{\rm LC}^{\rm mix}
+
2\mathbf L_{\rm IC}
+
2\mathbf L_{\rm IC}^{\rm mix}
\label{eq:L-bulk}
\end{equation}
Employing the equilibrium expectation values for $\mathbf L_{\rm LC}$, Eq.~\eqref{eq:L-LC-equilibrium}, and $\mathbf L_{\rm IC}$, Eq.~\eqref{eq:L-IC-equilibrium}, derived in the previous section, this operator yields for the equilibrium state

\begin{equation}
\langle \mathbf L_{\rm bulk}\rangle
=
-\frac{1}{V_{\rm BZ}}
\sum_n
\int_{\rm BZ}d\mathbf k\,
f_{n\mathbf k}\,
\operatorname{Im}
\left[
\langle\nabla_{\mathbf k}u_{n\mathbf k}|
\left(
H_{\mathbf k}
+
E_{n\mathbf k}
-
2\mu
\right)
\times
|\nabla_{\mathbf k}u_{n\mathbf k}\rangle
\right]
\label{eq:L-bulk-equilibrium}
\end{equation}
exactly the familiar form of the bulk orbital angular momentum in the modern theory of orbital magnetism
\cite{XiaoShiNiu2005,
ThonhauserCeresoliVanderbiltResta2005,
CeresoliThonhauserVanderbiltResta2006,
ShiVignaleXiaoNiu2007}.

{Out of equilibrium, the same BvK doubling applies to the factorized lever-arm part of the IC sector. The intrinsic bulk orbital angular momentum is therefore}

\begin{equation}
{
\left\langle\mathbf L_{\rm bulk}\right\rangle_{\rm int}
=
\left\langle\mathbf L_{\rm bulk}\right\rangle
-
2\left\langle\mathbf r_{\rm intra}\right\rangle
\times
\left\langle\mathbf v\right\rangle
}
\label{eq:L-bulk-intrinsic}
\end{equation}
%


\subsection{Non-equilibrium orbital angular momentum}

In the non-equilibrium case the coefficient section is no longer given by the equilibrium form considered above but rather provides the initial condition for each initially occupied band $n$,

\begin{equation}
{\mathbf c_{\mathbf q}^{(n)}(t_0)}
=
{\sqrt{f_{n\mathbf q}}\,\mathbf e_n}
\end{equation}
with each such coefficient section then evolved dynamically under the laser pulse. It is convenient at this point to introduce the full Berry-connection matrix

\begin{equation}
\boldsymbol{\xi}
=
\boldsymbol{\mathcal A}
+
\mathbf d
\end{equation}
where $\boldsymbol{\mathcal A}$ and $\mathbf d$ are respectively the band-diagonal and off-diagonal parts introduced above. {The standard time-dependent Schr\"odinger equation in the instantaneous band basis, with each trajectory labelled by its initial crystal momentum $\mathbf q\equiv\mathbf k(t_0)$, may then be written as}

\begin{equation}
i\partial_t
{\mathbf c_{\mathbf q}^{(n)}(t)}
=
\left[
E(\mathbf k(t))
-
\dot{\mathbf k}(t)\cdot
\boldsymbol{\xi}(\mathbf k(t))
\right]
{\mathbf c_{\mathbf q}^{(n)}(t)}
\label{TDSEc}
\end{equation}
where $\mathbf k(t)=\mathbf q+\mathbf A(t)/c$.

As the laser pulse generates transitions to the other bands an initially band-diagonal coefficient section will, in general, evolve into a superposition vector,

\begin{equation}
{\mathbf c_{\mathbf q}^{(n)}(t)}
=
\sum_m
{c_{m\mathbf q}^{(n)}(t)}\,
\mathbf e_m
\end{equation}

We now consider the general features of the resulting expectation values of each of the four angular-momentum contributions, $\mathbf L_{\rm LC}$, $\mathbf L_{\rm LC}^{\rm mix}$, $\mathbf L_{\rm IC}$, and $\mathbf L_{\rm IC}^{\rm mix}$, for such a non-equilibrium state. Evidently, the local circulation operators will, as they operate pointwise on {$\mathbf c_{\mathbf q}$}, behave very differently from the itinerant circulation operators, which involve the covariant derivative acting on the non-equilibrium coherent phase structures encoded in {$\mathbf c_{\mathbf q}$}.

In the former case, the key non-equilibrium feature will be the appearance of off-diagonal terms $n\neq m$ in band space. In the field-free post-pulse regime these acquire factors of the form $e^{\,i(E_{n\mathbf q}-E_{m\mathbf q})t}$ and, for {generic dispersive non-degenerate bands}, dephase under Brillouin-zone integration so that their contribution washes out as $t\rightarrow\infty$. They therefore give rise to generally oscillatory and transient contributions to the dynamical orbital angular momentum, analogous to the interband current familiar from ultrafast dynamics.

In contrast, the IC terms are expected to generate, through the action of the covariant derivative on non-equilibrium momentum-dependent phase textures, entirely new contributions. For $\mathbf L_{\rm IC}^{\rm mix}$ the only non-zero terms are band off-diagonal, which will again lead to dynamical-phase oscillatory terms in the post-pulse evolution, and so might again be expected to generate only a transient orbital angular momentum. For this operator there is, however, an additional subtlety: the covariant derivative may act on the free-evolution phase and generate terms proportional to $t\,e^{i(E_{n\mathbf q}-E_{m\mathbf q})t}$. For isolated non-degenerate stationary points of the interband phase, the corresponding Brillouin-zone integral generically decays as $t^{-d/2}$, where $d$ is the Brillouin-zone dimensionality, and thus in two dimensions the explicit factor of $t$ can generate an $O(1)$ oscillatory contribution rather than a term that, post-pulse, dephases to zero. 

The $\mathbf L_{\rm IC}$ operator, on the other hand, is band-diagonal and can therefore, as we now show, \emph{generically} give rise to a dynamically generated residual quasi-stationary orbital angular momentum, analogous to the post-pulse residual current much explored in ultrafast dynamics.
{After the pulse $\mathbf A(t)=0$ and hence $\mathbf k=\mathbf q$; for notational simplicity we therefore revert to $\mathbf k$ in the field-free expressions below.}
Considering a general post-pulse coefficient section evolved from an initially occupied band $n$, with its component in band $m$ written as

\begin{equation}
c_{m\mathbf k}^{(n)}(t)
=
\sqrt{p_{m\mathbf k}^{(n)}}\,
\exp\!\left[
-iE_{m\mathbf k}t
+i\eta_{m\mathbf k}^{(n)}
\right]
\end{equation}
the action of $i\mathbf D_{\mathbf k}$ yields, after summing over the initially occupied sections $n$ and their band components $m$,

\begin{equation}
\left\langle
\mathbf L_{\rm IC}
\right\rangle
=
\frac{1}{V_{\rm BZ}}
\sum_{n\in{\rm occ}}
\sum_m
\int_{\rm BZ}d\mathbf k\,
p_{m\mathbf k}^{(n)}
\boldsymbol{\nabla}_{\mathbf k}E_{m\mathbf k}
\times
\left(
\boldsymbol{\nabla}_{\mathbf k}\eta_{m\mathbf k}^{(n)}
-
\boldsymbol{\mathcal A}_{m\mathbf k}
\right)
\end{equation}
Here we have used the fact that the contribution from the dynamical phase vanishes identically as $
\boldsymbol{\nabla}_{\mathbf k}E_{m\mathbf k}
\times
\boldsymbol{\nabla}_{\mathbf k}E_{m\mathbf k}
=
0
$
while the amplitude-gradient contribution is proportional to

\begin{equation}
-\frac{i}{2}
\boldsymbol{\nabla}_{\mathbf k}E_{m\mathbf k}
\times
\boldsymbol{\nabla}_{\mathbf k}p_{m\mathbf k}^{(n)}
\end{equation}
and therefore is a total derivative whose integral vanishes over the BZ torus. What therefore remains, in addition to the Berry-connection contribution already present in equilibrium, is an entirely new contribution to $\mathbf L_{\rm IC}$ involving only the post-pulse $k$-dependent phase texture $\eta_{m\mathbf k}^{(n)}$ dynamically imprinted onto the band manifolds during laser excitation. To explore this physics we now consider a simple minimal model: gapped graphene.

\section{Numerical example: gapped graphene}\label{sec:graphene}

\subsection{Gapped graphene model}

Our model consists of the minimal two-band gapped-graphene model, whose Hamiltonian can be written as

\begin{equation}
H(\mathbf k)
=
\begin{pmatrix}
\Delta & f(\mathbf k)\\
f^\ast(\mathbf k) & -\Delta
\end{pmatrix}.
\label{H}
\end{equation}
{Here $f(\mathbf k)=-t\sum_{j=1}^{3}e^{i\mathbf k\cdot\boldsymbol{\delta}_j}$ is the standard nearest-neighbour honeycomb structure factor, with $t$ the nearest-neighbour hopping and $\boldsymbol{\delta}_j$ the three nearest-neighbour vectors.} This staggered-honeycomb Hamiltonian represents both a standard minimal model for inversion-broken honeycomb systems, for example monolayer hexagonal boron nitride~\cite{Fuchs2010TopologicalBerry}, as well as the simplest model that exhibits light--valley selection rules under circularly polarized excitation~\cite{Yao2008ValleyOptics}.
This generates a pair of symmetric bands

\begin{equation}
\pm\varepsilon(\mathbf k)
=
\pm\sqrt{\Delta^2+|f(\mathbf k)|^2}
\end{equation}
with the direct band gap at the Dirac points therefore $E_g=2|\Delta|$. The corresponding eigenvectors are given by

\begin{equation}
|u_{\tau}(\mathbf k)\rangle
=
\frac{1}{\sqrt{2}}
\begin{pmatrix}
g_{\tau}(\mathbf k)e^{i\theta(\mathbf k)/2}\\
\tau g_{-\tau}(\mathbf k)e^{-i\theta(\mathbf k)/2}
\end{pmatrix},
\label{eq:phase-gauge}
\end{equation}
where $\tau=\pm1$ labels the two bands, and where we have defined the pseudospin mixing function

\begin{equation}
g_{\tau}(\mathbf k)
=
\sqrt{1+\frac{\tau\Delta}{\varepsilon(\mathbf k)}},
\label{gmix}
\end{equation}
and the phase angle $\theta(\mathbf k)$,

\begin{equation}
e^{i\theta(\mathbf k)}
=
\frac{f(\mathbf k)}{|f(\mathbf k)|}.
\end{equation}
From these can be derived the full Berry-connection matrix required for evaluating the orbital angular momentum operators defined in the previous section; these standard results and their derivation are provided in Supplemental Information Sec.~S5.

\subsection{Equilibrium and dynamical OAM in gapped graphene}

The four orbital angular momentum operators for gapped graphene may readily be obtained for this model; details of the calculations may be found in Supplemental Information Sec.~S5. Using the Pauli matrices $\sigma_i$ to encode the band space structure of our results, we find for the LC sector

\begin{equation}
\mathbf L_{\rm LC}(\mathbf k)
=
-\frac{\Delta}{\varepsilon(\mathbf k)}
\left[
\nabla_{\mathbf k}\varepsilon(\mathbf k)
\times
\nabla_{\mathbf k}\theta(\mathbf k)
\right]\sigma_0
\end{equation}

\begin{equation}
\mathbf L_{\rm LC}^{\rm mix}(\mathbf k)
= 0,
\end{equation}
and so the local circulation orbital angular momentum operator $\mathbf L_{\rm LC}$ is proportional to $\sigma_0$, a fact that represents a general consequence of the two-band structure. This is in concordance with the equality of the intrinsic equilibrium single Bloch state orbital magnetic moments of the conduction- and valence-band eigenstates reported for two-band models~\cite{Yao2008ValleyOptics,
Bhowal2021OrbitalHall}.
The mixed intra- and interband local circulation operator, on the other hand, vanishes identically, a finding that follows from the $k$-independent band centre, Eq.~\ref{H}.

The itinerant circulation sector reveals a much richer band-space structure, with

\begin{equation}
\mathbf L_{\rm IC}(\mathbf k)
=
\frac{\Delta}{2\varepsilon(\mathbf k)}
\left[
\nabla_{\mathbf k}\varepsilon(\mathbf k)
\times
\nabla_{\mathbf k}\theta(\mathbf k)
\right]\sigma_0
-i\left[
\nabla_{\mathbf k}\varepsilon(\mathbf k)
\times
\nabla_{\mathbf k}
\right]\sigma_z.
\label{LICgg}
\end{equation}
This operator therefore differs from the local circulation operators in containing a differential term that (i) couples directly to dynamically generated phase texture via the coefficient section and, (ii), through its $\sigma_z$ structure distinguishes the valence- and conduction-band sectors in their response to light, in sharp contrast to the band-independent coupling of $\mathbf L_{\rm LC}$.

Finally, the mixed IC operator can be written as

\begin{equation}
\mathbf L_{\rm IC}^{\rm mix}(\mathbf k)
=
\begin{pmatrix}
0 & \bm{\Lambda}(\mathbf k) \\
\bm{\Lambda}^{\dagger}(\mathbf k) & 0
\end{pmatrix},
\end{equation}
where

\begin{equation}
\begin{aligned}
\bm{\Lambda}(\mathbf k)
={}&
\frac{\varepsilon(\mathbf k)}{2|f(\mathbf k)|}
\left[
\nabla_{\mathbf k}\varepsilon(\mathbf k)
\times
\nabla_{\mathbf k}\theta(\mathbf k)
\right]
\\
&+
\left[
i\frac{\Delta}{|f(\mathbf k)|}
\nabla_{\mathbf k}\varepsilon(\mathbf k)
-
|f(\mathbf k)|
\nabla_{\mathbf k}\theta(\mathbf k)
\right]
\times
\nabla_{\mathbf k}.
\end{aligned}
\end{equation}
This expression is valid on any smooth phase-gauge patch for which $|f(\mathbf k)|\neq0$, i.e. away from the Dirac points.

In equilibrium, the derivative term in $\mathbf L_{\rm IC}$ vanishes when acting on the equilibrium coefficient section, while $\mathbf L_{\rm IC}^{\rm mix}$ has no diagonal matrix elements, and so the remaining LC and IC contributions obey the pointwise identity

\begin{equation}
\left[
\mathbf L_{\rm LC}(\mathbf k)
\right]_{\tau\tau}
+
2
\left[
\mathbf L_{\rm IC}(\mathbf k)
\right]_{\tau\tau}
=
0
\label{eq:graphene-equilibrium-cancellation}
\end{equation}
wherever the phase-gauge expressions are defined. 
As noted in Sec.~\ref{4op}, $\mathbf{k}$-resolved expectation values in the IC sector are not themselves physical observables. Filled-band phase freedom,

\begin{equation}
c_{n\mathbf{k}}
\rightarrow
e^{i\eta_n(\mathbf{k})}c_{n\mathbf{k}},
\end{equation}
changes the $\mathbf{k}$-resolved expectation value of ${\mathbf L_{\rm IC}}$ by a total derivative,

\begin{equation}
\delta {\mathbf L_{\rm IC}}(\mathbf{k})
=
\nabla_{\mathbf{k}}E_n
\times
\nabla_{\mathbf{k}}\eta_n
=
\nabla_{\mathbf{k}}
\times
\left(
E_n\nabla_{\mathbf{k}}\eta_n
\right),
\end{equation}
whose integral over the Brillouin zone vanishes. The pointwise cancellation expressed by Eq.~\ref{eq:graphene-equilibrium-cancellation} therefore has no gauge-independent physical significance; its physical content is simply that the Brillouin-zone-integrated total orbital angular momentum vanishes in equilibrium, as previously reported for this model~\cite{Xiao2007Valley,
Fuchs2010TopologicalBerry,
Bhowal2021OrbitalHall}. {More generally, the momentum-resolved IC textures used below should be regarded as gauge-fixed diagnostics of the distribution of the response rather than independent observables; this qualification is especially important for filled-band phase textures, whereas the excited-state IC remains a useful diagnostic of where the laser-generated response is concentrated.} The "phase" gauge used here is therefore, in this sense, the representationally simplest gauge choice.

Out of equilibrium, however, this cancellation between the LC and IC sectors will evidently not occur: the bulk orbital angular momentum operator contains differential terms sensitive to a dynamically evolving state given by $2{\mathbf L_{\rm IC}} + 2{\mathbf L_{\rm IC}^{\rm mix}}$

\begin{equation}
\mathbf L_{\rm bulk}(\mathbf k)
=
-2i
\left[
\nabla_{\mathbf k}\varepsilon(\mathbf k)
\times
\nabla_{\mathbf k}
\right]\sigma_z
+
2\mathbf L_{\rm IC}^{\rm mix}(\mathbf k).
\label{eq:graphene-Lbulk-phase}
\end{equation}
{This contrast between the zero Brillouin-zone-integrated orbital angular momentum of the time-reversal-symmetric equilibrium state and the intrinsic non-equilibrium dynamics represented by Eq.~\ref{eq:graphene-Lbulk-phase} renders gapped graphene an ideal minimal model for probing orbital angular momentum dynamics out of equilibrium, and we now explore this model numerically.}

\subsection{Numerical method}

Our numerical method is based on direct time propagation of the time-dependent Schr\"odinger equation, Eq.~\ref{TDSEc}; however, it is numerically more convenient to transform from the instantaneous Bloch eigenbasis in which Eq.~\ref{TDSEc} is expressed to the fixed two-component Bloch basis in which the tight-binding Hamiltonian is written.

Defining

\begin{equation}
\mathbf a(\mathbf q,t)
=
U(\mathbf k(t))\,\mathbf c(\mathbf q,t)
\end{equation}
where the columns of $U$ are the instantaneous eigenvectors, the equation of motion becomes simply

\begin{equation}
i\partial_t\mathbf a(\mathbf q,t)
=
H(\mathbf k(t))\,\mathbf a(\mathbf q,t).
\label{eq:num-aprop}
\end{equation}
The propagation is therefore written entirely in terms of the field-free tight-binding Hamiltonian $H(\mathbf k)$ of Eq.~\ref{H}, with the initial condition for a state occupying band $n$ given by

\begin{equation}
\mathbf a(\mathbf q,0)=|u_{n\mathbf q}\rangle.
\end{equation}
For the driving field we employ a circularly polarized vector potential with Gaussian envelope given by

\begin{equation}
\mathbf A(t)
=
A_0
\exp\left[-\frac{(t-t_0)^2}{2\sigma^2}\right]
\left(
\sin(\omega t+\phi_g),
\cos(\omega t+\phi_g)
\right)
\end{equation}
where $A_0$ is the vector-potential amplitude, $\omega$ the central frequency, $\phi_g$ the carrier-envelope phase, $t_0$ the pulse centre, and $\mathrm{FWHM}=2\sqrt{2\ln2}\,\sigma$.

Calculation of orbital angular momentum additionally requires the covariant derivative, in turn requiring (i) the crystal-momentum derivative of the coefficient section and (ii) the band diagonal Berry connection. Our numerical strategy here will be to propagate the crystal-momentum derivative itself, rather than obtain this by
numerical differentiation of independently propagated trajectories, and to obtain the covariant derivative without explicit calculation of the gauge-dependent Berry connection. We now detail this approach.

The central object of interest is the crystal momentum derivative 

\begin{equation}
\mathbf G_i(\mathbf q,t)
=
\partial_{q_i}\mathbf a(\mathbf q,t),
\end{equation}
and differentiation of Eq.~\eqref{eq:num-aprop} then gives

\begin{equation}
i\partial_t\mathbf G_i
= H(\mathbf k(t))\,\mathbf G_i
+
\left(\partial_{k_i}H(\mathbf k(t))\right)\mathbf a,
\label{eq:num-Gprop}
\end{equation}
where we have used the fact that as $\mathbf k(t)=\mathbf q+\mathbf A(t)/c$, one has $\partial_{q_i}=\partial_{k_i}$ for the instantaneous band-space quantities.
For an initial condition

\begin{equation}
\mathbf G_i(\mathbf q,0)
=
\partial_{q_i}|u_{n\mathbf q}\rangle.
\end{equation}
we then propagate first Eq.~\ref{eq:num-aprop} and then subsequently Eq.~\ref{eq:num-Gprop}. To obtain the covariant derivative we note that from $\mathbf a=U\mathbf c$ and the full Berry connection $\boldsymbol{\xi}=\boldsymbol{\mathcal A}+\mathbf d$ we may write

\begin{equation}
iU^\dagger\partial_{k_i}U
=
\mathcal A_i+d_i,
\end{equation}
and so the derivative of the instantaneous-band coefficient section is given by

\begin{equation}
i\partial_{q_i}\mathbf c
=
iU^\dagger\mathbf G_i
-
(\mathcal A_i+d_i)\mathbf c.
\end{equation}
It then follows directly from Eq.~\eqref{eq:covariant-derivative} that

\begin{equation}
iD_i\mathbf c
= iU^\dagger\mathbf G_i - d_i\mathbf c,
\label{eq:num-iDc}
\end{equation}
and so the quantity required by the IC operators is evaluated directly from the propagated derivative $\mathbf G_i$ and the off-diagonal dipole matrix $d_i$; the diagonal Berry connection $\mathcal A_i$ need never be calculated separately.

While the "phase" gauge of Eq.~\ref{eq:phase-gauge} is useful for analytical work,
the Berry-connection and dipole objects entering the angular-momentum
operators are singular at the Dirac points and therefore unsuited
to numerical work. We instead introduce a ``smooth'' gauge (Supplemental Information Sec.~S6)

\begin{equation}
|u_+(\mathbf k)\rangle
=
\frac{1}{\sqrt{2\varepsilon(\varepsilon+\Delta)}}
\begin{pmatrix}
\varepsilon+\Delta\\
f^*
\end{pmatrix},
\qquad
|u_-(\mathbf k)\rangle
=
\frac{1}{\sqrt{2\varepsilon(\varepsilon+\Delta)}}
\begin{pmatrix}
-f\\
\varepsilon+\Delta
\end{pmatrix},
\label{eq:num-smooth-gauge}
\end{equation}
that is regular at $f(\mathbf k)=0$ for $\Delta>0$ (similar formulas may be obtained for the case $\Delta <0$).

There exists, however, an important subtlety in transforming gauge in this way: away from the zeros of
$f(\mathbf k)$, the "smooth" and "phase" gauges are related by

\begin{equation}
|u_+^{\rm sm}\rangle
=
e^{-i\theta(\mathbf k)/2}|u_+^{\rm ph}\rangle,
\qquad
|u_-^{\rm sm}\rangle
=
-e^{+i\theta(\mathbf k)/2}|u_-^{\rm ph}\rangle,
\end{equation}
and the corresponding compensating transformation of the coefficient section is therefore given by

\begin{equation}
c_+^{\rm sm}
=
e^{+i\theta(\mathbf k)/2}c_+^{\rm ph},
\qquad
c_-^{\rm sm}
=
-e^{-i\theta(\mathbf k)/2}c_-^{\rm ph}.
\end{equation}
Thus, given a smooth initial condition in the "phase" gauge, the transformation has simply
shifted the Dirac-point phase singularity from the basis functions into the initial condition of the coefficient section in the "smooth" gauge.

Here, however, we can appeal to the equilibrium phase freedom of a completely filled band. Multiplication of the occupied coefficient
section by an arbitrary phase $e^{i\eta(\mathbf k)}$ leaves all physical
equilibrium observables unchanged. After changing to the smooth Bloch basis, we may therefore \emph{reset this compensating coefficient phase to unity} rather than carry its singularity into the initial condition. In the present case, however, the transformation
$\eta(\mathbf k)$ is evidently singular, and the proof of equivalence requires a patchwise
application of Stokes' theorem: the Brillouin-zone manifold is
punctured at $K$ and $K'$, and the contribution from the branch cut
connecting the two valleys also retained. Details of this calculation are given in Supplemental Information Sec.~S7.

Having established this, we are free to use the smooth eigenstates of Eq.~\eqref{eq:num-smooth-gauge} together with the simple equilibrium initial condition $\mathbf c_{\mathbf q}^{(n)}(0)=\mathbf e_n
$ for each occupied band. The reset of the filled-band initial phase is also legitimate for the subsequent dynamics: it changes the initial Slater determinant only by an overall phase, and linear unitary time evolution preserves this physical equivalence for the initially occupied band. In this way phase singularities are removed entirely from the numerical treatment at almost no cost to the compactness of the underlying formalism.

\emph{Numerical parameters}: Finally, we report the essential numerical parameters of the TB calculations that follow. Calculation of ${\mathbf L_{\rm IC}}$ and the "mixed" intra- and inter-band IC, ${\mathbf L_{\rm IC}^{\rm mix}}$, turn out to require markedly different convergence of the k-mesh, with the former essentially converged at a grid of $150\times150$ in the BZ, while the latter requires $500\times500$. The time step for the dynamics we take to be 0.025~fs.

\subsection{Transient orbital angular momentum during ultrafast excitation}

\begin{figure*}[t]
\centering
\includegraphics[width=\textwidth]{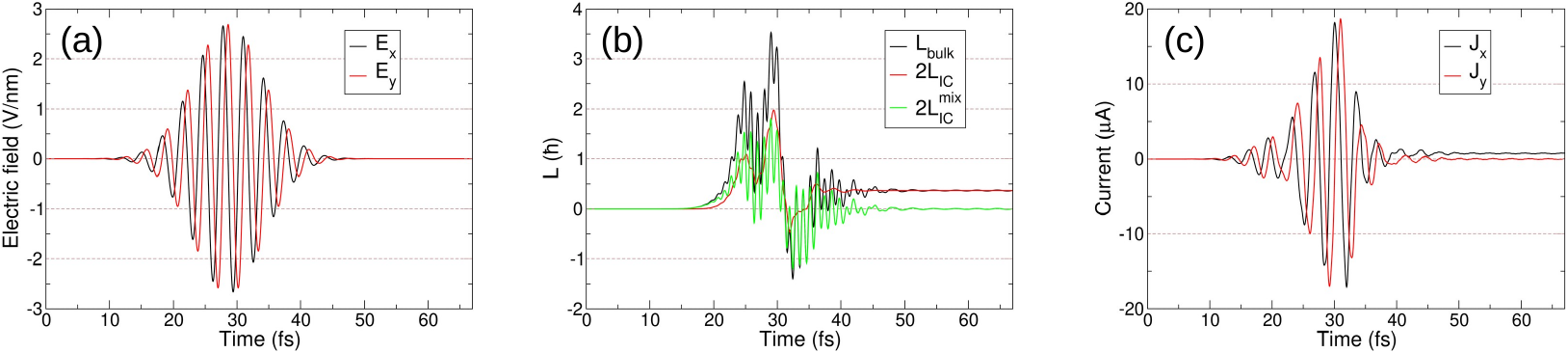}
\caption{{\textbf{Ultrafast generation of giant transient orbital angular momentum in gapped graphene:} A circularly polarized driving field, (a), generates a dynamically changing bulk orbital angular momentum in gapped graphene, (b), where we show both the bulk orbital angular momentum (${\mathbf L_{\rm bulk}}$) as well as its two components, the dynamical "itinerant circulation" (IC) ($2{\mathbf L_{\rm IC}}$) and the "mixed" intra- and inter-band IC ($2{\mathbf L_{\rm IC}^{\rm mix}}$). During the pulse-driven excitation a significant transient orbital angular momentum of nearly $4\hbar$ is generated, corresponding in magnitude to an orbital magnetic moment of $4\mu_B$, which, post-pulse, drops to nearly zero (i.e. back to the ground state orbital angular momentum which is identically zero in gapped graphene). (c) The corresponding total current, which again shows a significant transient peak during laser excitation.}}
\label{fig1}
\end{figure*}

We first consider the case of a gap of $1.28$~eV and apply a gap-tuned circularly polarized laser pulse with a pulse duration (full width at half maximum) of $12.95$~fs and a peak electric field of $2.69$~V/nm; the corresponding transient electric field is presented in Fig.~\ref{fig1}(a). In dramatic contrast to the zero bulk orbital angular momentum ${\mathbf L_{\rm bulk}}$ of the ground state, a significant light-induced growth of ${\mathbf L_{\rm bulk}}(t)$ is immediately apparent, peaking at approximately $4\hbar$ close to the pulse maximum. Subsequent to the pulse peak this rapidly decreases, leaving a small residual post-pulse ${\mathbf L_{\rm bulk}}$.

\begin{figure*}[t]
\centering
\includegraphics[width=0.7\textwidth]{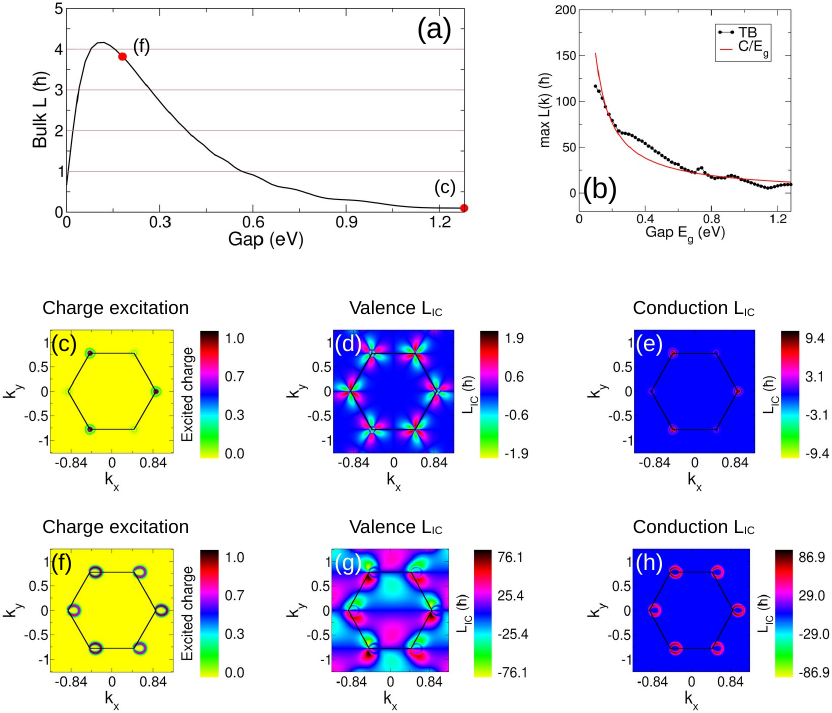}
\caption{\textbf{Gap control over light-created post-pulse orbital angular momentum:} Laser excitation generates a quasi-stationary post-pulse orbital angular momentum (OAM) that is strongly gap dependent, panel (a). Examination of the momentum resolved OAM texture reveals that the maximum post-pulse  itinerant orbital angular momentum ${\mathbf L_{\rm IC}}$ scales approximately as $C/E_g$, a fact that has its origin in the pseudospin texture of the gapped Dirac point, as described in the text. (c-e) Momentum-resolved excited charge (c), and the corresponding valence- and conduction-band itinerant orbital angular momentum ${\mathbf L_{\rm IC}}$ (d,e), for the large gap case marked (c) in figure panel (a).
Note that while the conduction band ${\mathbf L_{\rm IC}}$ follows closely the excited charge, this is not so for the valence band. This arises as, under the filled-band phase freedom discussed in Sec.~2.1, the  valence-band $k$-resolved IC texture is representation dependent, with only the integrated IC representation invariant. The valence texture shown here is thus imprinted with a "Houston phase" contribution that, as a total derivative, integrates to zero over the BZ torus.
Panels (f-h) present the same quantities for the small-gap case chosen at the limit of the single-cycle regime, showing the dramatic enhancement and restructuring of the light-written orbital texture as the gap is reduced.}
\label{fig2}
\end{figure*}

This transient signal exhibits pronounced oscillations, a feature that arises from the mixed itinerant contribution $2{\mathbf L_{\rm IC}^{\rm mix}}$, while the band-diagonal contribution $2{\mathbf L_{\rm IC}}$ presents a comparatively smooth temporal function. Comparison of Figs.~\ref{fig1}(b) and \ref{fig1}(c) reveals, furthermore, striking similarities between the temporal structures of ${\mathbf L_{\rm bulk}}(t)$ and the laser-driven current, with the current  obtained simply by multiplying the calculated two-dimensional sheet-current density by the lattice constant $a$.

Such transient behaviour is well known in ultrafast current dynamics, where the interband contribution to current typically generates a strongly oscillating and transient signal~\cite{MotlaghApalkov2021,Faisal2022Graphene}. As a consequence of interband coherence and post-pulse decay of the integrated observable via increasingly rapid BZ oscillation of $e^{i\Delta E t}$ phases as $t$ increases, such transient behaviour seems almost inevitable, and indeed signatures of transient light-induced orbital responses during femtosecond excitation have been reported in real-time tight-binding, density-matrix, and real-time TDDFT calculations~\cite{BuschZiolkowskiMertigHenk2023,
HeFrauenheimLi2024,MrudulOppeneer2026,
DarmawanGrunerPentcheva2026,
SharmaGillKrishnaHarrisLeeDewhurstShallcross2026}. The key distinction of the present approach and gapped graphene model is that the "local circulation" orbital angular momentum is identically zero throughout the dynamics, and so the transient signal is generated entirely by dynamical "itinerant circulation", which can further be identified with a dynamical \emph{coupling} of the intra- and inter-band sectors in $2{\mathbf L_{\rm IC}^{\rm mix}}$. 

\subsection{Light-created quasi-stationary orbital angular momentum}

The existence of a post-pulse residual in Fig.~\ref{fig1}, which we term \({\mathbf L_{\rm bulk}^{\rm res}}\), points towards the possibility of direct control of crystal orbital angular momentum by laser light. To explore this we now consider how \({\mathbf L_{\rm bulk}^{\rm res}}\) depends both on (i) the magnitude of the avoided crossing gap $E_g = 2|\Delta|$ and, (ii), the parameters of the light waveform that drive the excitation. 
To that end we vary the gap \(E_g\) from the value of \(1.28\)~eV down to \(1\)~meV with the central frequency tuned in each case to the sub-gap value $0.78E_g$, but the fluence (0.6~mJ/cm$^2$) and duration (8.2~fs) held fixed across the scan.
This reduction of the gap generates a pronounced increase in post-pulse orbital angular momentum \({\mathbf L_{\rm bulk}^{\rm res}}\), Fig.~\ref{fig2}(a), rising from a comparatively small value of $0.41\hbar$ at $E_g=1.28$~eV to more than $4\hbar$. 

The variation of the gap and, at the same time, the nominal central frequency $\omega$ of the gap tuned pulse, implies $\omega\to0$ as the gap closes. As the pulse fluence is held fixed this implies (i) a continuous decrease in the ratio of the pulse FWHM to the oscillation period $T$, and (ii) a compensating  increase in the vector potential amplitude to retain the fixed fluence. Thus Fig.~\ref{fig2}(a) inevitably entangles two effects: the intrinsic change in OAM due to the changing gap and the impact of the changing light waveform. In fact it is the latter that underpins the vanishing of OAM in the zero gap limit: in the sub-cycle limit circularly polarized light goes over to approximate linear polarization, and this cannot generate OAM. We therefore now address these two effects in turn, first establishing what turns out to be the dominant intrinsic gap dependence from the valley pseudospin texture, before examining the secondary role of the pulse form itself.

\subsubsection{Gap dependence of light created OAM}

As ${\mathbf L_{\rm IC}^{\rm mix}}$ dephases to zero in the field-free long-time limit, while the multiplicative part of $2{\mathbf L_{\rm IC}}$ cancels ${\mathbf L_{\rm LC}}$, the residual bulk orbital angular momentum is governed entirely by the dynamical part of ${\mathbf L_{\rm IC}}$ that couples to the covariant derivative. Our analysis may therefore focus on this object alone. Introducing momentum $\boldsymbol{\kappa}=\mathbf k-\mathbf K_\nu$ measured from valley $\nu$, the low-energy Dirac--Weyl Hamiltonian for one chosen valley can be written as

\begin{equation}
H_D(\boldsymbol{\kappa}) =
\mathbf h(\boldsymbol{\kappa})\cdot\boldsymbol{\sigma}
\end{equation}
where

\begin{equation}
\mathbf h(\boldsymbol{\kappa})
= \left(
v_F\kappa_x,\,
v_F\kappa_y,\,
\Delta
\right)
\end{equation}
and the low energy spectrum given by $\varepsilon(\boldsymbol{\kappa})=\sqrt{\Delta^2+v_F^2|\boldsymbol{\kappa}|^2}$. The conduction-band component of the differential part of the single-IC operator, Eq.~\ref{LICgg}, becomes

\begin{equation}
\left[{\mathbf L_{\rm IC}^{\rm diff}}\right]_{++,z}
=
-\mathrm{i}
\left[
\nabla_{\boldsymbol{\kappa}}\varepsilon
\times
\nabla_{\boldsymbol{\kappa}}
\right]_z
=
-\mathrm{i}
\frac{v_F^2}{\varepsilon(\boldsymbol{\kappa})}
\partial_\phi
\label{eq:LIC-dirac}
\end{equation}
where $\phi=\arg(\kappa_x+i\kappa_y)$. This yields immediately a characteristic valley orbital angular momentum scale of $2v_F^2/E_g$ that increases with decreasing gap. To probe this rule-of-thumb scale we fit {the maximum conduction-band ${\mathbf L_{\rm IC}}(\mathbf k)$} in the Brillouin zone by a $C/E_g$ function, Fig.~\ref{fig2}(b), revealing qualitatively good agreement, with $C=3.4v_F^2$, close to the natural scale of $2v_F^2$.

The physical origin of this enhancement can be made particularly transparent by considering the changes in Dirac-Weyl pseudospin for a dynamical trajectory that impinges on the low energy valley. Along, for example, $\kappa_y=0$, we have

\begin{equation}
\mathbf h(\kappa_x)
=
\left(
v_F\kappa_x,\,
0,\,
\Delta
\right)
\end{equation}
and thus as the laser pulse drives the crystal momentum through the valley region, the pseudospin rotates by a phase of $\pi$ with, as may readily be seen from the pseudospin mixing function Eq.~\ref{gmix}, this rotation confined to a momentum scale $v_F|\kappa_x|\sim|\Delta|=E_g/2$. The corresponding momentum-space texture therefore varies on the scale $|\Delta|/v_F$, so that its derivative supplies a characteristic length scale $v_F/|\Delta|=2v_F/E_g$. Sampling this by a band velocity of order $v_F$ then yields the orbital angular momentum scale $2v_F^2/E_g$, exactly as found directly from Eq.~\ref{eq:LIC-dirac}. The gap dependence of the post-pulse OAM thus has its origin in the increasingly rapid variation of valley pseudospin texture as $E_g\to0$ that the dynamical itinerant circulation captures via its covariant derivative, establishing a direct link between momentum space texture and light created orbital angular momentum.

{Interestingly, a longer-lived orbital-magnetization component has also recently been reported in real-time TDDFT calculations of optically driven SrTiO$_3$~\cite{DarmawanGrunerPentcheva2026}, suggesting a possible manifestation of the effect we describe here. In current state-of-the-art first-principles theory, however, the  non-equilibrium OAM is evaluated only as a muffin-tin-projected local quantity
\cite{SharmaGillKrishnaHarrisLeeDewhurstShallcross2026} rather than as a bulk periodic-solid operator.}

\begin{figure*}[t!]
\centering
\includegraphics[width=\textwidth]{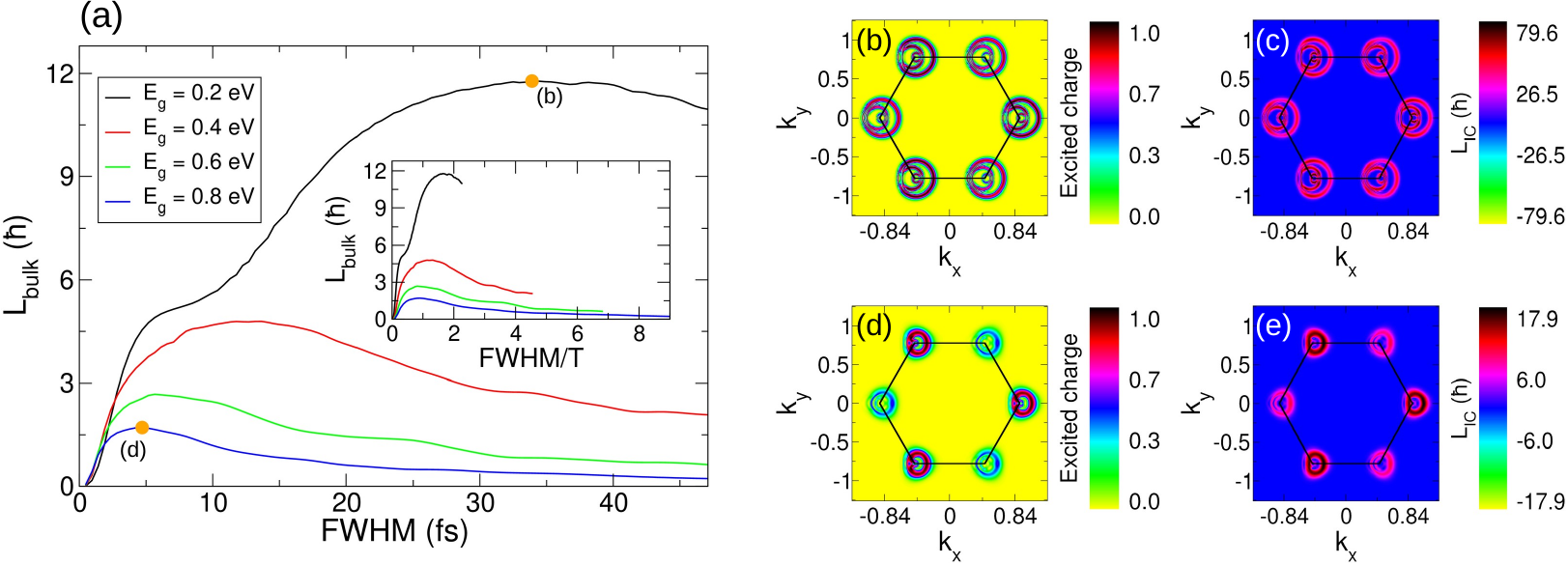}
\caption{\textbf{Control of light created orbital angular momentum via laser pulse parameters}. (a) Post-pulse bulk orbital angular momentum in gapped graphene ${\mathbf L_{\rm bulk}^{\rm res}}$ as a function of pulse full width at half maximum (FWHM) for four representative gaps: $E_g=0.2$, $0.4$, $0.6$, and $0.8$~eV. In the long time multi-cycle limit the light induced OAM substantially decreases. Maximum OAM creation is associated with the few cycle limit, as is made clear by the inset plot, in which the same data is presented against FWHM measured in optical cycles, FWHM/$T$. As the pulse fluence is held fixed over the FWHM scan the few cycle limit implies larger vector potential amplitude, allowing intense valley OAM to be transported out of the valley by dynamical evolution of non-valley crystal momenta through the valley active region. This underpins the very large $~12\hbar$ OAM found for the 0.2~eV gap, indicated by the orange point in panel (a), whose momentum resolved conduction band charge and "itinerant circulation" OAM show a clear imprint of such Houston trajectories, panels (b) and (c) respectively. A similar effect underpins the maximum OAM found for the 0.8~eV gap, also marked in panel (a), whose conduction band charge and OAM are presented in panels (d) and (e) respectively.}
\label{fig3}
\end{figure*}

In the ultrafast regime, laser light may possess a vector-potential amplitude \(\v A(t)/c\) comparable to Brillouin-zone dimensions, suggesting that such pulses can access a substantial fraction of the Brillouin zone. To see this we consider the momentum-resolved \(\mathbf L_{\rm IC}(\mathbf k)\) for two representative gaps, labelled by the points in Fig.~\ref{fig2}(a). For the large-gap case, which has a comparatively smaller \(\v A/c\), the excitation remains confined to one of the two inequivalent valleys, Fig.~\ref{fig2}(c-e), in which we show respectively the excited charge and \(\mathbf L_{\rm IC}(\mathbf k)\) in the valence and conduction manifolds. Reduction of the gap brings about a corresponding reduction in the pulse central frequency and concomitant increase in \(\v A/c\), yielding strikingly different momentum-resolved physics: each valley becomes surrounded by a loop of excited charge. These loops exactly track the Houston trajectories connecting points distant from the valley to the "hot" valley region. A corresponding loop of excited \({\mathbf L_{\rm IC}}\) in the conduction band, Fig.~\ref{fig2}(h), can clearly be seen. Although the detailed momentum-resolved IC texture is not by itself an observable, the fixed-gauge excited-state pattern provides a useful diagnostic showing that trajectories originating far from the valley can sample the large $v_F^2/|\Delta|$ valley scale and carry the associated response over an extended region of momentum space. 

Finally, we note a surprising -- and at first glance counter-intuitive -- feature of the momentum-resolved ${\mathbf L_{\rm IC}}$ presented in Fig.~\ref{fig2}: the valence-band texture possesses an approximately $C_3$ valley structure of alternating positive and negative lobes, extending far outside the regions in which charge excitation is confined. Even in the absence of interband excitation, laser-driven evolution of the crystal momentum imprints a phase texture on the occupied valence manifold which, post-pulse, may be written as

\begin{equation}
\eta_v(\mathbf q)
=
-\int_0^{t_f} dt\,E_v[\mathbf k(t)]
+\gamma_v(\mathbf q),
\label{eq:valence-phase-postpulse}
\end{equation}
where $t_f$ is a fixed post-pulse reference time and $\gamma_v(\mathbf q)$ is the Berry phase accumulated along the Houston trajectory. The corresponding filled-band reference contribution to ${\mathbf L_{\rm IC}}(\mathbf k)$ is, however, a total derivative,

\begin{equation}
\delta {\mathbf L_{\rm IC}}(\mathbf k)
=
\nabla_{\mathbf k}E_v\times\nabla_{\mathbf k}\eta_v
=
\nabla_{\mathbf k}\times
\left(
E_v\nabla_{\mathbf k}\eta_v
\right),
\end{equation}
and therefore vanishes upon integration over the Brillouin zone. The large valence-band textures visible in Fig.~\ref{fig2}(d) and (g) are thus fingerprints of this representation-dependent filled-band reference contribution to ${\mathbf L_{\rm IC}}$, and {their detailed momentum-space form carries no independent gauge-invariant physical meaning.}

\subsubsection{Pulse parameter dependence of light created OAM}

To disentangle the intrinsic effect of the gap on the light-induced orbital angular momentum from the physics of the light pulse itself, we now scan pulse duration for several representative gaps, Fig.~\ref{fig3}(a). In each case the same gap-tuned central frequency, $\hbar\omega=0.78E_g$, is retained and the pulse fluence is held fixed at $F=0.6~\mathrm{mJ\,cm^{-2}}$; increasing the FWHM therefore trades pulse amplitude for duration. As may be seen, each gap exhibits a broad maximum in the residual post-pulse orbital angular momentum ${\mathbf L_{\rm bulk}^{\rm res}}$, but both the magnitude of this maximum and the pulse duration at which it occurs increase strongly as the gap is reduced. The data presented in Fig.~\ref{fig2} correspond to a fixed-FWHM cut at $8.2$~fs through the pulse-duration dependence shown here. Particularly for smaller gaps, therefore, a substantially larger light-induced orbital angular momentum can be obtained by increasing the pulse duration towards its optimum value.

Interestingly, plotting the same data against the number of optical cycles contained within the pulse, FWHM/$T$, where $T$ is the period associated with the pulse central frequency, brings the maxima for the
four gaps into much closer alignment. This exposes a competition between two effects: as the pulse is shortened at fixed fluence, its increasing
vector-potential amplitude allows trajectories originating over a larger region of momentum space to reach the ``hot'' valley region. Increasing
the pulse duration, on the other hand, allows the driven state to sample the large valley orbital angular momentum scale over a longer interval.
The optimum pulse duration therefore reflects a balance between these effects, coupled to the intrinsic strength of the valley orbital angular momentum response: as this
intrinsic effect weakens with increasing gap, the optimum moves progressively towards the single-cycle regime.

\section{Outlook}\label{sec:outlook}

The central result of this work is that orbital angular momentum in a periodic solid admits a general operator formulation. The same operators that recover the modern theory of orbital magnetism in the equilibrium limit reveal, in a driven state, qualitatively new physics including a persistent light-induced change in OAM from its ground-state value. This latter effect arises as the OAM operators split, analogously to the expectation values of the modern equilibrium theory, into two fundamentally distinct sectors: local circulation (LC) operators that act {\it pointwise} in momentum space on the coefficient section of the Bloch bundle, while itinerant circulation (IC) operators that act through the covariant derivative on the coefficient section. This establishes a remarkable connection between the phase structures written onto a band manifold during coherent laser excitation and the resulting orbital angular momentum of the crystal, opening a route towards rich control of OAM in solids using ultrafast strong-field light pulses.

Employing gapped graphene as an example we find a direct connection between the momentum scale on which valley pseudospin texture changes, and the resulting dynamical IC orbital angular momentum. This manifests as a strong gap dependence to the magnitude of the light created OAM, with the small gap limit favouring OAM production. Optimization of this dynamically created OAM we show requires pulses of large vector potential, that allow evolution of a general BZ crystal momentum through this valley localized pseudospin texture.

The operator approach described here is formulated within the crystal momentum representation (CMR) introduced by Adams and Blount, in which the action of operators on quantum states of a periodic solid is transferred to an action on the coefficient section of the Bloch bundle, while preserving the underlying operator algebra. This approach is not restricted to topologically trivial bands: in the presence of a topological obstruction the position operator retains its local form $\mathbf r=i\mathbf D_{\mathbf k}+\mathbf d$ on each smooth patch, and the same remains true for composite operators containing $\mathbf r$ such as orbital angular momentum, with topology entering only through the global patching of these local representations. Given the intimate connection between dynamical IC orbital angular momentum and momentum space phase textures, this generalization may be useful in the calculation of laser induced OAM in topologically non-trivial materials.

A second consequence of the crystal momentum representation for OAM is that it admits a fully $U(N)$-covariant formulation, within which the LC/IC decomposition becomes a decomposition with respect to spectral subspaces: $\v d$, the "lever arm" of the LC sector, connects distinct spectral subspaces, while $i\mathbf D$, the centre-coordinate operator of the IC sector, parallel-transports within them. In this sense the operator level distinction between local and itinerant circulation is more fundamental than the intra- and inter-band language tied to a particular Bloch eigenstate gauge.

Finally, the operator viewpoint expounded here suggests several immediate directions for fruitful investigation beyond OAM itself. In particular, it may provide a kinematic foundation for a corresponding theory of orbital angular momentum currents, with OAM-current operators constructed directly from the underlying OAM operators themselves. Fundamental questions concerning the nature of orbital transport may therefore be addressed on this basis, alongside more experimentally immediate challenges -- not least why OAM currents appear to survive for timescales considerably longer than can presently be explained by theory.

\section*{Acknowledgements}

Sharma and Shallcross would like to thank the Leibniz Professorin Program (SAW P118/2021). The authors acknowledge the North-German Supercomputing Alliance (HLRN) for providing HPC resources that have contributed to the research results reported in this paper.


\clearpage

\setcounter{section}{0}
\setcounter{subsection}{0}
\setcounter{subsubsection}{0}
\setcounter{equation}{0}
\setcounter{figure}{0}
\setcounter{table}{0}
\renewcommand{\thesection}{S\arabic{section}}
\renewcommand{\theequation}{S\arabic{equation}}
\renewcommand{\thefigure}{S\arabic{figure}}
\renewcommand{\thetable}{S\arabic{table}}

\renewcommand{\v}[1]{\mathbf{#1}}
\renewcommand{\m}[1]{\bm{#1}}
\newcommand{\vk}{\mathbf k}
\newcommand{\vK}{\mathbf K}
\newcommand{\vL}{\mathbf L}
\newcommand{\vD}{\mathbf D}
\newcommand{\gradk}{\nabla_{\mathbf k}}
\newcommand{\Acal}{\boldsymbol{\mathcal A}}
\newcommand{\grad}{\nabla_{\vk}}

\begin{center}
{\Large\bfseries Orbital angular momentum in crystalline solids: non-equilibrium theory --- Supplemental information\par}
\vspace{1em}
Sangeeta Sharma$^{1,2}$, Peter Elliott$^{3}$, Samuel Shallcross$^{1*}$\\[0.5em]
{\small
$^1$Max-Born-Institute for Non-Linear Optics, Max-Born Strasse 2A, 12489 Berlin, Germany\\
$^2$Institute for Theoretical Solid-State Physics, Freie Universit\"at Berlin, Arnimallee 14, 14195 Berlin, Germany\\
$^3$Scientific Computing Department, Science and Technology Facilities Council UK Research and Innovation (STFC-UKRI), Rutherford Appleton Laboratory, Harwell Campus, Didcot OX11 0QX, United Kingdom\\
$^*$Corresponding author: phsss75@gmail.com}
\end{center}
\vspace{1em}

\section{Blount operator algebra}
\label{app:blount}

In the main paper we use the crystal momentum representation developed by Adams and Blount, in which quantum mechanical operators acting on a state expanded in Bloch states are instead represented as operators acting directly on the \emph{coefficient section} of the Bloch bundle \\cite{SI:Adams1953CrystalMomentum,SI:AdamsBlount1959,SI:Blount1962}. 

Three features of this approach are crucial here: (i) operator algebra is preserved, so that a sequence of operators $\hat O_1\cdots\hat O_N$ acting on the quantum state maps to a corresponding sequence $O_1\cdots O_N$ acting on the coefficient section, with the associated vector algebra inherited; (ii) the resulting operator representation is manifestly $U(1)$ gauge covariant; and (iii) while matrix elements involving Bloch states become ill defined when the operator sequence contains the position operator, the coefficient section representation avoids the direct use of such matrix elements.

Historically, Adams introduced the crystal momentum operator viewpoint, Adams and Blount used the same representation to derive anomalous velocity terms, and Blount subsequently presented the full operator calculus, including operator products and matrix valued bands \\cite{SI:Adams1953CrystalMomentum,SI:AdamsBlount1959,SI:Blount1962}. Closely related forms of the position operator and its generalized or covariant derivative have subsequently become standard in length gauge nonlinear optics and modern treatments of crystalline response \\cite{SI:AversaSipe1995,SI:Ventura2017,SI:Parker2019,SI:BonbienManchon2022}. 

Here we give an overview of the construction needed for orbital angular momentum in full: we introduce the formal operator map both for the position operator and for lattice translation invariant operators, establish closure under operator composition, derive the expectation value formula and the $U(1)$ gauge covariant structure used in the main text, and derive explicitly the velocity operator. The corresponding covariance under a general smooth $U(N)$ change of Bloch frame is treated separately in Sec.~\ref{app:UN}, while the extension to topologically non-trivial Bloch bundles and patchwise representations is presented in Sec.~\ref{app:chern}.

\subsection{Coefficient section representation}

Let

\begin{equation}
|\phi_{n\mathbf k}\rangle
=
e^{i\mathbf k\cdot\mathbf r}
|u_{n\mathbf k}\rangle
\label{app:eq:bloch-state}
\end{equation}
be a complete orthonormal set of Bloch eigenstates over the Brillouin zone. We adopt the normalization

\begin{equation}
\langle\phi_{n\mathbf k}|\phi_{m\mathbf k'}\rangle
=
V_{\rm BZ}
\delta_{nm}
\delta(\mathbf k-\mathbf k')
\label{app:eq:bloch-orthogonality}
\end{equation}
and therefore

\begin{equation}
\mathbf{1}
=
\frac{1}{V_{\rm BZ}}
\int_{\rm BZ}d\mathbf k
\sum_n
|\phi_{n\mathbf k}\rangle
\langle\phi_{n\mathbf k}|
\label{app:eq:bloch-completeness}
\end{equation}
so that an arbitrary state may be written as

\begin{equation}
|\Psi\rangle
=
\frac{1}{V_{\rm BZ}}
\int_{\rm BZ}d\mathbf k
\sum_n
|\phi_{n\mathbf k}\rangle
c_{n\mathbf k}
\label{app:eq:general-state}
\end{equation}
The basic observation of the Adams and Blount construction is that an operator acting on $|\Psi\rangle$ can be represented by an induced operator acting on $\mathbf c_{\mathbf k}$ \\cite{SI:Adams1953CrystalMomentum,SI:AdamsBlount1959,SI:Blount1962}. We will denote this correspondence by $B$.
Note that here we have introduced the shorthand notation of $\mathbf c_{\mathbf k}$ as the column vector -- the "coefficient section" -- whose components are $c_{n\mathbf k}$. Generally, we will employ a compact notation in which band indices are suppressed in favour of the implied indices of matrix-vector multiplication.

It should be emphasized that Blount's treatment is an operator representation rather than only a prescription for position matrix elements; any operator acting on $\ket{\Psi}$ induces an operator acting on the coefficient section, a distinction that is useful in the present context as orbital angular momentum is an antisymmetric product of the position and velocity operators.

For the position operator we begin from the derivative of the Bloch state,

\begin{equation}
\nabla_{\mathbf k}
|\phi_{n\mathbf k}\rangle
=
i\mathbf r
|\phi_{n\mathbf k}\rangle
+
e^{i\mathbf k\cdot\mathbf r}
\nabla_{\mathbf k}
|u_{n\mathbf k}\rangle
\label{app:eq:bloch-derivative}
\end{equation}
which gives

\begin{equation}
\mathbf r
|\phi_{n\mathbf k}\rangle
=
-i\nabla_{\mathbf k}
|\phi_{n\mathbf k}\rangle
+
e^{i\mathbf k\cdot\mathbf r}
i\nabla_{\mathbf k}
|u_{n\mathbf k}\rangle .
\label{app:eq:r-on-bloch}
\end{equation}
Insertion into Eq.~\eqref{app:eq:general-state} yields

\begin{eqnarray}
\mathbf r|\Psi\rangle
&=&
-\frac{i}{V_{\rm BZ}}
\int_{\rm BZ}d\mathbf k
\sum_n
\left(
\nabla_{\mathbf k}
|\phi_{n\mathbf k}\rangle
\right)
c_{n\mathbf k}
\nonumber
\\
&&+
\frac{1}{V_{\rm BZ}}
\int_{\rm BZ}d\mathbf k
\sum_n
e^{i\mathbf k\cdot\mathbf r}
i\nabla_{\mathbf k}|u_{n\mathbf k}\rangle
c_{n\mathbf k},
\label{app:eq:r-state}
\end{eqnarray}
and integrating the first term by parts gives

\begin{eqnarray}
-\frac{i}{V_{\rm BZ}}
\int_{\rm BZ}d\mathbf k
\sum_n
\left(
\nabla_{\mathbf k}
|\phi_{n\mathbf k}\rangle
\right)
c_{n\mathbf k}
&=&
\frac{i}{V_{\rm BZ}}
\int_{\rm BZ}d\mathbf k
\sum_n
|\phi_{n\mathbf k}\rangle
\nabla_{\mathbf k}c_{n\mathbf k}
\label{app:eq:integration-by-parts}
\end{eqnarray}
where the boundary term cancels between opposite faces of the Brillouin zone torus for a smooth periodic section. Here the relevant periodic object is the physical crystal momentum section $\sum_n|\phi_{n\mathbf k}\rangle c_{n\mathbf k}$; neither the coefficient section nor the Bloch frame is required separately to provide a globally smooth representative. When a globally smooth frame is obstructed, the same construction is understood patchwise, see Sec.~\ref{app:chern}; all local operator identities below remain unchanged.

The derivative of the periodic part may be resolved in the complete basis of periodic functions,

\begin{equation}
i\nabla_{\mathbf k}
|u_{n\mathbf k}\rangle
=
\sum_m
|u_{m\mathbf k}\rangle
\boldsymbol{\xi}_{mn}(\mathbf k)
\label{app:eq:xi-resolution}
\end{equation}
with the full Berry connection matrix

\begin{equation}
\boldsymbol{\xi}_{mn}(\mathbf k)
=
\langle u_{m\mathbf k}|
i\nabla_{\mathbf k}
|u_{n\mathbf k}\rangle .
\label{app:eq:xi-definition}
\end{equation}
Using Eq.~\eqref{app:eq:xi-resolution} in Eq.~\eqref{app:eq:r-state}, and combining the result with Eq.~\eqref{app:eq:integration-by-parts}, one then obtains

\begin{equation}
\mathbf r|\Psi\rangle
=
\frac{1}{V_{\rm BZ}}
\int_{\rm BZ}d\mathbf k
\sum_m
|\phi_{m\mathbf k}\rangle
\left[
i\nabla_{\mathbf k}\mathbf c_{\mathbf k}
+
\boldsymbol{\xi}(\mathbf k)\mathbf c_{\mathbf k}
\right]_m .
\label{app:eq:r-induced-full}
\end{equation}
The position operator is thus represented on the coefficient section by

\begin{equation}
B(\hat{\mathbf r})
=
i\nabla_{\mathbf k}
+
\boldsymbol{\xi}(\mathbf k).
\label{app:eq:r-blount-xi}
\end{equation}
It is useful to split the full connection into its diagonal and off diagonal parts,

\begin{equation}
\boldsymbol{\xi}(\mathbf k)
=
\boldsymbol{\mathcal A}(\mathbf k)
+
\mathbf d(\mathbf k)
\label{app:eq:xi-split}
\end{equation}
where

\begin{eqnarray}
\boldsymbol{\mathcal A}_{nm}(\mathbf k)
&=&
\delta_{nm}
\langle u_{n\mathbf k}|
i\nabla_{\mathbf k}
|u_{n\mathbf k}\rangle
\label{app:eq:A-definition}
\\
\mathbf d_{nm}(\mathbf k)
&=&
(1-\delta_{nm})
\langle u_{n\mathbf k}|
i\nabla_{\mathbf k}
|u_{m\mathbf k}\rangle .
\label{app:eq:d-definition}
\end{eqnarray}
Defining the band diagonal covariant derivative by

\begin{equation}
i\mathbf D_{\mathbf k}
=
i\mathbf{1}\nabla_{\mathbf k}
+
\boldsymbol{\mathcal A}(\mathbf k)
\label{app:eq:D-definition}
\end{equation}
the position operator takes the form used throughout the main text,

\begin{equation}
B(\hat{\mathbf r})
=
i\mathbf D_{\mathbf k}
+
\mathbf d(\mathbf k).
\label{app:eq:r-final}
\end{equation}
The distinction between these two pieces will be essential for orbital angular momentum: $\mathbf d$ acts multiplicatively at fixed $\mathbf k$, whereas $i\mathbf D_{\mathbf k}$ is a differential operator acting on everything to its right.

\subsection{Lattice translation invariant operators}

Consider now a Hilbert space operator $\hat O$ satisfying

\begin{equation}
[\hat O,\hat T_{\mathbf R}]
=
0
\label{app:eq:O-translation-invariant}
\end{equation}
for every lattice vector $\mathbf R$. Since

\begin{equation}
\hat T_{\mathbf R}
|\phi_{n\mathbf k}\rangle
=
e^{-i\mathbf k\cdot\mathbf R}
|\phi_{n\mathbf k}\rangle
\label{app:eq:translation-eigenvalue}
\end{equation}
we have

\begin{equation}
\begin{aligned}
\hat T_{\mathbf R}\hat O|\phi_{n\mathbf k}\rangle
&=\hat O\hat T_{\mathbf R}|\phi_{n\mathbf k}\rangle \\
&=e^{-i\mathbf k\cdot\mathbf R}\hat O|\phi_{n\mathbf k}\rangle ,
\end{aligned}
\label{app:eq:O-preserve-k-1}
\end{equation}
and thus $\hat O|\phi_{n\mathbf k}\rangle$ belongs to the same crystal momentum eigenspace, and may be expanded as

\begin{equation}
\hat O
|\phi_{n\mathbf k}\rangle
=
\sum_m
|\phi_{m\mathbf k}\rangle
O_{mn}(\mathbf k).
\label{app:eq:O-on-bloch}
\end{equation}
Acting on the general state then gives

\begin{equation}
\hat O|\Psi\rangle
=
\frac{1}{V_{\rm BZ}}
\int_{\rm BZ}d\mathbf k
\sum_m
|\phi_{m\mathbf k}\rangle
\left[
O(\mathbf k)\mathbf c_{\mathbf k}
\right]_m
\label{app:eq:O-general-action}
\end{equation}
and therefore we have

\begin{equation}
B(\hat O)
=
O(\mathbf k).
\label{app:eq:O-blount}
\end{equation}
Equivalently, after removing the Bloch phase one may write

\begin{equation}
O_{nm}(\mathbf k)
=
\langle u_{n\mathbf k}|
\hat O_{\mathbf k}
|u_{m\mathbf k}\rangle
\label{app:eq:O-matrix-element}
\end{equation}
with

\begin{equation}
\hat O_{\mathbf k}
=
e^{-i\mathbf k\cdot\hat{\mathbf r}}
\hat O
e^{i\mathbf k\cdot\hat{\mathbf r}} .
\label{app:eq:Ok-definition}
\end{equation}
For the specific case of the Hamiltonian,

\begin{equation}
\hat H_{\mathbf k}
|u_{n\mathbf k}\rangle
=
E_{n\mathbf k}
|u_{n\mathbf k}\rangle
\label{app:eq:Hk-eigenproblem}
\end{equation}
and hence, in the Hamiltonian eigenstate basis,

\begin{equation}
B(\hat H)
=
E(\mathbf k)
\label{app:eq:H-blount}
\end{equation}
with

\begin{equation}
E_{nm}(\mathbf k)
=
\delta_{nm}E_{n\mathbf k}.
\label{app:eq:E-diagonal}
\end{equation}
The elementary rules required in the main text can then be summarized as

\begin{center}
\begin{tabular}{c|c}
Hilbert space operator & Coefficient section operator \\
\hline
$\hat{\mathbf r}$ & $i\mathbf D_{\mathbf k}+\mathbf d(\mathbf k)$ \\
$\hat H$ & $E(\mathbf k)$ \\
$\hat O$ with $[\hat O,\hat T_{\mathbf R}]=0$ & $O(\mathbf k)$
\end{tabular}
\end{center}

\subsection{Closure under operator composition}

The crucial additional step is that the correspondence preserves operator products. (More precisely, this statement holds on the common domain on which the product $\hat O_1\hat O_2$ is defined.) Let $\hat O_1$ and $\hat O_2$ induce the coefficient section operators $O_1$ and $O_2$. If

\begin{equation}
\hat O_2|\Psi\rangle
=
|\Psi'\rangle
\label{app:eq:composition-1}
\end{equation}
then the coefficient section of $|\Psi'\rangle$ is

\begin{equation}
\mathbf c'_{\mathbf k}
=
O_2\mathbf c_{\mathbf k}.
\label{app:eq:composition-2}
\end{equation}
Acting next with $\hat O_1$ gives

\begin{equation}
\mathbf c''_{\mathbf k}
=
O_1\mathbf c'_{\mathbf k}
\label{app:eq:composition-3}
\end{equation}
and therefore

\begin{equation}
\mathbf c''_{\mathbf k}
=
O_1
\left(
O_2\mathbf c_{\mathbf k}
\right).
\label{app:eq:composition-4}
\end{equation}
Since the same state is obtained from $\hat O_1\hat O_2|\Psi\rangle$, the induced operator of the product is

\begin{equation}
B(\hat O_1\hat O_2)
=
B(\hat O_1)
B(\hat O_2).
\label{app:eq:composition-homomorphism}
\end{equation}
The product on the right hand side is operator composition. This point is important whenever the left factor contains $i\mathbf D_{\mathbf k}$: the derivative acts on both the $\mathbf k$ dependent matrices and the coefficient section appearing to its right. For example,

\begin{equation}
i\mathbf D_{\mathbf k}
\left(
O\mathbf c_{\mathbf k}
\right)
=
i
\left(
\nabla_{\mathbf k}O
\right)
\mathbf c_{\mathbf k}
+
O
i\nabla_{\mathbf k}\mathbf c_{\mathbf k}
+
\boldsymbol{\mathcal A}
O\mathbf c_{\mathbf k}.
\label{app:eq:D-product}
\end{equation}
This closure property is what permits composite observables such as $\hat{\mathbf r}\times\hat{\mathbf v}$ to be constructed directly on the coefficient section without introducing separate matrix element formulas for each composite operator. Position containing composite observables remain an active technical issue in periodic systems: related constructions have been developed for orbital angular momentum and orbital magnetization~\\cite{SI:Desmarais2023}, for proper spin current operators~\\cite{SI:Ma2024ProperSpin}, and, most recently, for a systematic evaluation of composite operators containing position in the Bloch representation~\\cite{SI:An2026PositionOperators}.

\subsection{Expectation values}

As the induced operator is defined by its action on the complete Bloch expansion, expectation values follow directly from Bloch orthogonality. If

\begin{equation}
\hat O|\Psi\rangle
=
\frac{1}{V_{\rm BZ}}
\int_{\rm BZ}d\mathbf k
\sum_n
|\phi_{n\mathbf k}\rangle
\left[
O\mathbf c_{\mathbf k}
\right]_n
\label{app:eq:O-action-for-expectation}
\end{equation}
then

\begin{equation}
\langle\Psi|\hat O|\Psi\rangle
=
\frac{1}{V_{\rm BZ}}
\int_{\rm BZ}d\mathbf k\,
\mathbf c_{\mathbf k}^{\dagger}
O
\mathbf c_{\mathbf k}.
\label{app:eq:expectation-value}
\end{equation}
When $O$ contains derivatives, these act on the coefficient section and on all $\mathbf k$ dependent objects to their right before the Brillouin zone integral is taken. The smooth periodic domain employed above ensures that integration by parts produces no boundary contribution, and a Hermitian Hilbert space operator therefore has the corresponding Hermitian realization with respect to the Brillouin zone inner product in Eq.~\eqref{app:eq:expectation-value}.

\subsection{\texorpdfstring{$U(1)$}{U(1)} gauge covariance}

The coefficient section representation evidently must not depend on the phase convention chosen for the Bloch eigenstates, i.e. it must be gauge covariant. Under a smooth band phase transformation,

\begin{eqnarray}
|u_{n\mathbf k}\rangle
&\rightarrow&
e^{i\chi_{n\mathbf k}}
|u_{n\mathbf k}\rangle
\label{app:eq:gauge-u}
\\
c_{n\mathbf k}
&\rightarrow&
e^{-i\chi_{n\mathbf k}}
c_{n\mathbf k}
\label{app:eq:gauge-c}
\end{eqnarray}
the physical state in Eq.~\eqref{app:eq:general-state} is unchanged. Introducing the diagonal unitary matrix

\begin{equation}
U_{\chi}(\mathbf k)
=
{\rm diag}
\left(
e^{i\chi_{1\mathbf k}},
e^{i\chi_{2\mathbf k}},
\ldots
\right)
\label{app:eq:Uchi}
\end{equation}
the coefficient section transforms as

\begin{equation}
\mathbf c_{\mathbf k}
\rightarrow
U_{\chi}^{\dagger}(\mathbf k)
\mathbf c_{\mathbf k},
\label{app:eq:c-matrix-gauge}
\end{equation}
and the full Berry connection matrix transforms as

\begin{equation}
\boldsymbol{\xi}
\rightarrow
U_{\chi}^{\dagger}
\boldsymbol{\xi}
U_{\chi}
+
iU_{\chi}^{\dagger}
\nabla_{\mathbf k}U_{\chi}.
\label{app:eq:xi-gauge}
\end{equation}
For the diagonal and off diagonal pieces this then gives

\begin{eqnarray}
\boldsymbol{\mathcal A}
&\rightarrow&
\boldsymbol{\mathcal A}
-
\nabla_{\mathbf k}\chi
\label{app:eq:A-gauge}
\\
\mathbf d
&\rightarrow&
U_{\chi}^{\dagger}
\mathbf d
U_{\chi},
\label{app:eq:d-gauge}
\end{eqnarray}
where $\nabla_{\mathbf k}\chi$ denotes the diagonal matrix with entries $\nabla_{\mathbf k}\chi_{n\mathbf k}$. Acting on the transformed coefficient section one finds

\begin{equation}
i\mathbf D'_{\mathbf k}
\mathbf c'_{\mathbf k}
=
U_{\chi}^{\dagger}
i\mathbf D_{\mathbf k}
\mathbf c_{\mathbf k}
\label{app:eq:D-gauge-covariance}
\end{equation}
and hence

\begin{equation}
\left(
i\mathbf D'_{\mathbf k}
+
\mathbf d'
\right)
\mathbf c'_{\mathbf k}
=
U_{\chi}^{\dagger}
\left(
i\mathbf D_{\mathbf k}
+
\mathbf d
\right)
\mathbf c_{\mathbf k}.
\label{app:eq:r-gauge-covariance}
\end{equation}
Thus the position operator is represented gauge covariantly on the coefficient section. Any translation invariant operator transforms pointwise as

\begin{equation}
O(\mathbf k)
\rightarrow
U_{\chi}^{\dagger}(\mathbf k)
O(\mathbf k)
U_{\chi}(\mathbf k)
\label{app:eq:O-gauge}
\end{equation}
and therefore the closure property in Eq.~\eqref{app:eq:composition-homomorphism} ensures covariance of every composite operator constructed from these elementary objects. The extension of this statement to a general smooth $U(N)$ change of Bloch frame is given in Sec.~\ref{app:UN}.

\subsection{Active lattice translations}

Gauge transformations change the representation of the same physical state. An active lattice translation changes the state itself. For a lattice vector $\mathbf a$,

\begin{equation}
\hat T_{\mathbf a}
=
e^{-i\hat{\mathbf p}\cdot\mathbf a}
\label{app:eq:active-T}
\end{equation}
and the Bloch property gives

\begin{equation}
\hat T_{\mathbf a}
|\phi_{n\mathbf k}\rangle
=
e^{-i\mathbf k\cdot\mathbf a}
|\phi_{n\mathbf k}\rangle,
\label{app:eq:active-T-bloch}
\end{equation}
and thus the translated state can be represented in the original Bloch basis by the coefficient section

\begin{equation}
\mathbf c_{\mathbf k}
\rightarrow
e^{-i\mathbf k\cdot\mathbf a}
\mathbf c_{\mathbf k}.
\label{app:eq:active-c}
\end{equation}
(equivalently the translation operator acting on coefficient sections is $T_{\v a} = e^{-i\v k\cdot\v a}$).
The multiplicative operators $\mathbf d$ and $E$ are unaffected,

\begin{eqnarray}
\mathbf d(\mathbf k)
&\xrightarrow{T_{\mathbf a}}&
\mathbf d(\mathbf k)
\label{app:eq:active-d}
\\
E(\mathbf k)
&\xrightarrow{T_{\mathbf a}}&
E(\mathbf k)
\label{app:eq:active-E}
\end{eqnarray}
while, by contrast, acting with the covariant derivative on the translated section gives

\begin{eqnarray}
i\mathbf D_{\mathbf k}
\left(
e^{-i\mathbf k\cdot\mathbf a}
\mathbf c_{\mathbf k}
\right)
&=&
e^{-i\mathbf k\cdot\mathbf a}
\left(
i\mathbf D_{\mathbf k}
+
\mathbf a
\right)
\mathbf c_{\mathbf k}
\label{app:eq:active-D-action}
\end{eqnarray}
so that

\begin{equation}
i\mathbf D_{\mathbf k}
\xrightarrow{T_{\mathbf a}}
i\mathbf D_{\mathbf k}
+
\mathbf a.
\label{app:eq:active-D}
\end{equation}

This is the origin of the distinct transformation properties used in the main text: $\mathbf d$ is cell internal, whereas $i\mathbf D_{\mathbf k}$ \emph{carries the unit cell position} under an active lattice translation.

\subsection{Velocity operator}

The velocity operator follows directly from the Blount representation of the position operator. Direct substitution into the commutator yields

\begin{align}
\mathbf v
&=-i[i\mathbf D_{\mathbf k}+\mathbf d,E]
\nonumber\\
&=\nabla_{\mathbf k}E-i[\mathbf d,E],
\label{app:eq:velocity-final}
\end{align}
where $[i\mathbf D_{\mathbf k},E]=i\nabla_{\mathbf k}E$ in the Hamiltonian eigenstate frame. Thus the diagonal and off diagonal parts are respectively the band velocity $\nabla_{\mathbf k}E$ and the interband contribution $-i[\mathbf d,E]$. The derivative acting on the coefficient section cancels in the commutator, and so $\mathbf v$ is multiplicative in $\mathbf k$.

The equations above establish all elements of the Blount operator algebra used in the main text. In particular, the position operator is represented by the sum of a covariant differential part and an interband multiplicative part, lattice translation invariant operators act multiplicatively in crystal momentum, the correspondence is closed under operator composition, and the velocity operator follows from the same algebra without any additional prescription.

\section{\texorpdfstring{$U(N)$}{U(N)} covariance of the Blount operator algebra}
\label{app:UN}

Sec.~\ref{app:blount} develops the Blount construction in the Hamiltonian eigenstate frame, in which the Hamiltonian is diagonal and the residual smooth gauge freedom is a band dependent phase. Operator level covariant differentiation, in which the covariant derivative acts on a general band space operator through a commutator, is explicit in Ref.~\\cite{SI:Ventura2017}; fully non-Abelian formulations under arbitrary smooth $U(N)$ changes of Bloch frame are given in Refs.~\\cite{SI:Parker2019,SI:BonbienManchon2022}. This is stronger than covariance of the position operator alone and is the aspect required for composite observables. Here we apply that general operator covariance specifically to orbital angular momentum, showing that the construction is covariant under an arbitrary smooth change of Bloch frame within an isolated set of bands and that the four orbital angular momentum sectors used in the main text possess frame independent definitions. {This statement is exact for the complete band space under consideration, including a complete finite band model. If a larger Hilbert space is deliberately truncated to a proper subspace, the same construction instead represents the corresponding projected operators, and projection need not commute with forming a composite observable.}

Let \(\lvert u\rangle\) denote the matrix whose columns are the cell periodic Bloch states spanning the chosen \(N\) dimensional subspace. Under a smooth change of frame,

\begin{eqnarray}
\lvert u'\rangle
&=&
\lvert u\rangle U(\mathbf k)
\label{app:eq:UN-frame}
\\
\mathbf c'
&=&
U^\dagger(\mathbf k)\mathbf c
\label{app:eq:UN-coeff}
\end{eqnarray}
with \(U(\mathbf k)\in U(N)\). A band space operator therefore transforms as

\begin{equation}
O' = U^\dagger O U .
\label{app:eq:UN-O}
\end{equation}
The full non-Abelian Bloch connection is

\begin{equation}
\boldsymbol{\xi}_i
=
i\langle u|\partial_i u\rangle
\label{app:eq:UN-xi-def}
\end{equation}
and transforms according to

\begin{equation}
\boldsymbol{\xi}_i'
=
U^\dagger\boldsymbol{\xi}_iU
+
iU^\dagger\partial_iU .
\label{app:eq:UN-xi}
\end{equation}
It follows immediately that the full Blount position operator

\begin{equation}
r_i
=
i\partial_i
+
\boldsymbol{\xi}_i
\label{app:eq:UN-R}
\end{equation}
is covariant on coefficient sections,

\begin{equation}
r_i'\mathbf c'
=
U^\dagger r_i\mathbf c .
\label{app:eq:UN-R-cov}
\end{equation}
Thus the position operator itself requires no preferred Bloch frame.

\subsection{Covariant differentiation and spectral projectors}

For a band space operator \(O(\mathbf k)\) define the covariant derivative associated with the full connection by

\begin{equation}
\mathcal D_i^{\xi}O
=
\partial_iO
-
i[\boldsymbol{\xi}_i,O].
\label{app:eq:UN-covder}
\end{equation}
Using Eqs.~\eqref{app:eq:UN-O} and \eqref{app:eq:UN-xi}, together with

\begin{equation}
\partial_iU^\dagger
=
-
U^\dagger
(\partial_iU)
U^\dagger ,
\label{app:eq:UN-dUdag}
\end{equation}
one finds

\begin{eqnarray}
(\mathcal D_i^{\xi}O)'
&=&
\partial_i(U^\dagger O U)
-
i
[
U^\dagger\boldsymbol{\xi}_iU
+
iU^\dagger\partial_iU
,
U^\dagger O U
]
\label{app:eq:UN-covder-proof1}
\\
&=&
U^\dagger
\left(
\partial_iO
-
i[\boldsymbol{\xi}_i,O]
\right)
U .
\label{app:eq:UN-covder-proof2}
\end{eqnarray}
Hence

\begin{equation}
(\mathcal D_i^{\xi}O)'
=
U^\dagger
(\mathcal D_i^{\xi}O)
U .
\label{app:eq:UN-covder-cov}
\end{equation}
The derivative \(\mathcal D_i^\xi\) therefore maps covariant band space operators to covariant band space operators.

To recover the LC/IC decomposition without choosing the Hamiltonian gauge, write the spectral representation of the Hamiltonian as

\begin{equation}
H
=
\sum_\alpha
\varepsilon_\alpha
P_\alpha ,
\label{app:eq:UN-spectral-H}
\end{equation}
where \(P_\alpha\) projects onto an isolated spectral subspace. A degenerate eigenspace is treated as one block. The projectors obey

\begin{eqnarray}
P_\alpha P_\beta
&=&
\delta_{\alpha\beta}P_\alpha
\label{app:eq:UN-projector-orth}
\\
\sum_\alpha P_\alpha
&=&
\mathbf{1}
\label{app:eq:UN-projector-complete}
\\
P_\alpha'
&=&
U^\dagger P_\alpha U .
\label{app:eq:UN-projector-cov}
\end{eqnarray}
We now define the inter subspace part of the position operator by

\begin{equation}
d_i
=
\frac{i}{2}
\sum_\alpha
[
\mathcal D_i^\xi P_\alpha
,
P_\alpha
].
\label{app:eq:UN-d}
\end{equation}
{Since both factors in each commutator are Hermitian, each commutator is anti Hermitian; multiplication by $i$ therefore makes \(d_i\) Hermitian.} Since \(\mathcal D_i^\xi P_\alpha\) is covariant, Eq.~\eqref{app:eq:UN-d} also gives

\begin{equation}
d_i'
=
U^\dagger d_i U .
\label{app:eq:UN-d-cov}
\end{equation}
Using

\begin{equation}
\mathcal D_i^\xi(P_\alpha^2)
=
\mathcal D_i^\xi P_\alpha
\label{app:eq:UN-projector-product}
\end{equation}
and

\begin{equation}
\sum_\alpha
\mathcal D_i^\xi P_\alpha
=
0
\label{app:eq:UN-projector-sum}
\end{equation}
Eq.~\eqref{app:eq:UN-d} may equivalently be written as

\begin{equation}
d_i
=
i
\sum_\alpha
(\mathcal D_i^\xi P_\alpha)
P_\alpha .
\label{app:eq:UN-d2}
\end{equation}
In Hamiltonian gauge the projectors are \(\mathbf k\) independent. Equation~\eqref{app:eq:UN-d2} then becomes

\begin{equation}
d_i
=
\boldsymbol{\xi}_i
-
\sum_\alpha
P_\alpha
\boldsymbol{\xi}_i
P_\alpha .
\label{app:eq:UN-d-Hgauge}
\end{equation}
Thus \(d_i\) is precisely the part of the Bloch connection connecting distinct spectral subspaces. For non-degenerate bands this reduces to the familiar off diagonal dipole matrix,

\begin{equation}
(d_i)_{nm}
=
(1-\delta_{nm})
(\xi_i)_{nm}.
\label{app:eq:UN-d-nondegenerate}
\end{equation}

\subsection{Covariant LC and IC sectors}

Define the complementary connection by

\begin{equation}
\mathcal A_i
=
\boldsymbol{\xi}_i
-
d_i
\label{app:eq:UN-A}
\end{equation}
and the corresponding differential operator by

\begin{equation}
iD_i
=
i\partial_i
+
\mathcal A_i .
\label{app:eq:UN-D}
\end{equation}
Since \(d_i\) transforms homogeneously, \(\mathcal A_i\) inherits the connection transformation law,

\begin{equation}
\mathcal A_i'
=
U^\dagger\mathcal A_iU
+
iU^\dagger\partial_iU .
\label{app:eq:UN-A-transform}
\end{equation}
Consequently

\begin{equation}
(iD_i)'\mathbf c'
=
U^\dagger iD_i\mathbf c .
\label{app:eq:UN-D-cov}
\end{equation}
and the full position operator again separates as

\begin{equation}
r_i
=
iD_i
+
d_i .
\label{app:eq:UN-R-split}
\end{equation}
The physical meaning of the two pieces is fixed by the spectral decomposition rather than by a particular basis within each spectral block.

To make this explicit, define

\begin{equation}
Q_{\alpha i}
=
\mathcal D_i^\xi P_\alpha .
\label{app:eq:UN-Q}
\end{equation}
Orthogonality and completeness imply

\begin{equation}
[d_i,P_\alpha]
=
iQ_{\alpha i}.
\label{app:eq:UN-dP}
\end{equation}
Hence the projectors are parallel with respect to the complementary connection,

\begin{equation}
\mathcal D_i^{\mathcal A}P_\alpha
=
\partial_iP_\alpha
-
i[\mathcal A_i,P_\alpha]
=
0 .
\label{app:eq:UN-parallel}
\end{equation}
Thus \(\mathcal A_i\) parallel transports each isolated spectral subspace into itself, whereas \(d_i\) connects distinct spectral subspaces. This is the frame independent content of the intraband/interband split used in the Hamiltonian gauge of the main text.

The covariant generalization of \(\partial_iE\) is then

\begin{equation}
g_i
=
\mathcal D_i^{\mathcal A}H
\label{app:eq:UN-g}
\end{equation}
or explicitly

\begin{equation}
g_i
=
\partial_iH
-
i[\mathcal A_i,H].
\label{app:eq:UN-g-expanded}
\end{equation}
Using Eq.~\eqref{app:eq:UN-parallel} in the spectral representation \eqref{app:eq:UN-spectral-H} gives

\begin{equation}
g_i
=
\sum_\alpha
(\partial_i\varepsilon_\alpha)
P_\alpha .
\label{app:eq:UN-g2}
\end{equation}
Therefore

\begin{equation}
g_i'
=
U^\dagger g_iU
\label{app:eq:UN-g-cov}
\end{equation}
and in non-degenerate Hamiltonian gauge

\begin{equation}
\mathbf g
=
\nabla_{\mathbf k}E .
\label{app:eq:UN-g-Hgauge}
\end{equation}

\subsection{Velocity and orbital angular momentum}

The velocity operator follows from the same commutator used in Sec.~\ref{app:blount},

\begin{equation}
\mathbf v
=
-i[\mathbf r,H].
\label{app:eq:UN-v-start}
\end{equation}
Using Eq.~\eqref{app:eq:UN-R-split} gives the frame independent form

\begin{equation}
\mathbf v
=
\mathbf g
-
i[\mathbf d,H].
\label{app:eq:UN-v}
\end{equation}
Equation~\eqref{app:eq:UN-g2} also implies

\begin{equation}
\mathcal D_i^{\mathcal A}g_j
=
\sum_\alpha
(\partial_i\partial_j\varepsilon_\alpha)
P_\alpha .
\label{app:eq:UN-Dg}
\end{equation}
The right hand side is symmetric under interchange of \(i\) and \(j\), and therefore

\begin{equation}
\boldsymbol{\mathcal D}^{\mathcal A}
\times
\mathbf g
=
0 .
\label{app:eq:UN-curlg}
\end{equation}
Equivalently,

\begin{equation}
i\mathbf D
\times
\mathbf g
=
-
\mathbf g
\times
i\mathbf D .
\label{app:eq:UN-curl}
\end{equation}
This identity is the \(U(N)\)-covariant version of the relation used in the main text to establish the Hermitian form of the pure IC operator.

We now insert

\begin{equation}
\mathbf r
=
i\mathbf D
+
\mathbf d
\label{app:eq:UN-R-OAM}
\end{equation}
and Eq.~\eqref{app:eq:UN-v} into

\begin{equation}
\mathbf L
=
\frac{1}{2}
\left(
\mathbf r\times\mathbf v
-
\mathbf v\times\mathbf r
\right).
\label{app:eq:UN-L-def}
\end{equation}
Using Eq.~\eqref{app:eq:UN-curl}, the four terms are

\begin{eqnarray}
\mathbf L_{\rm LC}
&=&
\frac{i}{2}
\left(
[\mathbf d,H]\times\mathbf d
-
\mathbf d\times[\mathbf d,H]
\right)
\label{app:eq:UN-L-LC}
\\
\mathbf L_{\rm LC}^{\rm mix}
&=&
\frac{1}{2}
\left(
\mathbf d\times\mathbf g
-
\mathbf g\times\mathbf d
\right)
\label{app:eq:UN-L-LCmix}
\\
\mathbf L_{\rm IC}
&=&
-
\mathbf g\times i\mathbf D
\label{app:eq:UN-L-IC}
\\
\mathbf L_{\rm IC}^{\rm mix}
&=&
-\frac{i}{2}
\left(
i\mathbf D\times[\mathbf d,H]
-
[\mathbf d,H]\times i\mathbf D
\right).
\label{app:eq:UN-L-ICmix}
\end{eqnarray}
Every multiplicative object in Eqs.~\eqref{app:eq:UN-L-LC}--\eqref{app:eq:UN-L-ICmix} transforms homogeneously, while \(i\mathbf D\) transforms covariantly as a differential operator acting on coefficient sections. Therefore each contribution is separately \(U(N)\)-covariant,

\begin{equation}
\mathbf L_X'\mathbf c'
=
U^\dagger
\mathbf L_X
\mathbf c
\label{app:eq:UN-L-cov}
\end{equation}
for

\begin{equation}
X
\in
\left\{
{\rm LC},
\;
{\rm LC}^{\rm mix},
\;
{\rm IC},
\;
{\rm IC}^{\rm mix}
\right\}.
\label{app:eq:UN-X-set}
\end{equation}
In non-degenerate Hamiltonian gauge one has

\begin{eqnarray}
H
&=&
E
\label{app:eq:UN-reduction-H}
\\
\mathbf g
&=&
\nabla_{\mathbf k}E
\label{app:eq:UN-reduction-g}
\end{eqnarray}
while \(\mathcal A_i\) and \(d_i\) reduce respectively to the diagonal and off diagonal parts of the Bloch connection. Equations~\eqref{app:eq:UN-L-LC}--\eqref{app:eq:UN-L-ICmix} then reduce exactly to the four orbital angular momentum operators used in the main text. For a degenerate eigenspace the same formulas hold with diagonal/off diagonal replaced by the corresponding block diagonal/inter block decomposition.

Finally, the LC/IC classification itself is independent of Bloch frame choice. Under an active translation by a lattice vector \(\mathbf a\),

\begin{eqnarray}
\mathbf d
&\rightarrow&
\mathbf d
\label{app:eq:UN-active-d}
\\
\mathbf g
&\rightarrow&
\mathbf g
\label{app:eq:UN-active-g}
\\
i\mathbf D
&\rightarrow&
i\mathbf D
+
\mathbf a .
\label{app:eq:UN-active-D}
\end{eqnarray}
It follows that the total local and itinerant sectors transform as

\begin{eqnarray}
\mathbf L_{\rm LC}^{\rm tot}
&\rightarrow&
\mathbf L_{\rm LC}^{\rm tot}
\label{app:eq:UN-active-LC}
\\
\mathbf L_{\rm IC}^{\rm tot}
&\rightarrow&
\mathbf L_{\rm IC}^{\rm tot}
+
\mathbf a\times\mathbf v .
\label{app:eq:UN-active-IC}
\end{eqnarray}
Thus both the LC/IC decomposition and the translation properties used in the bulk construction are properties of the spectral subspaces themselves rather than artifacts of Hamiltonian gauge. The construction applies whenever the Hamiltonian admits a smooth spectral decomposition into fixed rank isolated subspaces. At isolated gap closing points at which such a decomposition fails, the construction applies away from the point; any limiting treatment must be considered separately.

\section{Chern bands, singular gauges, and the domain of the position operator}
\label{app:chern}

The purpose of this section is to place the position operator on the same footing for a topologically non-trivial Bloch bundle as in the original crystal momentum representation, and then to carry the same construction over to the position containing orbital angular momentum operator. The starting point is Blount's Appendix A analysis of $\hat{\v r}$ \\cite{SI:Blount1962}, which separates two logically distinct questions: the domain of the unbounded position operator, and the possible surface terms generated when its crystal momentum representation is obtained by integration by parts.  

We retain Blount's ordinary Fourier transform argument for the regularity of the physical crystal momentum section, and express it in weak derivative form.  We then formulate the same construction patchwise on the Bloch bundle, using the covariant derivative viewpoint developed in modern treatments of crystalline response \\cite{SI:Ventura2017,SI:Parker2019,SI:BonbienManchon2022}, so that Chern singularities are treated as singularities of the chosen frame rather than of the physical section.  Finally, we extend the domain and boundary term analysis from $\hat{\v r}$ to the composite operators $r_i v_j$ and $v_jr_i$ entering orbital angular momentum, in the setting of recent work on position containing observables in periodic systems \\cite{SI:Desmarais2023,SI:Ma2024ProperSpin,SI:An2026PositionOperators}.  The result is that a Chern obstruction changes the global patching, and can reappear as a finite boundary contribution when covariant expressions are rewritten in terms of Berry curvature, but it does not require an additional local position or orbital angular momentum operator.

\subsection{Domain of position and the physical crystal momentum section}

Consider a state

\begin{equation}
\ket{\Psi}
=
\frac{1}{V_{\rm BZ}}
\int_{\rm BZ}d\v k
\sum_n
\ket{\phi_{n\v k}}c_n(\v k),
\end{equation}
and define

\begin{equation}
{\cal F}_{\v k}(\v r)
=
\sum_n
c_n(\v k)\phi_{n\v k}(\v r).
\label{app:eq:physical-k-section}
\end{equation}
The important object for the surface term question is ${\cal F}_{\v k}$, rather than the Bloch frame or coefficient section separately.

Following Blount, this can be seen directly by returning to the ordinary Fourier transform of the physical wavefunction. Choosing the normalization of $\widetilde\Psi$ to match the Brillouin zone convention above, write

\begin{equation}
\Psi(\v r)
=
\frac{1}{V_{\rm BZ}}
\int_{\mathbb R^d}d\v q\,
\widetilde\Psi(\v q)e^{i\v q\cdot\v r}.
\end{equation}
Decomposing

\begin{equation}
\v q=\v k+\v G,
\end{equation}
with $\v k$ restricted to one Brillouin zone and $\v G$ a reciprocal lattice vector, gives

\begin{equation}
{\cal F}_{\v k}(\v r)
=
\sum_{\v G}
\widetilde\Psi(\v k+\v G)
e^{i(\v k+\v G)\cdot\v r}.
\label{app:eq:Blount-Fourier-folding}
\end{equation}
Equation~\eqref{app:eq:Blount-Fourier-folding} is simply the ordinary Fourier representation folded into the Brillouin zone. It contains no Bloch gauge choice. A singularity or discontinuity may therefore be shifted between $c_n(\v k)$ and $\phi_{n\v k}$ by changing Bloch frame, while their physical combination is fixed by $\widetilde\Psi$.

The domain of one Cartesian component of the position operator is

\begin{equation}
{\rm Dom}(\hat r_i)
=
\left\{
\Psi\in{\cal H}:r_i\Psi\in{\cal H}
\right\}.
\label{app:eq:r-domain}
\end{equation}
Normalizability of $\Psi$ alone does not imply this condition. By the ordinary Fourier transform,

\begin{equation}
r_i\Psi
\longleftrightarrow
i\partial_{q_i}\widetilde\Psi,
\end{equation}
so that, by Plancherel (up to the fixed normalization factor set by our Fourier convention),

\begin{equation}
\Psi\in{\rm Dom}(\hat r_i)
\qquad\Longleftrightarrow\qquad
\partial_{q_i}\widetilde\Psi\in L^2
\label{app:eq:r-domain-fourier}
\end{equation}
in the weak sense. Requiring this for all Cartesian components is the usual $H^1$ condition on the ordinary Fourier amplitude.

The folding step makes this statement completely explicit. Define

\begin{equation}
a_{\v G}(\v k)=\widetilde\Psi(\v k+\v G).
\end{equation}
Then reciprocal lattice reindexing gives

\begin{eqnarray}
\int_{\rm BZ}d\v k\sum_{\v G}|a_{\v G}(\v k)|^2
&=&
\int_{\mathbb R^d}d\v q\,|\widetilde\Psi(\v q)|^2,
\label{app:eq:fold-L2}
\\
\int_{\rm BZ}d\v k\sum_{\v G}|\partial_{k_i}a_{\v G}(\v k)|^2
&=&
\int_{\mathbb R^d}d\v q\,|\partial_{q_i}\widetilde\Psi(\v q)|^2.
\label{app:eq:fold-H1}
\end{eqnarray}
The same fixed convention dependent Fourier factor multiplies both sides if a different normalization is chosen and is irrelevant for domain membership. Thus the folded Fourier amplitudes form an $H^1({\rm BZ};\ell^2_{\v G})$ section whenever $\Psi\in{\rm Dom}(\v r)$. By Parseval on a unit cell, the $\ell^2_{\v G}$ norm is exactly the fibre norm of the cell periodic physical section $e^{-i\v k\cdot\v r}{\cal F}_{\v k}$; multiplication by the smooth factor $e^{i\v k\cdot\v r}$ then gives the corresponding Hilbert space valued weak regularity of ${\cal F}_{\v k}$ itself. No regularity of the Bloch frame coefficients $c_n$ is required separately. This is the modern functional analytic form of Blount's original Fourier argument \\cite{SI:Blount1962}: his reciprocal lattice sum is exactly this folding construction, used to infer regularity of the physical crystal momentum representation from the ordinary Fourier amplitude.

Requiring Eq.~\eqref{app:eq:r-domain-fourier} for all Cartesian components excludes a genuine finite jump of the physical section across any codimension one surface in $\v k$ space: the derivative normal to that surface would contain a delta function contribution and would not be square integrable. No corresponding restriction need hold on $c_n$ or $\phi_{n\v k}$ separately, since these are representation dependent quantities. Blount's exceptional plane discontinuity case is the old language counterpart of the same statement \\cite{SI:Blount1962}: a non-shrinking codimension one discontinuity is accompanied by a divergent position moment in the normal direction.

A convenient modern formulation is therefore to establish the differential representation first on a smooth dense core of the position operator, where the integration by parts and all boundary limits are literal, and then extend the representation by closure to the full domain in Eq.~\eqref{app:eq:r-domain}. In this way no pointwise continuity assumption beyond that required on the smooth core is needed.

\subsection{Integration by parts and the Blount position operator}

Using

\begin{equation}
\hat{\v r}\ket{\phi_{n\v k}}
=
-i\m\nabla_{\v k}\ket{\phi_{n\v k}}
+
e^{i\v k\cdot\v r}
i\m\nabla_{\v k}\ket{u_{n\v k}},
\end{equation}
the action of position contains the term

\begin{equation}
-i
\int_{\rm BZ}d\v k\,
\sum_n
\m\nabla_{\v k}\ket{\phi_{n\v k}}\,
c_n(\v k).
\end{equation}
On the smooth core, integration by parts transfers the derivative to the coefficient section. The boundary contributions from opposite faces of the Brillouin zone cancel because ${\cal F}_{\v k}$ in Eq.~\eqref{app:eq:Blount-Fourier-folding} is periodic under reciprocal lattice reindexing. Resolving $i\m\nabla_{\v k}\ket{u_{n\v k}}$ in the periodic Bloch basis then gives

\begin{equation}
\hat{\v r}
\longleftrightarrow
i\m\nabla_{\v k}
+
\boldsymbol{\xi}(\v k)
=
i\v D_{\v k}
+
\v d(\v k).
\label{app:eq:r-domain-Blount}
\end{equation}
The general statement follows by closure. Equation~\eqref{app:eq:r-domain-Blount} is therefore an equality of unbounded operators on their appropriate domain, rather than an identity acting on arbitrary square integrable coefficient sections.

This distinction is important: the absence of an additional integration by parts surface term establishes the differential representation of the position operator; it does not by itself establish that a particular state belongs to its domain. The latter requires the result of applying Eq.~\eqref{app:eq:r-domain-Blount} to be square integrable.

\subsection{Patchwise formulation and Chern bands}

For a topologically trivial isolated band a globally smooth periodic Bloch frame may be chosen. For a Chern band this is impossible, but the failure is one of representation rather than of the local operator.  This is the natural setting of the bundle and covariant derivative formulations of the Bloch problem \\cite{SI:Parker2019,SI:BonbienManchon2022}.  Cover the Brillouin zone by smooth patches $U_\alpha$. On an overlap,

\begin{equation}
\ket{u^{(\beta)}}
=
e^{i\chi_{\alpha\beta}}
\ket{u^{(\alpha)}},
\qquad
c^{(\beta)}
=
e^{-i\chi_{\alpha\beta}}
c^{(\alpha)},
\label{app:eq:chern-transition}
\end{equation}
and therefore

\begin{equation}
{\cal F}_{\v k}^{(\beta)}
=
{\cal F}_{\v k}^{(\alpha)}.
\label{app:eq:physical-section-patch}
\end{equation}
Integration by parts may consequently be performed patch by patch. Surface contributions on a common patch boundary occur with opposite orientations and cancel because the physical section has the same trace on the two sides. This is the bundle theoretic version of Blount's observation that a discontinuity may be placed in the Bloch functions rather than in their coefficients without creating a discontinuity of the physical combination \\cite{SI:Blount1962}. The local position operator on every patch is therefore still

\begin{equation}
\v r
=
i\v D+\v d.
\end{equation}

The same conclusion may be seen directly in a singular gauge representation. The obstruction of a two dimensional isolated Chern band may be represented by one or more phase vortices; for simplicity consider a vortex at $\v k_v$. Locally, with $q=|\v k-\v k_v|$,

\begin{equation}
\m{\mathcal A}
=
\m{\mathcal A}_{\rm reg}
+
w\,\m\nabla_{\v k}\theta,
\qquad
\m\nabla_{\v k}\theta
\sim
\frac{\hat{\v\theta}}{q},
\end{equation}
where $w$ is the winding number. This is a singularity of the chosen Bloch frame, not of the Hamiltonian or spectral projector.

Remove a disk $D_\epsilon$ around $\v k_v$ and perform the integration by parts on

\begin{equation}
M_\epsilon
=
{\rm BZ}\setminus D_\epsilon.
\end{equation}
For a smooth state in the operator core, the possible additional term is

\begin{equation}
\ket{B_i(\epsilon)}
=
i
\oint_{\partial D_\epsilon}
ds\,n_i\,
{\cal F}_{\v k}.
\label{app:eq:chern-boundary-r}
\end{equation}
Since the physical section is regular on this core and $ds=\epsilon d\theta$,

\begin{equation}
\|B_i(\epsilon)\|
=
O(\epsilon)
\longrightarrow0.
\label{app:eq:chern-boundary-r-zero}
\end{equation}
This is precisely the shrinking surface mechanism identified by Blount \\cite{SI:Blount1962}, translated into modern bundle language: an isolated point or line singularity of a chosen band representation does not generate an additional position operator because its enclosing surface shrinks to zero measure. In contrast, Blount's plane discontinuity example is codimension one and the enclosing surface does not shrink; this is the same case already excluded above by the position domain regularity condition. The general domain statement follows from the equivalent patchwise weak formulation.

It is useful to emphasize again that Eq.~\eqref{app:eq:chern-boundary-r-zero} and membership of ${\rm Dom}(\v r)$ are separate statements. For example, in a singular one band gauge with

\begin{equation}
\m{\mathcal A}
\sim
\frac{\hat{\v\theta}}{q},
\end{equation}
the formal choice $c(\v k)=1$ has a singular contribution

\begin{equation}
\v r c
=
\m{\mathcal A}c+\v d\,c
\sim
\m{\mathcal A}
\sim
\frac{1}{q},
\end{equation}
where the interband part $\v d$ remains bounded for the isolated gapped band.

In two dimensions,

\begin{equation}
\int_0 dq\,q\,|\v r c|^2
\sim
\int_0\frac{dq}{q},
\end{equation}
which diverges. Thus this particular coefficient section does not lie in the domain of $\v r$, despite the fact that the shrinking puncture does not generate an additional surface operator. A change of Bloch frame must of course be accompanied by the compensating transformation of the coefficient section.

\subsection{Application to orbital angular momentum}

The same logic applies to the position containing composite operators entering orbital angular momentum.  Such operators require care precisely because the position operator is unbounded; related formulations for orbital and spin observables, and a recent systematic treatment of position containing composite operators in the Bloch representation, are given in Refs.~\\cite{SI:Desmarais2023,SI:Ma2024ProperSpin,SI:An2026PositionOperators}.  Here the point needed for orbital angular momentum is especially direct: the inner occurrence of $\v r$ in $\v r\times[\v r,H]$ combines algebraically into the physical velocity,

\begin{equation}
\v v
=
-i[\v r,H]
=
\m\nabla_{\v k}E
-
i[\v d,E]
\label{app:eq:chern-v}
\end{equation}
in Hamiltonian gauge. The quantities $\v d$, $[\v d,E]$, and $\v v$ transform homogeneously under a change of Bloch frame and therefore do not acquire the $1/q$ gauge singularity of $\m{\mathcal A}$.

The composite operator domain should be stated separately from the boundary calculation. In the finite isolated band manifold considered here, $v_j$ is a bounded covariant multiplication operator: in each smooth local frame it is a smooth matrix, and compactness of the Brillouin zone bounds its operator norm. Consequently
\begin{eqnarray}
{\rm Dom}(r_i v_j)
&=&
\left\{\Psi\in{\cal H}:v_j\Psi\in{\rm Dom}(r_i)\right\},
\label{app:eq:Rv-domain}
\\
{\rm Dom}(v_jr_i)
&=&
{\rm Dom}(r_i),
\label{app:eq:vR-domain}
\end{eqnarray}
and the symmetrized OAM operator is defined on the corresponding common domain. Because $v_j$ and its local crystal momentum derivatives are bounded on the isolated manifold, multiplication by $v_j$ preserves ${\rm Dom}(r_i)$. Hence ${\rm Dom}(r_i)\subseteq{\rm Dom}(r_i v_j)$ while ${\rm Dom}(v_jr_i)={\rm Dom}(r_i)$, so the natural common domain is ${\rm Dom}(r_i)$. This makes precise the same distinction stressed by Blount for $x$ itself: first specify the domain on which the composite operator exists, and only then perform the integration by parts representation.

The physical crystal momentum section of $v_j\ket{\Psi}$ is

\begin{equation}
{\cal F}_{\v k}^{(v_j)}(\v r)
=
\sum_n
\phi_{n\v k}(\v r)
(v_jc)_n(\v k).
\end{equation}
The Blount argument may therefore be applied directly to this state. In a singular gauge the possible puncture contribution is

\begin{equation}
\ket{B_{ij}(\epsilon)}
=
i\oint_{\partial D_\epsilon}
ds\,n_i\,
\sum_n
\ket{\phi_{n\v k}}
(v_jc)_n,
\end{equation}
and vanishes on the common smooth core as $\epsilon\rightarrow0$. Since multiplication by the smooth bounded velocity preserves this core, the same Blount argument applies to the $r_i v_j$ part; the oppositely ordered term $v_jr_i$ contains no new integration by parts. The operator identities are therefore established on the common core and interpreted on the domain on which the symmetrized composite is defined. Hence the Chern obstruction generates no additional local position or orbital angular momentum operator.

Topology instead enters through the global patching of the Bloch frame and coefficient section. For one isolated band,

\begin{equation}
\langle\v L_{\rm IC}\rangle
=
-\frac{1}{V_{\rm BZ}}
\int d\v k\,
\m\nabla_{\v k}E
\times
\left[
ic^*\m\nabla_{\v k}c
+
|c|^2\m{\mathcal A}
\right].
\label{app:eq:chernIC}
\end{equation}
Under a change of local Bloch frame,

\begin{equation}
\ket{u'}=e^{i\chi}\ket{u},
\qquad
c'=e^{-i\chi}c,
\qquad
\m{\mathcal A}'
=
\m{\mathcal A}-\m\nabla_{\v k}\chi,
\end{equation}
one has

\begin{equation}
ic'^*\m\nabla_{\v k}c'
+
|c'|^2\m{\mathcal A}'
=
ic^*\m\nabla_{\v k}c
+
|c|^2\m{\mathcal A}.
\label{app:eq:chernIC-patch}
\end{equation}
Thus the coefficient phase and Berry connection pieces separately depend on the chosen representation, while their covariant sum is a globally well defined physical quantity. In particular, if $c=1$ in one local representative, the same physical section after a gauge change is represented by $c'=e^{-i\chi}$, not by setting $c'=1$ again.

A subtlety appears only if this covariant expression is rewritten in terms of Berry curvature.  The distinction between a globally covariant derivative and a connection defined only in local frames is standard in the bundle formulation \\cite{SI:Parker2019,SI:BonbienManchon2022}; here it fixes where the topology enters the orbital angular momentum expression. On a smooth patch,

\begin{equation}
\m\nabla_{\v k}\times
(E\m{\mathcal A})
=
\m\nabla_{\v k}E
\times
\m{\mathcal A}
+
E\m\Omega .
\end{equation}
For a Chern band the integral of this total derivative cannot be discarded globally. In a singular gauge $\m{\mathcal A}\sim1/q$, and therefore on a puncture boundary the line integral scales as

\begin{equation}
\oint_{\partial D_\epsilon}
E\m{\mathcal A}\cdot d\boldsymbol{\ell}
\sim
\int d\theta\,
E(\v k_v)
\frac{1}{\epsilon}
\epsilon
=
O(1).
\end{equation}
A finite patch or puncture contribution may therefore remain. This is not an additional term in the fundamental Blount position operator. It is the boundary term generated when the globally covariant expression is rewritten in terms of a connection which is not itself globally defined.

We therefore arrive at the same logical structure as Blount, expressed in the modern language of the Bloch bundle. The domain of the position operator determines which physical states admit the action of $\v r$; the ordinary Fourier transform supplies the corresponding weak crystal momentum regularity; and integration by parts then yields the covariant operator

\begin{equation}
\v r=i\v D+\v d
\end{equation}
without an additional singular contribution. A non-zero Chern number obstructs a globally smooth Bloch frame, but does not modify this local operator. Its effects reside in the global patching data and in boundary terms that arise when the fundamental covariant expressions are rewritten in non-global forms.

Finally, the gauge transformation considered here should be distinguished from the occupied state phase freedom discussed in Sec.~\ref{app:gg-patchwise-stokes}. Here a genuine topological obstruction forces a patchwise or singular representation of the Bloch frame, and the Bloch frame and coefficient section transform together; the covariant position expectation, and hence the IC integrand in Eq.~\ref{app:eq:chernIC}, is therefore invariant pointwise wherever a local frame is defined. In Sec.~\ref{app:gg-patchwise-stokes} the gapped graphene band itself admits a smooth gauge: the patching arises only because a particular half-angle phase representative is singular. There the reference Bloch frame is held fixed while the occupied representative is rephased, so the IC density changes locally by a phase gradient term although its Brillouin zone integral is unchanged. The same patch transition bookkeeping is then required, but for a different reason.

\section{Equilibrium occupied state phase freedom in metals}
\label{app:metal-phase-freedom}

The distinction between local and itinerant circulation in the equilibrium theory of orbital magnetization was introduced in the Wannier formulation of Refs.~\\cite{SI:ThonhauserCeresoliVanderbiltResta2005,SI:CeresoliThonhauserVanderbiltResta2006}, and the extension of the equilibrium magnetization to metals and finite temperature was established in Refs.~\\cite{SI:CeresoliThonhauserVanderbiltResta2006,SI:ShiVignaleXiaoNiu2007}.  The purpose of this section is narrower.  We verify the corresponding occupied state phase freedom directly for the LC and IC operators derived in this work.  For a filled band the LC contribution is pointwise unchanged, while the change of the IC contribution is a Brillouin zone total derivative.  We then show that the same cancellation survives partial occupation in an equilibrium metal because the occupation has the form $f_{n\vk}=f(E_{n\vk})$.

\subsection{Filled band}

For a completely filled isolated band the occupied projector

\begin{equation}
 P_{n\vk}=|\phi_{n\vk}\rangle\langle\phi_{n\vk}|
\end{equation}
is unchanged by a smooth rephasing of the occupied orbital. In a fixed reference Bloch frame this freedom may be represented as

\begin{equation}
 c_{n\vk}\longrightarrow e^{i\eta_n(\vk)}c_{n\vk}.
 \label{metal:eq:phasefreedom}
\end{equation}
This is the phase freedom of a filled band projector, not an arbitrary rephasing of the coefficient section of a coherent wavepacket.

The pure LC operator contains no derivative acting on the coefficient section and is therefore pointwise unchanged by Eq.~\eqref{metal:eq:phasefreedom}.  For the pure IC operator,

\begin{equation}
 \vL_{\rm IC}=-\gradk E\times i\vD_{\vk},
 \label{metal:eq:LIC}
\end{equation}
one has on a smooth patch

\begin{equation}
 c^\dagger i\vD_{\vk}c
 \longrightarrow
 c^\dagger i\vD_{\vk}c
 -|c|^2\gradk\eta .
 \label{metal:eq:iDphase}
\end{equation}
For a normalized filled band representative, $|c|^2=1$, and hence

\begin{equation}
 \delta \vL_{{\rm IC},n}(\vk)
 =\gradk E_{n\vk}\times\gradk\eta_n(\vk)
 =\gradk\times\left(E_{n\vk}\gradk\eta_n\right).
 \label{metal:eq:filled-totalderivative}
\end{equation}
For a smooth phase factor whose one form $d\eta_n$ is globally defined on the Brillouin zone torus, the Brillouin zone integral therefore vanishes.  Thus a filled band phase change produces no physical change in either the LC contribution or the integrated IC contribution.  This is the form taken, for the present LC/IC operators, by the standard gauge independence of equilibrium orbital magnetization \\cite{SI:ThonhauserCeresoliVanderbiltResta2005,SI:CeresoliThonhauserVanderbiltResta2006}.

\subsection{Partially occupied equilibrium band}

For a noninteracting equilibrium state,

\begin{equation}
 \hat\rho_{\rm eq}
 =\frac{1}{V_{\rm BZ}}
 \sum_n\int_{\rm BZ}d\vk\,
 f(E_{n\vk})
 |\phi_{n\vk}\rangle\langle\phi_{n\vk}| ,
 \label{metal:eq:rhoeq}
\end{equation}
so the same bandwise phase freedom leaves the density matrix unchanged.  The equilibrium extension to metals is standard \\cite{SI:CeresoliThonhauserVanderbiltResta2006,SI:ShiVignaleXiaoNiu2007}; the point needed here is that the occupation depends on momentum through the band energy alone.  Equation~\eqref{metal:eq:iDphase} then gives

\begin{equation}
 \delta \vL_{{\rm IC},n}(\vk)
 =f_{n\vk}\,\gradk E_{n\vk}\times\gradk\eta_n(\vk).
 \label{metal:eq:deltaLIC}
\end{equation}
Introduce

\begin{equation}
 G_{n\vk}
 =-\int_{E_{n\vk}}^{\mu}d\epsilon\,f(\epsilon),
 \qquad
 \gradk G_{n\vk}
 =f_{n\vk}\gradk E_{n\vk}.
 \label{metal:eq:Gdef}
\end{equation}
Then

\begin{equation}
 \delta \vL_{{\rm IC},n}(\vk)
 =\gradk G_{n\vk}\times\gradk\eta_n
 =\gradk\times\left(G_{n\vk}\gradk\eta_n\right),
 \label{metal:eq:totalderivative}
\end{equation}
and therefore

\begin{equation}
 \delta\langle\vL_{\rm IC}\rangle
 =\frac{1}{V_{\rm BZ}}
 \sum_n\int_{\rm BZ}d\vk\,
 \gradk\times\left(G_{n\vk}\gradk\eta_n\right)
 =0.
 \label{metal:eq:mainresult}
\end{equation}

Thus the filled band phase freedom argument extends unchanged in structure to a partially occupied equilibrium band: the only replacement is $E\rightarrow G(E)$.  This step uses equilibrium essentially.  If instead the occupation has explicit momentum dependence, $f=f(\vk,E)$, then $\gradk G$ acquires an additional explicit-$\vk$ derivative and Eq.~\eqref{metal:eq:totalderivative} does not follow in general.

At zero temperature,

\begin{equation}
 G_{n\vk}
 =(E_{n\vk}-\mu)\,\Theta(\mu-E_{n\vk}),
 \label{metal:eq:GzeroT}
\end{equation}
so for a regular Fermi surface one may equivalently apply Stokes' theorem to the occupied Fermi sea $\mathcal V_n$,

\begin{equation}
 \delta\langle\vL_{\rm IC}\rangle_n
 =\frac{1}{V_{\rm BZ}}
 \int_{\mathcal V_n}d\vk\,
 \gradk\times\left[(E_{n\vk}-\mu)\gradk\eta_n\right].
 \label{metal:eq:FSstokes}
\end{equation}
The Fermi surface contribution vanishes because $E_{n\vk}-\mu=0$ on $\partial\mathcal V_n$.  The same $(E-\mu)$ weight appears in the equilibrium metallic orbital magnetization formula \\cite{SI:CeresoliThonhauserVanderbiltResta2006,SI:ShiVignaleXiaoNiu2007}.  At a singular Fermi surface the torus proof in Eq.~\eqref{metal:eq:mainresult} is the cleaner statement.

For a band diagonal equilibrium state the mixed LC and IC expectation values vanish because those operators are interband (more generally, inter spectral block), while the pure LC contribution is pointwise invariant as above.  Equation~\eqref{metal:eq:mainresult} therefore completes the occupied state phase freedom check for the full equilibrium orbital angular momentum.  At an exact degeneracy the corresponding statement is blockwise and follows from the $U(N)$ covariance of Sec.~\ref{app:UN}.

An active lattice translation has the same phase gradient algebra but a different physical meaning.  With $e^{-i\vk\cdot\mathbf a}$ one has $\gradk\eta=-\mathbf a$, and Eq.~\eqref{metal:eq:deltaLIC} reproduces the local lever arm shift of the IC sector.  Its integrated equilibrium contribution vanishes by Eq.~\eqref{metal:eq:mainresult}.  Singular or patchwise phase representatives require the corresponding transition boundary terms in Stokes' theorem; the filled band gapped graphene example is treated explicitly in Sec.~\ref{app:gg-patchwise-stokes}.

\section{Orbital angular momentum in the two band model}
\label{app:twoband}

Here we provide a detailed derivation of the orbital angular momentum (OAM) operators for gapped graphene, used in the main paper as a numerical example of dynamical OAM. Consider the Hamiltonian

\begin{equation}
H(\v k) = \begin{pmatrix}
\Delta & f(\v k) \\
f^\ast(\v k) & -\Delta
\end{pmatrix}
\end{equation}
This two band Hamiltonian possesses a symmetric pair of bands

\begin{equation}
\pm\varepsilon(\v k) = \pm\sqrt{\Delta^2+|f(\v k)|^2}
\end{equation}
with the eigenvector $\ket{u_{\sigma}(\v k)}$ corresponding to eigenvalue $\sigma\varepsilon(\v k)$ given by

\begin{equation}
\ket{u_{\sigma}(\v k)} = \frac{1}{\sqrt{2}}
\begin{pmatrix}
g_{\sigma}(\v k) e^{i\theta(\v k)/2} \\
\sigma g_{-\sigma}(\v k) e^{-i\theta(\v k)/2}
\end{pmatrix}
\end{equation}
where

\begin{equation}
g_{\sigma}(\v k) = \sqrt{1+\frac{\sigma\Delta}{\varepsilon(\v k)}}
\end{equation}
and the phase angle $\theta(\v k)$ is defined by

\begin{equation}
e^{i\theta(\v k)} = \frac{f(\v k)}{|f(\v k)|}.
\end{equation}
This particle-hole-symmetric two band structure contains the standard massive Dirac description of inversion broken graphene and is also the basic orbital sector of the minimal two band valley models widely used for group VI transition metal dichalcogenides \\cite{SI:Xiao2007Valley,SI:Xiao2012SpinValley,SI:Xu2014SpinPseudospin}.

Using the $\v k$-derivative of the eigenvector

\begin{equation}
\m\nabla_{\v k} u_{\sigma}(\v k) =
-\frac{\sigma\Delta}{2\varepsilon(\v k)|f(\v k)|}
\m\nabla_{\v k}\varepsilon(\v k) \ket{u_{-\sigma}(\v k)}
+i\frac{\m\nabla_{\v k}\theta(\v k)}{2}\frac{1}{\sqrt{2}}
\begin{pmatrix}
g_{\sigma}(\v k) e^{i\theta(\v k)/2} \\
-\sigma g_{-\sigma}(\v k) e^{-i\theta(\v k)/2}
\end{pmatrix}
\end{equation}
we can obtain the dipole matrix elements, the off diagonal elements of the full Berry connection matrix,

\begin{equation}
\v d(\v k) = \begin{pmatrix}
0 & \v p(\v k) \\ \v p^\ast(\v k) & 0
\end{pmatrix},
\end{equation}
where the matrix element $\v p(\v k)$ is given by

\begin{equation}
\v p(\v k) = i\frac{\Delta\m\nabla_{\v k}\varepsilon(\v k)}{2\varepsilon(\v k)|f(\v k)|}
-\frac{|f(\v k)|\m\nabla_{\v k}\theta(\v k)}{2\varepsilon(\v k)}.
\label{appC:p}
\end{equation}
The diagonal elements give the Berry connection

\begin{equation}
\m{\mathcal{A}}(\v k) = \begin{pmatrix}
\v A(\v k) & 0 \\ 0 &-\v A(\v k)
\end{pmatrix},
\end{equation}
where

\begin{equation}
\v A(\v k) = -\frac{\Delta}{2\varepsilon(\v k)}\m\nabla_{\v k}\theta(\v k).
\label{appC:A}
\end{equation}
The remaining objects entering the four orbital angular momentum operators are therefore

\begin{equation}
E(\v k)=\begin{pmatrix}
\varepsilon(\v k)&0\\
0&-\varepsilon(\v k)
\end{pmatrix},
\qquad
\m\nabla_{\v k}E(\v k)=\begin{pmatrix}
\m\nabla_{\v k}\varepsilon(\v k)&0\\
0&-\m\nabla_{\v k}\varepsilon(\v k)
\end{pmatrix},
\label{appC:E}
\end{equation}
and

\begin{equation}
i\v D_{\v k}
=i\mathbf{1}\m\nabla_{\v k}+\m{\mathcal A}(\v k)
=\begin{pmatrix}
i\m\nabla_{\v k}+\v A(\v k)&0\\
0&i\m\nabla_{\v k}-\v A(\v k)
\end{pmatrix}.
\label{appC:D}
\end{equation}

\subsection{Four orbital angular momentum operators}

We now evaluate directly the four orbital angular momentum operators defined in the main text for this two band Hamiltonian. The basic commutator entering both interband terms is

\begin{equation}
[\v d(\v k),E(\v k)]
=2\varepsilon(\v k)
\begin{pmatrix}
0&-\v p(\v k)\\
\v p^\ast(\v k)&0
\end{pmatrix}.
\label{appC:dE}
\end{equation}

\subsubsection{Local circulation}

For the pure local circulation operator,

\begin{equation}
\v L_{LC}(\v k)
=\frac{i}{2}\left(
[\v d(\v k),E(\v k)]\times\v d(\v k)
-\v d(\v k)\times[\v d(\v k),E(\v k)]
\right),
\end{equation}
Eq.~\ref{appC:dE} gives

\begin{equation}
\v L_{LC}(\v k)
=-2i\varepsilon(\v k)
\left[\v p(\v k)\times\v p^\ast(\v k)\right]\mathbf{1}.
\label{appC:LCp}
\end{equation}
From Eq.~\ref{appC:p},

\begin{equation}
\v p(\v k)\times\v p^\ast(\v k)
=-i\frac{\Delta}{2\varepsilon(\v k)^2}
\left[
\m\nabla_{\v k}\varepsilon(\v k)
\times
\m\nabla_{\v k}\theta(\v k)
\right],
\end{equation}
and hence

\begin{equation}\v L_{LC}(\v k)
=-\frac{\Delta}{\varepsilon(\v k)}
\left[
\m\nabla_{\v k}\varepsilon(\v k)
\times
\m\nabla_{\v k}\theta(\v k)
\right]\mathbf{1}
\label{appC:LC}
\end{equation}
Thus the pure local circulation operator is proportional to the identity in the two band space. This is not a consequence of the particle-hole symmetry $E_{\pm}(\v k)=\pm\varepsilon(\v k)$. More generally, for an arbitrary exact two band spectrum $E_+(\v k),E_-(\v k)$ with the same off diagonal form of $\v d(\v k)$ one obtains

\begin{equation}
\v L_{LC}(\v k)
=-i\left[E_+(\v k)-E_-(\v k)\right]
\left[\v p(\v k)\times\v p^\ast(\v k)\right]\mathbf{1}.
\end{equation}
The identity structure is therefore a general property of an exact two band Hilbert space. The equality of the upper- and lower band orbital moments in a two band model was noted explicitly by Yao, Xiao and Niu \\cite{SI:Yao2008ValleyOptics}; for the massive Dirac model of gapped graphene, Bhowal and Vignale obtain an orbital moment operator proportional to the identity in the conduction/valence band space \\cite{SI:Bhowal2021OrbitalHall}.

The mixed local circulation operator is

\begin{equation}
\v L_{LC}^{mix}(\v k)
=\frac{1}{2}\left[
\v d(\v k)\times\m\nabla_{\v k}E(\v k)
-\m\nabla_{\v k}E(\v k)\times\v d(\v k)
\right].
\end{equation}
Using Eq.~\ref{appC:E}, the upper off diagonal element is

\begin{equation}
\frac{1}{2}\left[
\v p(\v k)\times\left(-\m\nabla_{\v k}\varepsilon(\v k)\right)
-\m\nabla_{\v k}\varepsilon(\v k)\times\v p(\v k)
\right]=0,
\end{equation}
and the lower off diagonal element vanishes identically in the same way. Consequently

\begin{equation}
\v L_{LC}^{mix}(\v k)=0.
\label{appC:LCmix}
\end{equation}
Unlike the identity structure of Eq.~\ref{appC:LC}, this vanishing does use the symmetric spectrum $E_+(\v k)+E_-(\v k)=0$. For a general two band Hamiltonian the upper off diagonal element is

\begin{equation}
[\v L_{LC}^{mix}(\v k)]_{+-}
=\frac{1}{2}\v p(\v k)\times
\m\nabla_{\v k}\left[E_+(\v k)+E_-(\v k)\right].
\end{equation}

Writing $E_\pm(\v k)=\bar E(\v k)\pm\varepsilon(\v k)$ makes the structure transparent: the pure LC term depends on the band splitting $E_+-E_-=2\varepsilon$, whereas $L_{LC}^{mix}$ depends on the gradient of the band centre, $\m\nabla_{\v k}(E_++E_-)=2\m\nabla_{\v k}\bar E$. Consequently a constant shift of the energy origin, $H\rightarrow H+C\mathbf{1}$, changes neither operator. Only a momentum dependent band centre term $\bar E(\v k)$ represents a genuine breaking of this traceless two band structure.

\subsubsection{Itinerant circulation}

For the pure itinerant circulation operator,

\begin{equation}
\v L_{IC}(\v k)
=-\m\nabla_{\v k}E(\v k)\times i\v D_{\v k},
\end{equation}
Eqs.~\ref{appC:A}-\ref{appC:D} give directly

\begin{equation}
\v L_{IC}(\v k)
=\begin{pmatrix}
-\m\nabla_{\v k}\varepsilon(\v k)\times
\left[i\m\nabla_{\v k}+\v A(\v k)\right]&0\\
0&
\m\nabla_{\v k}\varepsilon(\v k)\times
\left[i\m\nabla_{\v k}-\v A(\v k)\right]
\end{pmatrix},
\end{equation}
and therefore

\begin{equation}\v L_{IC}(\v k)
=
\frac{\Delta}{2\varepsilon(\v k)}
\left[
\m\nabla_{\v k}\varepsilon(\v k)
\times
\m\nabla_{\v k}\theta(\v k)
\right]\mathbf{1}
-i\left[
\m\nabla_{\v k}\varepsilon(\v k)
\times\m\nabla_{\v k}
\right]
\begin{pmatrix}
1&0\\0&-1
\end{pmatrix}.
\label{appC:IC}
\end{equation}
The first term is multiplicative in $\v k$, while the second is a differential operator acting on the coefficient section to its right.

Finally consider

\begin{equation}
\v L_{IC}^{mix}(\v k)
=-\frac{i}{2}\left[
 i\v D_{\v k}\times[\v d(\v k),E(\v k)]
-[\v d(\v k),E(\v k)]\times i\v D_{\v k}
\right].
\end{equation}
Writing only for this intermediate step

\begin{equation}
\v K_{\pm}(\v k)=i\m\nabla_{\v k}\pm\v A(\v k),
\end{equation}
so that $i\v D_{\v k}=\operatorname{diag}(\v K_+(\v k),\v K_-(\v k))$, the upper off diagonal element becomes

\begin{equation}
[\v L_{IC}^{mix}(\v k)]_{+-}
=i\left\{
\v K_+(\v k)\times[\varepsilon(\v k)\v p(\v k)]
-[\varepsilon(\v k)\v p(\v k)]\times\v K_-(\v k)
\right\}.
\end{equation}
For a multiplicative vector function $\v F(\v k)$ the differential operator identity

\begin{equation}
(i\m\nabla_{\v k})\times\v F(\v k)
=i\m\nabla_{\v k}\times\v F(\v k)
-i\v F(\v k)\times\m\nabla_{\v k}
\label{appC:diffidentity}
\end{equation}
is understood as an operator identity acting on the coefficient section to the right. The terms containing $\v A(\v k)$ cancel between the two products, leaving

\begin{equation}
[\v L_{IC}^{mix}(\v k)]_{+-}
=-\m\nabla_{\v k}\times[\varepsilon(\v k)\v p(\v k)]
+2\varepsilon(\v k)\v p(\v k)\times\m\nabla_{\v k}.
\label{appC:Lmixupper}
\end{equation}
Similarly,

\begin{equation}
[\v L_{IC}^{mix}(\v k)]_{-+}
=\m\nabla_{\v k}\times[\varepsilon(\v k)\v p^\ast(\v k)]
-2\varepsilon(\v k)\v p^\ast(\v k)\times\m\nabla_{\v k}.
\label{appC:Lmixlower}
\end{equation}
Thus a compact form in the original variables introduced above is

\begin{equation}\v L_{IC}^{mix}(\v k)=
\begin{pmatrix}
0 & \bm{\Lambda}(\v k)\\
\bm{\Lambda}^{\dagger}(\v k)&0
\end{pmatrix},
\label{appC:ICmix}
\end{equation}
with

\begin{eqnarray}
\bm{\Lambda}(\v k)
&=&-\m\nabla_{\v k}\times\left[
 i\frac{\Delta}{2|f(\v k)|}\m\nabla_{\v k}\varepsilon(\v k)
-\frac{|f(\v k)|}{2}\m\nabla_{\v k}\theta(\v k)
\right]\nonumber\\
&&+\left[
 i\frac{\Delta}{|f(\v k)|}\m\nabla_{\v k}\varepsilon(\v k)
-|f(\v k)|\m\nabla_{\v k}\theta(\v k)
\right]\times\m\nabla_{\v k},
\label{appC:Lambda}
\end{eqnarray}
and $\bm{\Lambda}^{\dagger}(\v k)$ is equivalently given explicitly by Eq.~\ref{appC:Lmixlower}. This form keeps the differential operator character and hermiticity of $\v L_{IC}^{mix}(\v k)$ manifest.

On a smooth patch on which $\m\nabla_{\v k}\times\m\nabla_{\v k}\theta(\v k)=0$, and for constant $\Delta$, one may simplify

\begin{equation}
\m\nabla_{\v k}\times
\left[\frac{\m\nabla_{\v k}\varepsilon(\v k)}{|f(\v k)|}\right]=0,
\qquad
\m\nabla_{\v k}|f(\v k)|
=\frac{\varepsilon(\v k)}{|f(\v k)|}\m\nabla_{\v k}\varepsilon(\v k),
\end{equation}
to obtain

\begin{eqnarray}
\bm{\Lambda}(\v k)
&=&
\frac{\varepsilon(\v k)}{2|f(\v k)|}
\left[
\m\nabla_{\v k}\varepsilon(\v k)
\times\m\nabla_{\v k}\theta(\v k)
\right]\nonumber\\
&&+\left[
 i\frac{\Delta}{|f(\v k)|}\m\nabla_{\v k}\varepsilon(\v k)
-|f(\v k)|\m\nabla_{\v k}\theta(\v k)
\right]\times\m\nabla_{\v k}.
\label{appC:Lambdasimple}
\end{eqnarray}
The corresponding lower off diagonal element is

\begin{eqnarray}
\bm{\Lambda}^{\dagger}(\v k)
&=&-
\frac{\varepsilon(\v k)}{2|f(\v k)|}
\left[
\m\nabla_{\v k}\varepsilon(\v k)
\times\m\nabla_{\v k}\theta(\v k)
\right]\nonumber\\
&&+\left[
 i\frac{\Delta}{|f(\v k)|}\m\nabla_{\v k}\varepsilon(\v k)
+|f(\v k)|\m\nabla_{\v k}\theta(\v k)
\right]\times\m\nabla_{\v k}.
\label{appC:Lambdadagsimple}
\end{eqnarray}
At a phase singularity of $\theta(\v k)$ these last patchwise forms should be understood by taking the appropriate smooth limiting expression; Eqs.~\ref{appC:Lmixupper}-\ref{appC:ICmix} remain the more fundamental operator statements.

\subsubsection{Pauli matrix form and structural summary}

The preceding derivation requires no Pauli matrix notation. It is nevertheless useful, after the explicit two band matrices have been established, to introduce

\begin{equation}
\sigma_x=\begin{pmatrix}0&1\\1&0\end{pmatrix},\qquad
\sigma_y=\begin{pmatrix}0&-i\\i&0\end{pmatrix},\qquad
\sigma_z=\begin{pmatrix}1&0\\0&-1\end{pmatrix},
\end{equation}
solely to expose the band space structure. Equations~\ref{appC:p}-\ref{appC:D} then become

\begin{eqnarray}
\v d(\v k)
&=&-\frac{|f(\v k)|}{2\varepsilon(\v k)}
\m\nabla_{\v k}\theta(\v k)\,\sigma_x
-\frac{\Delta}{2\varepsilon(\v k)|f(\v k)|}
\m\nabla_{\v k}\varepsilon(\v k)\,\sigma_y,
\\
\m{\mathcal A}(\v k)
&=&-\frac{\Delta}{2\varepsilon(\v k)}
\m\nabla_{\v k}\theta(\v k)\,\sigma_z,
\\
E(\v k)&=&\varepsilon(\v k)\sigma_z.
\end{eqnarray}
The four orbital angular momentum operators can therefore be summarized as

\begin{equation}\v L_{LC}(\v k)
=-\frac{\Delta}{\varepsilon(\v k)}
\left[
\m\nabla_{\v k}\varepsilon(\v k)
\times\m\nabla_{\v k}\theta(\v k)
\right]\mathbf{1}
\end{equation}
\begin{equation}
\v L_{LC}^{mix}(\v k)=0
\end{equation}
\begin{equation}\v L_{IC}(\v k)
=
\frac{\Delta}{2\varepsilon(\v k)}
\left[
\m\nabla_{\v k}\varepsilon(\v k)
\times\m\nabla_{\v k}\theta(\v k)
\right]\mathbf{1}
-i\left[
\m\nabla_{\v k}\varepsilon(\v k)\times\m\nabla_{\v k}
\right]\sigma_z
\end{equation}
\begin{equation}
\begin{split}
\v L_{IC}^{mix}(\v k)
={}&i\frac{\Delta}{|f(\v k)|}
\left[
\m\nabla_{\v k}\varepsilon(\v k)\times\m\nabla_{\v k}
\right]\sigma_x\\
&+i\left\{
\frac{\varepsilon(\v k)}{2|f(\v k)|}
\left[
\m\nabla_{\v k}\varepsilon(\v k)\times\m\nabla_{\v k}\theta(\v k)
\right]
-|f(\v k)|
\left[
\m\nabla_{\v k}\theta(\v k)\times\m\nabla_{\v k}
\right]
\right\}\sigma_y.
\end{split}
\label{appC:fourL}
\end{equation}
The repeated multiplicative vector structure in these expressions is therefore

\begin{equation}
\m\nabla_{\v k}\varepsilon(\v k)
\times\m\nabla_{\v k}\theta(\v k),
\end{equation}
while the itinerant operators additionally contain derivatives acting on the coefficient section.

In equilibrium, when the derivative part of $i\v D_{\v k}$ acts on a constant coefficient section, the diagonal elements are

\begin{equation}
[\v L_{LC}(\v k)]_{++}
=[\v L_{LC}(\v k)]_{--}
=-\frac{\Delta}{\varepsilon(\v k)}
\left[
\m\nabla_{\v k}\varepsilon(\v k)
\times\m\nabla_{\v k}\theta(\v k)
\right],
\end{equation}
and

\begin{equation}
[\v L_{IC}(\v k)]_{++}
=[\v L_{IC}(\v k)]_{--}
=\frac{\Delta}{2\varepsilon(\v k)}
\left[
\m\nabla_{\v k}\varepsilon(\v k)
\times\m\nabla_{\v k}\theta(\v k)
\right].
\end{equation}
Thus, for the traceless two band Hamiltonian considered here, the equilibrium bulk combination obeys the pointwise identity

{
\begin{equation}
[\v L_{LC}(\v k)]_{\sigma\sigma}+2[\v L_{IC}(\v k)]_{\sigma\sigma}=0,\qquad \sigma=\pm
\end{equation}
on either diagonal equilibrium band section, independently of the band occupation. The identity is unaffected by an arbitrary constant shift of the energy origin, $H\rightarrow H+C\mathbf{1}$. It is the form, within the Blount construction, of the cancellation between the band orbital moment and the Berry curvature contribution in the modern theory of equilibrium orbital magnetization \\cite{SI:Xiao2010BerryRMP}. This provides a direct analytical check on the two band expressions above.
}

{The pointwise cancellation is a property of the phase gauge representation used here; its gauge independent physical content is the vanishing of the Brillouin zone integrated equilibrium orbital angular momentum.}

It is useful to go one step further and consider the complete bulk operator derived in the main text.

\begin{equation}
\v L_{bulk}(\v k)
=
\v L_{LC}(\v k)+\v L_{LC}^{mix}(\v k)
+2\v L_{IC}(\v k)+2\v L_{IC}^{mix}(\v k).
\label{appC:Lbulkdef}
\end{equation}
For the present particle-hole-symmetric two band Hamiltonian, $\v L_{LC}^{mix}(\v k)=0$, while the multiplicative part of $2\v L_{IC}(\v k)$ cancels $\v L_{LC}(\v k)$ exactly. The complete bulk operator therefore reduces to

\begin{equation}\v L_{bulk}(\v k)
=
-2i\left[
\m\nabla_{\v k}\varepsilon(\v k)\times\m\nabla_{\v k}
\right]\sigma_z
+2\v L_{IC}^{mix}(\v k).
\label{appC:LbulkNEQ}
\end{equation}
Equation~\ref{appC:LbulkNEQ} is a particularly simple property of the particle-hole-symmetric two band model. The first term is diagonal in band space but is a differential operator acting on the $\v k$ dependence of the coefficient section. It therefore vanishes for the constant equilibrium coefficient section, but can be finite for an excited state carrying a non-trivial coefficient texture in the Brillouin zone. For a post-pulse coefficient section of the form discussed in the main text, the dynamical phase produces no contribution because $\m\nabla_{\v k}\varepsilon(\v k)\times\m\nabla_{\v k}\varepsilon(\v k)=0$; the remaining excitation induced phase texture can therefore provide a quasi-stationary contribution after the pulse.

The second term in Eq.~\ref{appC:LbulkNEQ} is purely off diagonal in band space. Its expectation value consequently requires interband coherence. Under field free evolution such coherence carries the relative phase between the two bands and hence gives the familiar oscillatory contribution with frequency set locally by the band splitting $2\varepsilon(\v k)$.

{Thus, for the traceless two band Hamiltonian considered here, the equilibrium bulk integrand vanishes pointwise in the present phase gauge on either diagonal band section, independently of occupation and independently of any constant choice of energy origin. This pointwise statement is representation dependent; its gauge independent content is the vanishing of the Brillouin zone integrated equilibrium orbital angular momentum. The non-vanishing bulk operator content separates into two intrinsically non-equilibrium structures: a diagonal coefficient texture contribution, which may remain quasi-stationary after excitation, and an off diagonal interband coherence contribution, which is generally oscillatory. A momentum dependent band centre term $\bar E(\v k)\mathbf{1}$ is a genuine departure from this traceless structure and generally removes the simple pointwise cancellation.}

\section{Smooth gauge for the two band model}
\label{app:smooth2band}

The phase angle representation of the two band eigenstates is convenient analytically but becomes singular at an avoided crossing centre $f(\v k)=0$. This section reformulates the same two band problem in a smooth gauge and derives expressions for the Blount operator ingredients and orbital angular momentum sectors directly in terms of $f(\v k)$ and its derivatives. These forms remain regular at the avoided crossing and are used for the numerical calculations.

We consider again

\begin{equation}
H(\v k) = \begin{pmatrix}
\Delta & f(\v k) \\
f^\ast(\v k) & -\Delta
\end{pmatrix}
\end{equation}
with

\begin{equation}
\pm\varepsilon(\v k)=\pm\sqrt{\Delta^2+|f(\v k)|^2}.
\end{equation}
Taking $\Delta>0$, a smooth choice of eigenvectors is

\begin{equation}
\ket{u_+(\v k)}=
\frac{1}{\sqrt{2\varepsilon(\v k)[\varepsilon(\v k)+\Delta]}}
\begin{pmatrix}
\varepsilon(\v k)+\Delta \\
f^\ast(\v k)
\end{pmatrix},
\label{appD:uplus}
\end{equation}
and

\begin{equation}
\ket{u_-(\v k)}=
\frac{1}{\sqrt{2\varepsilon(\v k)[\varepsilon(\v k)+\Delta]}}
\begin{pmatrix}
-f(\v k) \\
\varepsilon(\v k)+\Delta
\end{pmatrix}.
\label{appD:uminus}
\end{equation}
Since $\varepsilon(\v k)+\Delta\geq2\Delta>0$, this gauge is smooth wherever $f(\v k)$ is smooth, including at $f(\v k)=0$. For $\Delta<0$ the complementary smooth gauge obtained by interchanging the two pole choices may be used.

A useful identity is

\begin{equation}
\m\nabla_{\v k}\varepsilon(\v k)
=
\frac{
f^\ast(\v k)\m\nabla_{\v k}f(\v k)
+f(\v k)\m\nabla_{\v k}f^\ast(\v k)
}{2\varepsilon(\v k)}.
\label{appD:gradeps}
\end{equation}
The off diagonal elements of the full Berry connection again define

\begin{equation}
\v d(\v k)=\begin{pmatrix}
0&\v p(\v k)\\
\v p^\ast(\v k)&0
\end{pmatrix},
\label{appD:d}
\end{equation}
where direct differentiation of Eqs.~\ref{appD:uplus}-\ref{appD:uminus} gives

\begin{equation}\v p(\v k)
=
i\frac{
f(\v k)\m\nabla_{\v k}\varepsilon(\v k)
-[\varepsilon(\v k)+\Delta]\m\nabla_{\v k}f(\v k)
}{
2\varepsilon(\v k)[\varepsilon(\v k)+\Delta]
}.
\label{appD:p}
\end{equation}
The diagonal elements give

\begin{equation}
\m{\mathcal A}(\v k)=\begin{pmatrix}
\v A(\v k)&0\\
0&-\v A(\v k)
\end{pmatrix},
\label{appD:Afull}
\end{equation}
with

\begin{equation}\v A(\v k)
=
i\frac{
f(\v k)\m\nabla_{\v k}f^\ast(\v k)
-f^\ast(\v k)\m\nabla_{\v k}f(\v k)
}{
4\varepsilon(\v k)[\varepsilon(\v k)+\Delta]
}.
\label{appD:A}
\end{equation}
Both quantities are regular at $f(\v k)=0$. In particular,

\begin{equation}
\v p(\v k)\longrightarrow
-\frac{i}{2\Delta}\m\nabla_{\v k}f(\v k),
\qquad
\v A(\v k)\longrightarrow0,
\qquad f(\v k)\rightarrow0.
\label{appD:limits}
\end{equation}
The remaining objects entering the four orbital angular momentum operators are

\begin{equation}
E(\v k)=\begin{pmatrix}
\varepsilon(\v k)&0\\
0&-\varepsilon(\v k)
\end{pmatrix},
\qquad
\m\nabla_{\v k}E(\v k)=\begin{pmatrix}
\m\nabla_{\v k}\varepsilon(\v k)&0\\
0&-\m\nabla_{\v k}\varepsilon(\v k)
\end{pmatrix},
\label{appD:E}
\end{equation}
and

\begin{equation}
i\v D_{\v k}
=i\mathbf{1}\m\nabla_{\v k}+\m{\mathcal A}(\v k)
=
\begin{pmatrix}
i\m\nabla_{\v k}+\v A(\v k)&0\\
0&i\m\nabla_{\v k}-\v A(\v k)
\end{pmatrix}.
\label{appD:D}
\end{equation}
Where the phase $\theta(\v k)$ of Sec.~\ref{app:twoband} is well defined, the repeated geometrical structure of that section can be written in the explicitly regular form

\begin{equation}
\m\nabla_{\v k}\varepsilon(\v k)
\times\m\nabla_{\v k}\theta(\v k)
=
\frac{i}{2\varepsilon(\v k)}
\left[
\m\nabla_{\v k}f(\v k)
\times\m\nabla_{\v k}f^\ast(\v k)
\right].
\label{appD:textureidentity}
\end{equation}
The right hand side remains regular at $f(\v k)=0$ and will be used below.

\subsection{Four orbital angular momentum operators}

The basic commutator entering both interband terms is again

\begin{equation}
[\v d(\v k),E(\v k)]
=2\varepsilon(\v k)
\begin{pmatrix}
0&-\v p(\v k)\\
\v p^\ast(\v k)&0
\end{pmatrix}.
\label{appD:dE}
\end{equation}

\subsubsection{Local circulation}

For the pure local circulation operator,

\begin{equation}
\v L_{LC}(\v k)
=\frac{i}{2}\left(
[\v d(\v k),E(\v k)]\times\v d(\v k)
-\v d(\v k)\times[\v d(\v k),E(\v k)]
\right),
\end{equation}
Eq.~\ref{appD:dE} gives

\begin{equation}
\v L_{LC}(\v k)
=-2i\varepsilon(\v k)
\left[\v p(\v k)\times\v p^\ast(\v k)\right]\mathbf{1}.
\label{appD:LCp}
\end{equation}
Using Eq.~\ref{appD:p},

\begin{equation}
\v p(\v k)\times\v p^\ast(\v k)
=
\frac{\Delta}{4\varepsilon(\v k)^3}
\left[
\m\nabla_{\v k}f(\v k)
\times\m\nabla_{\v k}f^\ast(\v k)
\right],
\label{appD:pxp}
\end{equation}
and hence

\begin{equation}\v L_{LC}(\v k)
=-i\frac{\Delta}{2\varepsilon(\v k)^2}
\left[
\m\nabla_{\v k}f(\v k)
\times\m\nabla_{\v k}f^\ast(\v k)
\right]\mathbf{1}.
\label{appD:LC}
\end{equation}
Using Eq.~\ref{appD:textureidentity}, this is exactly the result of Sec.~\ref{app:twoband}. In particular, the identity structure of the pure local circulation operator is unchanged by the gauge choice.

The mixed local circulation operator is

\begin{equation}
\v L_{LC}^{mix}(\v k)
=\frac{1}{2}\left[
\v d(\v k)\times\m\nabla_{\v k}E(\v k)
-\m\nabla_{\v k}E(\v k)\times\v d(\v k)
\right].
\end{equation}
The upper off diagonal element is

\begin{equation}
\frac{1}{2}\left[
\v p(\v k)\times(-\m\nabla_{\v k}\varepsilon(\v k))
-\m\nabla_{\v k}\varepsilon(\v k)\times\v p(\v k)
\right]=0,
\end{equation}
and the lower off diagonal element vanishes identically in the same way. Consequently,

\begin{equation}
\v L_{LC}^{mix}(\v k)=0.
\label{appD:LCmix}
\end{equation}
As in Sec.~\ref{app:twoband}, this vanishing follows from the symmetric spectrum $E_+(\v k)+E_-(\v k)=0$ and is independent of the gauge choice.

\subsubsection{Itinerant circulation}

For the pure itinerant circulation operator,

\begin{equation}
\v L_{IC}(\v k)
=-\m\nabla_{\v k}E(\v k)\times i\v D_{\v k},
\end{equation}
Eqs.~\ref{appD:A}-\ref{appD:D} give

\begin{equation}
\v L_{IC}(\v k)=
\begin{pmatrix}
-\m\nabla_{\v k}\varepsilon(\v k)\times
[i\m\nabla_{\v k}+\v A(\v k)]&0\\
0&
\m\nabla_{\v k}\varepsilon(\v k)\times
[i\m\nabla_{\v k}-\v A(\v k)]
\end{pmatrix}.
\label{appD:ICmatrix}
\end{equation}
From Eqs.~\ref{appD:gradeps} and \ref{appD:A},

\begin{equation}
\m\nabla_{\v k}\varepsilon(\v k)\times\v A(\v k)
=
i\frac{\varepsilon(\v k)-\Delta}{4\varepsilon(\v k)^2}
\left[
\m\nabla_{\v k}f(\v k)
\times\m\nabla_{\v k}f^\ast(\v k)
\right],
\label{appD:gradExA}
\end{equation}
and therefore

\begin{equation}\v L_{IC}(\v k)
=
-i\frac{\varepsilon(\v k)-\Delta}{4\varepsilon(\v k)^2}
\left[
\m\nabla_{\v k}f(\v k)
\times\m\nabla_{\v k}f^\ast(\v k)
\right]\mathbf{1}
-i\left[
\m\nabla_{\v k}\varepsilon(\v k)\times\m\nabla_{\v k}
\right]\sigma_z.
\label{appD:IC}
\end{equation}
The multiplicative and differential pieces differ from their allocation in Sec.~\ref{app:twoband} because the decomposition $i\v D=i\m\nabla_{\v k}+\m{\mathcal A}$ is gauge dependent. The complete operator acting on the correspondingly transformed coefficient section is gauge covariant.

Finally consider

\begin{equation}
\v L_{IC}^{mix}(\v k)
=-\frac{i}{2}\left[
 i\v D_{\v k}\times[\v d(\v k),E(\v k)]
-[\v d(\v k),E(\v k)]\times i\v D_{\v k}
\right].
\label{appD:ICmixdef}
\end{equation}
Writing only for this intermediate step

\begin{equation}
\v K_\pm(\v k)=i\m\nabla_{\v k}\pm\v A(\v k),
\end{equation}
the upper off diagonal element is

\begin{equation}
[\v L_{IC}^{mix}(\v k)]_{+-}
=i\left\{
\v K_+(\v k)\times[\varepsilon(\v k)\v p(\v k)]
-[\varepsilon(\v k)\v p(\v k)]\times\v K_-(\v k)
\right\}.
\end{equation}
Using the differential operator identity of Eq.~\ref{appC:diffidentity}, the terms containing $\v A(\v k)$ again cancel and one obtains

\begin{equation}
[\v L_{IC}^{mix}(\v k)]_{+-}
=-\m\nabla_{\v k}\times[\varepsilon(\v k)\v p(\v k)]
+2\varepsilon(\v k)\v p(\v k)\times\m\nabla_{\v k},
\label{appD:ICmixupper}
\end{equation}
and

\begin{equation}
[\v L_{IC}^{mix}(\v k)]_{-+}
=\m\nabla_{\v k}\times[\varepsilon(\v k)\v p^\ast(\v k)]
-2\varepsilon(\v k)\v p^\ast(\v k)\times\m\nabla_{\v k}.
\label{appD:ICmixlower}
\end{equation}
These forms are already regular everywhere and may be used directly in a numerical implementation. They can also be written explicitly in terms of the smooth Hamiltonian function $f(\v k)$. Equation~\ref{appD:p} gives

\begin{equation}
\varepsilon(\v k)\v p(\v k)
=\frac{i}{2}\left[
\frac{f(\v k)}{\varepsilon(\v k)+\Delta}
\m\nabla_{\v k}\varepsilon(\v k)
-\m\nabla_{\v k}f(\v k)
\right].
\label{appD:epsp}
\end{equation}
Since $\m\nabla_{\v k}\varepsilon$ and $\m\nabla_{\v k}f$ are gradients,

\begin{equation}
-\m\nabla_{\v k}\times[\varepsilon(\v k)\v p(\v k)]
=
\frac{i}{2[\varepsilon(\v k)+\Delta]}
\left[
\m\nabla_{\v k}\varepsilon(\v k)
\times\m\nabla_{\v k}f(\v k)
\right].
\label{appD:curlepsp1}
\end{equation}
Using Eq.~\ref{appD:gradeps},

\begin{equation}
\m\nabla_{\v k}\varepsilon(\v k)
\times\m\nabla_{\v k}f(\v k)
=-\frac{f(\v k)}{2\varepsilon(\v k)}
\left[
\m\nabla_{\v k}f(\v k)
\times\m\nabla_{\v k}f^\ast(\v k)
\right],
\label{appD:gradExgradf}
\end{equation}
so that

\begin{equation}
\v L_{IC}^{mix}(\v k)=
\begin{pmatrix}
0&\bm{\Lambda}(\v k)\\
\bm{\Lambda}^\dagger(\v k)&0
\end{pmatrix},
\label{appD:ICmixmatrix}
\end{equation}
with

\begin{eqnarray}
\bm{\Lambda}(\v k)
&=&
-i\frac{f(\v k)}{4\varepsilon(\v k)[\varepsilon(\v k)+\Delta]}
\left[
\m\nabla_{\v k}f(\v k)
\times\m\nabla_{\v k}f^\ast(\v k)
\right]
\nonumber\\
&&+i\left[
\frac{f(\v k)}{\varepsilon(\v k)+\Delta}
\m\nabla_{\v k}\varepsilon(\v k)
-\m\nabla_{\v k}f(\v k)
\right]\times\m\nabla_{\v k}.
\label{appD:Lambda}
\end{eqnarray}
The corresponding lower off diagonal element is

\begin{eqnarray}
\bm{\Lambda}^\dagger(\v k)
&=&
+i\frac{f^\ast(\v k)}{4\varepsilon(\v k)[\varepsilon(\v k)+\Delta]}
\left[
\m\nabla_{\v k}f(\v k)
\times\m\nabla_{\v k}f^\ast(\v k)
\right]
\nonumber\\
&&+i\left[
\frac{f^\ast(\v k)}{\varepsilon(\v k)+\Delta}
\m\nabla_{\v k}\varepsilon(\v k)
-\m\nabla_{\v k}f^\ast(\v k)
\right]\times\m\nabla_{\v k}.
\label{appD:Lambdadag}
\end{eqnarray}
No factor $1/|f(\v k)|$ and no derivative of a phase angle appears in Eqs.~\ref{appD:Lambda}-\ref{appD:Lambdadag}. At $f(\v k)=0$, Eq.~\ref{appD:gradeps} gives $\m\nabla_{\v k}\varepsilon(\v k)=0$, and therefore

\begin{equation}
\bm{\Lambda}(\v k)\longrightarrow
-i\m\nabla_{\v k}f(\v k)\times\m\nabla_{\v k},
\qquad
\bm{\Lambda}^\dagger(\v k)\longrightarrow
-i\m\nabla_{\v k}f^\ast(\v k)\times\m\nabla_{\v k}.
\label{appD:Lambdalimit}
\end{equation}
Thus the mixed itinerant circulation operator remains finite at the avoided crossing centre.

\subsubsection{Pauli matrix form and structural summary}

After the explicit matrices have been established, it is again useful to introduce

\begin{equation}
\sigma_x=\begin{pmatrix}
0&1\\
1&0
\end{pmatrix},
\qquad
\sigma_y=\begin{pmatrix}
0&-i\\
i&0
\end{pmatrix},
\qquad
\sigma_z=\begin{pmatrix}
1&0\\
0&-1
\end{pmatrix},
\label{appD:pauli}
\end{equation}
solely to expose the band space structure. The basic objects take the form

\begin{equation}
\v d(\v k)
=\operatorname{Re}\v p(\v k)\,\sigma_x
-\operatorname{Im}\v p(\v k)\,\sigma_y,
\qquad
\m{\mathcal A}(\v k)=\v A(\v k)\sigma_z,
\qquad
E(\v k)=\varepsilon(\v k)\sigma_z.
\label{appD:pauliobjects}
\end{equation}
The four orbital angular momentum operators are therefore summarized as

\begin{equation}\v L_{LC}(\v k)
=-i\frac{\Delta}{2\varepsilon(\v k)^2}
\left[
\m\nabla_{\v k}f(\v k)
\times\m\nabla_{\v k}f^\ast(\v k)
\right]\mathbf{1},
\end{equation}

\begin{equation}
\v L_{LC}^{mix}(\v k)=0,
\end{equation}

\begin{equation}\v L_{IC}(\v k)
=
-i\frac{\varepsilon(\v k)-\Delta}{4\varepsilon(\v k)^2}
\left[
\m\nabla_{\v k}f(\v k)
\times\m\nabla_{\v k}f^\ast(\v k)
\right]\mathbf{1}
-i\left[
\m\nabla_{\v k}\varepsilon(\v k)\times\m\nabla_{\v k}
\right]\sigma_z,
\end{equation}
and

\begin{equation}\v L_{IC}^{mix}(\v k)=
\begin{pmatrix}
0&\bm{\Lambda}(\v k)\\
\bm{\Lambda}^\dagger(\v k)&0
\end{pmatrix},
\end{equation}
where Eqs.~\ref{appD:Lambda}-\ref{appD:Lambdadag} give the explicitly regular differential operators $\bm{\Lambda}$ and $\bm{\Lambda}^\dagger$.

The complete bulk operator is

\begin{equation}
\v L_{bulk}(\v k)
=
\v L_{LC}(\v k)
+\v L_{LC}^{mix}(\v k)
+2\v L_{IC}(\v k)
+2\v L_{IC}^{mix}(\v k),
\end{equation}
and hence

\begin{equation}\begin{split}
\v L_{bulk}(\v k)
={}&
-\frac{i}{2\varepsilon(\v k)}
\left[
\m\nabla_{\v k}f(\v k)
\times\m\nabla_{\v k}f^\ast(\v k)
\right]\mathbf{1}
\\
&-2i\left[
\m\nabla_{\v k}\varepsilon(\v k)\times\m\nabla_{\v k}
\right]\sigma_z
+2\v L_{IC}^{mix}(\v k).
\end{split}
\label{appD:Lbulk}
\end{equation}
The additional multiplicative term in Eq.~\ref{appD:Lbulk}, relative to the particularly compact expression obtained in Sec.~\ref{app:twoband}, reflects only the redistribution between the Berry connection and coefficient derivative pieces under the gauge transformation. Away from a zero of $f(\v k)$ the smooth eigenvectors are related to those of Sec.~\ref{app:twoband} by

\begin{equation}
\ket{u_+^{\rm sm}(\v k)}
=e^{-i\theta(\v k)/2}\ket{u_+^{\rm ph}(\v k)},
\qquad
\ket{u_-^{\rm sm}(\v k)}
=-e^{+i\theta(\v k)/2}\ket{u_-^{\rm ph}(\v k)},
\label{appD:gaugetrans}
\end{equation}
so that the same physical coefficient section is represented by

\begin{equation}
c_+^{\rm sm}(\v k)
=e^{+i\theta(\v k)/2}c_+^{\rm ph}(\v k),
\qquad
c_-^{\rm sm}(\v k)
=-e^{-i\theta(\v k)/2}c_-^{\rm ph}(\v k).
\label{appD:coeftrans}
\end{equation}
Using Eq.~\ref{appD:textureidentity}, the additional multiplicative term in Eq.~\ref{appD:Lbulk} is cancelled exactly by the derivative of these phases. Thus Eqs.~\ref{appD:Lbulk} and \ref{appC:LbulkNEQ} represent the same bulk operator on the correspondingly transformed coefficient section.

For numerical evaluation, the regular expressions are therefore Eqs.~\ref{appD:gradeps}, \ref{appD:p}, \ref{appD:A}, \ref{appD:ICmixupper}-\ref{appD:ICmixlower}, or equivalently the fully explicit Eqs.~\ref{appD:Lambda}-\ref{appD:Lambdadag}. All quantities are expressed directly in terms of the smooth Hamiltonian function $f(\v k)$ and its derivatives; no phase angle $\theta(\v k)$, division by $|f(\v k)|$, or removal of the avoided crossing point from the numerical mesh is required.


\section{Filled band phase freedom in gapped graphene: a patchwise Stokes proof}
\label{app:gg-patchwise-stokes}

For a completely filled isolated band, a momentum dependent phase function applied to each state leaves the occupied projector, and hence the physical filled band state, unchanged. Working in a fixed smooth Bloch frame, we represent this freedom as

\begin{equation}
c_{\mathbf k}\rightarrow e^{i\eta(\mathbf k)}c_{\mathbf k}.
\end{equation}
The local IC density need not be invariant under this transformation, but its Brillouin zone integral must be. For a smooth $U(1)$ phase whose differential $d\eta$ is globally defined on the Brillouin zone torus, this follows immediately because the induced change in IC is a total derivative.

The only subtlety for gapped graphene arises from the half-angle phase relating the conventional phase gauge to a smooth gauge. Although the band itself is topologically trivial and therefore admits a globally smooth Bloch frame, the phase

\begin{equation}
e^{i\theta(\mathbf k)}=\frac{f(\mathbf k)}{|f(\mathbf k)|}
\end{equation}
is singular at the zeros of $f(\mathbf k)$, so that $\eta=\mp\theta/2$ is necessarily branch dependent. The total derivative argument must therefore be applied patchwise. As we show below, the apparent puncture contributions are exactly cancelled by the patch transition contribution, leaving the Brillouin zone integrated IC invariant.

\subsection{Phase gauge and smooth gauge}

For

\begin{equation}
 H(\vk)=
 \begin{pmatrix}
 \Delta & f(\vk)\\
 f^*(\vk)&-\Delta
 \end{pmatrix},
 \qquad
 \varepsilon(\vk)=\sqrt{\Delta^2+|f(\vk)|^2},
\end{equation}
a convenient phase gauge is

\begin{equation}
 |u_\sigma^{\rm ph}(\vk)\rangle
 =\frac{1}{\sqrt2}
 \begin{pmatrix}
 g_\sigma(\vk)e^{i\theta(\vk)/2}\\
 \sigma g_{-\sigma}(\vk)e^{-i\theta(\vk)/2}
 \end{pmatrix},
\end{equation}
whereas for $\Delta>0$ a smooth gauge is

\begin{align}
 |u_+^{\rm sm}(\vk)\rangle
 &=\frac{1}{\sqrt{2\varepsilon(\varepsilon+\Delta)}}
 \begin{pmatrix}
 \varepsilon+\Delta\\ f^*
 \end{pmatrix},\\
 |u_-^{\rm sm}(\vk)\rangle
 &=\frac{1}{\sqrt{2\varepsilon(\varepsilon+\Delta)}}
 \begin{pmatrix}
 -f\\ \varepsilon+\Delta
 \end{pmatrix}.
\end{align}
Away from a zero of $f$ the two are related by

\begin{equation}
 |u_+^{\rm sm}\rangle=e^{-i\theta/2}|u_+^{\rm ph}\rangle,
 \qquad
 |u_-^{\rm sm}\rangle=-e^{+i\theta/2}|u_-^{\rm ph}\rangle.
\label{gg:eq:gaugerel}
\end{equation}
Thus the same physical state is represented in the smooth gauge by the coefficient section

\begin{equation}
 c_+^{\rm sm}=e^{+i\theta/2}c_+^{\rm ph},
 \qquad
 c_-^{\rm sm}=-e^{-i\theta/2}c_-^{\rm ph}.
\label{gg:eq:coefrel}
\end{equation}
For the occupied valence band the gauge transformation therefore identifies the half-angle phase

\begin{equation}
 \eta(\vk)=-\frac{\theta(\vk)}{2},
\label{gg:eq:eta}
\end{equation}

up to an irrelevant constant $\pi$. Equations~\eqref{gg:eq:gaugerel} and \eqref{gg:eq:coefrel} themselves describe an ordinary Bloch frame gauge transformation, for which the basis and coefficients transform together and the covariant IC integrand is pointwise unchanged, as in Sec.~\ref{app:chern}. In what follows we instead use Eq.~\eqref{gg:eq:eta} to identify the phase texture and ask what happens when that occupied state representative is rephased while the smooth reference frame is held fixed.

The smooth gauge itself is completely regular at $f=0$; the singularity belongs only to the chosen half-angle phase representative. Unlike the Chern band case of Sec.~\ref{app:chern}, no topological obstruction forces this singularity. The singular half-angle representative itself need not lie in the domain of the position operator; the argument below is understood as the limit of smooth rephasings.

\subsection{Local filled band phase change of IC}

For one filled band the covariant position expectation entering the IC sector is

\begin{equation}
 c_{\vk}^\dagger i\mathbf D_{\vk}c_{\vk}.
\end{equation}
If, on a smooth patch, $c_{\vk}\rightarrow e^{i\eta(\vk)}c_{\vk}$, then

\begin{equation}
 c_{\vk}^\dagger i\mathbf D_{\vk}c_{\vk}
 \longrightarrow
 c_{\vk}^\dagger i\mathbf D_{\vk}c_{\vk}-|c_{\vk}|^2\grad\eta.
\end{equation}
For the normalized filled band representative used here, $|c_{\vk}|^2=1$. Hence the local change of the pure IC density is

\begin{equation}
 \delta \ell_{\rm IC}(\vk)
 =\grad E(\vk)\times\grad\eta(\vk).
\label{gg:eq:localchange}
\end{equation}
In two dimensions it is useful to write this as a differential form,

\begin{equation}
 \delta \ell_{\rm IC}\,d^2k
 =dE\wedge d\eta
 =d\!\left(E\,d\eta\right),
\label{gg:eq:exact}
\end{equation}
where the last equality is valid on every patch on which $\eta$ is smooth. For a smooth phase factor with globally defined $d\eta$, Eq.~\eqref{gg:eq:exact} immediately gives zero after integration over the Brillouin zone torus.

\subsection{Role of the patch transition contribution}

For gapped graphene, Eq.~\eqref{gg:eq:eta} is not globally smooth.  Around a zero of $f$, $\theta$ winds by $\pm2\pi$, and therefore $\eta=-\theta/2$ changes by
$\mp\pi$.  One must therefore choose branches, or equivalently cover the Brillouin zone by patches.

If one removes small disks around the zeros of $f$, applies Stokes' theorem to the remaining region, and retains only the small circle boundaries, one obtains terms of the schematic form

\begin{equation}
 \sum_i E(\vK_i)\oint_{C_i}d\eta.
\label{gg:eq:spurious}
\end{equation}
For the two graphene valleys the windings are opposite.  If $E(\vK)=E(\vK')$, Eq.~\eqref{gg:eq:spurious} may cancel accidentally.  But such a cancellation cannot be the fundamental proof: filled band phase freedom cannot depend on equality of the valley energies.

The missing piece is the contribution obtained when the branch cut is regularized by a narrow interpolation strip between the two branches.

\subsection{Patchwise Stokes proof}

Choose branch cuts joining the singular points so that $\eta$ is single valued on each patch.  On every smooth patch $U_\alpha$,

\begin{equation}
 \int_{U_\alpha} dE\wedge d\eta_\alpha
 =\oint_{\partial U_\alpha} E\,d\eta_\alpha.
\label{gg:eq:patchstokes}
\end{equation}
Summing over all patches, there is no external boundary because the Brillouin zone is a torus.  What remains are internal patch boundaries.  The two sides of an ordinary internal boundary cancel.  At a branch cut, however, the local representatives of the phase differ by a finite jump

\begin{equation}
 \Delta\eta=\eta_+-\eta_-.
\end{equation}
For the half-angle phase of gapped graphene, $\Delta\eta=\pm\pi$.

To keep this contribution, consider smooth rephasings that interpolate between the two branches across a narrow strip and then take the zero width limit. In local coordinates $(s,n)$, with $s$ along the cut and $n$ across it, the rapid variation is

\begin{equation}
 \partial_n\eta\longrightarrow \Delta\eta\,\delta(n).
\end{equation}
The corresponding contribution to Eq.~\eqref{gg:eq:localchange} is therefore

\begin{equation}
 \delta I_{\Gamma}
 =\Delta\eta\int_{\Gamma} ds\,\partial_s E
 =\Delta\eta\,[E(\Gamma_{\rm end})-E(\Gamma_{\rm start})],
\label{gg:eq:cutterm}
\end{equation}
with the sign fixed by the orientation of the cut and the chosen branch jump.

Exactly the same endpoint energies occur in the small circle terms generated by Eq.~\eqref{gg:eq:patchstokes}, with the opposite sign. Thus the cut term Eq.~\eqref{gg:eq:cutterm} cancels the puncture contribution identically.  In particular, the cancellation does \emph{not} require $E(\vK)=E(\vK')$.

Equivalently, once both the smooth patch derivative and the interpolation strip contribution are included, $d\eta$ is the complete derivative of the phase representative on the closed torus.  Hence

\begin{equation}
 \int_{\rm BZ} d^2k\,
 \grad E\times\grad\eta
 =\int_{T^2} d(E\,d\eta)
 =0.
\label{gg:eq:mainresult}
\end{equation}
Equation~\eqref{gg:eq:mainresult} is simply the smooth phase total derivative
argument, now written correctly for a phase which requires patches.

\subsection{Consequences for the four orbital angular momentum terms}

The filled band phase acts only on the coefficient section.  The two LC operators contain no derivative acting on that section, and therefore remain pointwise unchanged. The IC sector contains $i\mathbf D_{\vk}$ and hence its local momentum space density is shifted by the local lever arm term Eq.~\eqref{gg:eq:localchange}.  For the singular half-angle phase of gapped
graphene that local change must be interpreted patchwise, but its complete Brillouin zone integral vanishes by Eq.~\eqref{gg:eq:mainresult}.

Thus the gapped graphene singularity does not spoil filled band phase
invariance.  It only invalidates the naive step in which one treats
$\eta=-\theta/2$ as though it were a globally smooth scalar and then discards
the branch cut contribution.

\subsection{Consistency checks}

\paragraph{Unequal valley energies.}
If one artificially allows $E(\vK)\neq E(\vK')$, the puncture terms alone are
nonzero.  The branch cut term is then also nonzero and cancels them.  This is a
useful diagnostic that the proof has retained all patch data.

\paragraph{Projector form.}
For a completely filled isolated band the occupied projector

\begin{equation}
 P_{\vk}=|u_{\vk}\rangle\langle u_{\vk}|
\end{equation}
is unchanged by any rephasing of the occupied orbital.  The standard equilibrium IC expression written in terms of Berry curvature is therefore manifestly phase independent.  The patchwise Stokes proof above is the coefficient section version of the same statement within the Blount construction.

\paragraph{Relation to active translation.}
For the active translation phase factor $e^{-i\vk\cdot\mathbf a}$, with globally defined $d\eta=-\mathbf a\cdot d\vk$, the local change reduces to the familiar lever arm shift $\mathbf a\times\mathbf v(\vk)$.  In equilibrium its Brillouin zone integral vanishes either as the same total derivative on the torus or, equivalently, because the filled band has $\langle\mathbf v\rangle=0$.

Taken together, these results show that the half-angle phase $\eta=\mp\theta/2$ introduces no ambiguity in the filled band observable. Away from the patch boundaries the IC change is the exact form $dE\wedge d\eta=d(E\,d\eta)$, while the apparent puncture terms generated by patchwise Stokes are cancelled exactly by the branch cut transition term. The Brillouin zone integrated equilibrium IC is therefore invariant without any assumption that the two valley energies are equal.



\begin{thebibliography}{10}
\expandafter\ifx\csname url\endcsname\relax
  \def\url#1{\burl{#1}}\fi
\expandafter\ifx\csname urlprefix\endcsname\relax\def\urlprefix{URL }\fi
\providecommand{\bibinfo}[2]{#2}
\providecommand{\eprint}[2][]{\url{#2}}
\providecommand{\doi}[1]{\url{https://doi.org/#1}}
\bibcommenthead

\bibitem{Choi2023OHE}
\bibinfo{author}{Choi, Y.-G.} \emph{et~al.}
\newblock \bibinfo{title}{Observation of the orbital {Hall} effect in a light
  metal {Ti}}.
\newblock \emph{\bibinfo{journal}{Nature}} \textbf{\bibinfo{volume}{619}},
  \bibinfo{pages}{52--56} (\bibinfo{year}{2023}).

\bibitem{Seifert2023BallisticOAM}
\bibinfo{author}{Seifert, T.~S.} \emph{et~al.}
\newblock \bibinfo{title}{Time-domain observation of ballistic
  orbital-angular-momentum currents with giant relaxation length in tungsten}.
\newblock \emph{\bibinfo{journal}{Nature Nanotechnology}}
  \textbf{\bibinfo{volume}{18}}, \bibinfo{pages}{1132--1138}
  (\bibinfo{year}{2023}).

\bibitem{GoLee2020OrbitalTorque}
\bibinfo{author}{Go, D.} \& \bibinfo{author}{Lee, H.-W.}
\newblock \bibinfo{title}{Orbital torque: Torque generation by orbital current
  injection}.
\newblock \emph{\bibinfo{journal}{Physical Review Research}}
  \textbf{\bibinfo{volume}{2}}, \bibinfo{pages}{013177} (\bibinfo{year}{2020}).

\bibitem{Ding2020OrbitalSpin}
\bibinfo{author}{Ding, S.} \emph{et~al.}
\newblock \bibinfo{title}{Harnessing orbital-to-spin conversion of interfacial
  orbital currents for efficient spin-orbit torques}.
\newblock \emph{\bibinfo{journal}{Physical Review Letters}}
  \textbf{\bibinfo{volume}{125}}, \bibinfo{pages}{177201}
  (\bibinfo{year}{2020}).

\bibitem{Kim2021OrbitalInjection}
\bibinfo{author}{Kim, J.} \emph{et~al.}
\newblock \bibinfo{title}{Nontrivial torque generation by orbital angular
  momentum injection in ferromagnetic-metal/{Cu}/{Al$_2$O$_3$} trilayers}.
\newblock \emph{\bibinfo{journal}{Physical Review B}}
  \textbf{\bibinfo{volume}{103}}, \bibinfo{pages}{L020407}
  (\bibinfo{year}{2021}).

\bibitem{Lee2021OrbitalTorque}
\bibinfo{author}{Lee, D.} \emph{et~al.}
\newblock \bibinfo{title}{Orbital torque in magnetic bilayers}.
\newblock \emph{\bibinfo{journal}{Nature Communications}}
  \textbf{\bibinfo{volume}{12}}, \bibinfo{pages}{6710} (\bibinfo{year}{2021}).

\bibitem{GoJoLeeKlaeuiMokrousov2021}
\bibinfo{author}{Go, D.}, \bibinfo{author}{Jo, D.}, \bibinfo{author}{Lee,
  H.-W.}, \bibinfo{author}{Kl{\"a}ui, M.} \& \bibinfo{author}{Mokrousov, Y.}
\newblock \bibinfo{title}{Orbitronics: Orbital currents in solids}.
\newblock \emph{\bibinfo{journal}{EPL (Europhysics Letters)}}
  \textbf{\bibinfo{volume}{135}}, \bibinfo{pages}{37001}
  (\bibinfo{year}{2021}).

\bibitem{Park2011RashbaOAM}
\bibinfo{author}{Park, S.~R.}, \bibinfo{author}{Kim, C.~H.},
  \bibinfo{author}{Yu, J.}, \bibinfo{author}{Han, J.~H.} \&
  \bibinfo{author}{Kim, C.}
\newblock \bibinfo{title}{Orbital-angular-momentum based origin of
  {Rashba}-type surface band splitting}.
\newblock \emph{\bibinfo{journal}{Physical Review Letters}}
  \textbf{\bibinfo{volume}{107}}, \bibinfo{pages}{156803}
  (\bibinfo{year}{2011}).

\bibitem{Park2012ChiralOAM}
\bibinfo{author}{Park, S.~R.} \emph{et~al.}
\newblock \bibinfo{title}{Chiral orbital-angular momentum in the surface states
  of {Bi$_2$Se$_3$}}.
\newblock \emph{\bibinfo{journal}{Physical Review Letters}}
  \textbf{\bibinfo{volume}{108}}, \bibinfo{pages}{046805}
  (\bibinfo{year}{2012}).

\bibitem{XiaoShiNiu2005}
\bibinfo{author}{Xiao, D.}, \bibinfo{author}{Shi, J.} \& \bibinfo{author}{Niu,
  Q.}
\newblock \bibinfo{title}{Berry phase correction to electron density of states
  in solids}.
\newblock \emph{\bibinfo{journal}{Physical Review Letters}}
  \textbf{\bibinfo{volume}{95}}, \bibinfo{pages}{137204}
  (\bibinfo{year}{2005}).

\bibitem{ThonhauserCeresoliVanderbiltResta2005}
\bibinfo{author}{Thonhauser, T.}, \bibinfo{author}{Ceresoli, D.},
  \bibinfo{author}{Vanderbilt, D.} \& \bibinfo{author}{Resta, R.}
\newblock \bibinfo{title}{Orbital magnetization in periodic insulators}.
\newblock \emph{\bibinfo{journal}{Physical Review Letters}}
  \textbf{\bibinfo{volume}{95}}, \bibinfo{pages}{137205}
  (\bibinfo{year}{2005}).

\bibitem{CeresoliThonhauserVanderbiltResta2006}
\bibinfo{author}{Ceresoli, D.}, \bibinfo{author}{Thonhauser, T.},
  \bibinfo{author}{Vanderbilt, D.} \& \bibinfo{author}{Resta, R.}
\newblock \bibinfo{title}{Orbital magnetization in crystalline solids:
  Multi-band insulators, {Chern} insulators, and metals}.
\newblock \emph{\bibinfo{journal}{Physical Review B}}
  \textbf{\bibinfo{volume}{74}}, \bibinfo{pages}{024408}
  (\bibinfo{year}{2006}).

\bibitem{ShiVignaleXiaoNiu2007}
\bibinfo{author}{Shi, J.}, \bibinfo{author}{Vignale, G.},
  \bibinfo{author}{Xiao, D.} \& \bibinfo{author}{Niu, Q.}
\newblock \bibinfo{title}{Quantum theory of orbital magnetization and its
  generalization to interacting systems}.
\newblock \emph{\bibinfo{journal}{Physical Review Letters}}
  \textbf{\bibinfo{volume}{99}}, \bibinfo{pages}{197202}
  (\bibinfo{year}{2007}).

\bibitem{SouzaVanderbilt2008}
\bibinfo{author}{Souza, I.} \& \bibinfo{author}{Vanderbilt, D.}
\newblock \bibinfo{title}{Dichroic f-sum rule and the orbital magnetization of
  crystals}.
\newblock \emph{\bibinfo{journal}{Physical Review B}}
  \textbf{\bibinfo{volume}{77}}, \bibinfo{pages}{054438}
  (\bibinfo{year}{2008}).

\bibitem{ZhongMooreSouza2016}
\bibinfo{author}{Zhong, S.}, \bibinfo{author}{Moore, J.~E.} \&
  \bibinfo{author}{Souza, I.}
\newblock \bibinfo{title}{Gyrotropic magnetic effect and the magnetic moment on
  the {Fermi} surface}.
\newblock \emph{\bibinfo{journal}{Physical Review Letters}}
  \textbf{\bibinfo{volume}{116}}, \bibinfo{pages}{077201}
  (\bibinfo{year}{2016}).

\bibitem{YodaYokoyamaMurakami2018}
\bibinfo{author}{Yoda, T.}, \bibinfo{author}{Yokoyama, T.} \&
  \bibinfo{author}{Murakami, S.}
\newblock \bibinfo{title}{Orbital {Edelstein} effect as a condensed-matter
  analog of solenoids}.
\newblock \emph{\bibinfo{journal}{Nano Letters}} \textbf{\bibinfo{volume}{18}},
  \bibinfo{pages}{916--920} (\bibinfo{year}{2018}).

\bibitem{HaraBahramyMurakami2020}
\bibinfo{author}{Hara, D.}, \bibinfo{author}{Bahramy, M.~S.} \&
  \bibinfo{author}{Murakami, S.}
\newblock \bibinfo{title}{Current-induced orbital magnetization in systems
  without inversion symmetry}.
\newblock \emph{\bibinfo{journal}{Physical Review B}}
  \textbf{\bibinfo{volume}{102}}, \bibinfo{pages}{184404}
  (\bibinfo{year}{2020}).

\bibitem{JohanssonGobelHenkBibesMertig2021}
\bibinfo{author}{Johansson, A.}, \bibinfo{author}{G{\"o}bel, B.},
  \bibinfo{author}{Henk, J.}, \bibinfo{author}{Bibes, M.} \&
  \bibinfo{author}{Mertig, I.}
\newblock \bibinfo{title}{Spin and orbital {Edelstein} effects in a
  two-dimensional electron gas: Theory and application to interfaces}.
\newblock \emph{\bibinfo{journal}{Physical Review Research}}
  \textbf{\bibinfo{volume}{3}}, \bibinfo{pages}{013275} (\bibinfo{year}{2021}).

\bibitem{BaekHanCheonLee2024}
\bibinfo{author}{Baek, I.}, \bibinfo{author}{Han, S.}, \bibinfo{author}{Cheon,
  S.} \& \bibinfo{author}{Lee, H.-W.}
\newblock \bibinfo{title}{Nonlinear orbital and spin {Edelstein} effect in
  centrosymmetric metals}.
\newblock \emph{\bibinfo{journal}{npj Spintronics}}
  \textbf{\bibinfo{volume}{2}}, \bibinfo{pages}{33} (\bibinfo{year}{2024}).

\bibitem{GoLeeOppeneerBlugelMokrousov2024}
\bibinfo{author}{Go, D.}, \bibinfo{author}{Lee, H.-W.},
  \bibinfo{author}{Oppeneer, P.~M.}, \bibinfo{author}{Bl{\"u}gel, S.} \&
  \bibinfo{author}{Mokrousov, Y.}
\newblock \bibinfo{title}{First-principles calculation of orbital {Hall} effect
  by {Wannier} interpolation: Role of orbital dependence of the anomalous
  position}.
\newblock \emph{\bibinfo{journal}{Physical Review B}}
  \textbf{\bibinfo{volume}{109}}, \bibinfo{pages}{174435}
  (\bibinfo{year}{2024}).

\bibitem{AtenciaArovasCulcer2024}
\bibinfo{author}{Burgos~Atencia, R.}, \bibinfo{author}{Arovas, D.~P.} \&
  \bibinfo{author}{Culcer, D.}
\newblock \bibinfo{title}{Intrinsic torque on the orbital angular momentum in
  an electric field}.
\newblock \emph{\bibinfo{journal}{Physical Review B}}
  \textbf{\bibinfo{volume}{110}}, \bibinfo{pages}{035427}
  (\bibinfo{year}{2024}).

\bibitem{LiuCullenArovasCulcer2025}
\bibinfo{author}{Liu, H.}, \bibinfo{author}{Cullen, J.~H.},
  \bibinfo{author}{Arovas, D.~P.} \& \bibinfo{author}{Culcer, D.}
\newblock \bibinfo{title}{Quantum correction to the orbital {Hall} effect}.
\newblock \emph{\bibinfo{journal}{Physical Review Letters}}
  \textbf{\bibinfo{volume}{134}}, \bibinfo{pages}{036304}
  (\bibinfo{year}{2025}).

\bibitem{HanKoOhLeeLeeKim2025}
\bibinfo{author}{Han, S.} \emph{et~al.}
\newblock \bibinfo{title}{Orbital pumping incorporating both orbital angular
  momentum and position}.
\newblock \emph{\bibinfo{journal}{Physical Review Letters}}
  \textbf{\bibinfo{volume}{134}}, \bibinfo{pages}{036305}
  (\bibinfo{year}{2025}).

\bibitem{CullenArovasRaimondiCulcer2025}
\bibinfo{author}{Cullen, J.~H.}, \bibinfo{author}{Arovas, D.~P.},
  \bibinfo{author}{Raimondi, R.} \& \bibinfo{author}{Culcer, D.}
\newblock \bibinfo{title}{Quantum geometry and dipolar dynamics in the orbital
  magneto-electric effect}.
\newblock \emph{\bibinfo{journal}{arXiv preprint arXiv:2505.02911}}
  (\bibinfo{year}{2025}).

\bibitem{XiaoRenXiong2021}
\bibinfo{author}{Xiao, C.}, \bibinfo{author}{Ren, Y.} \&
  \bibinfo{author}{Xiong, B.}
\newblock \bibinfo{title}{Adiabatically induced orbital magnetization}.
\newblock \emph{\bibinfo{journal}{Physical Review B}}
  \textbf{\bibinfo{volume}{103}}, \bibinfo{pages}{115432}
  (\bibinfo{year}{2021}).

\bibitem{ToppTormaKennesMitra2022}
\bibinfo{author}{Topp, G.~E.}, \bibinfo{author}{T{\"o}rm{\"a}, P.},
  \bibinfo{author}{Kennes, D.~M.} \& \bibinfo{author}{Mitra, A.}
\newblock \bibinfo{title}{Orbital magnetization of {Floquet} topological
  systems}.
\newblock \emph{\bibinfo{journal}{Physical Review B}}
  \textbf{\bibinfo{volume}{105}}, \bibinfo{pages}{195426}
  (\bibinfo{year}{2022}).

\bibitem{DongLinGuDuan2024}
\bibinfo{author}{Dong, J.}, \bibinfo{author}{Lin, Z.}, \bibinfo{author}{Gu,
  B.-L.} \& \bibinfo{author}{Duan, W.}
\newblock \bibinfo{title}{Floquet formalism for tunable optically induced
  orbital magnetization in twisted bilayers}.
\newblock \emph{\bibinfo{journal}{Physical Review B}}
  \textbf{\bibinfo{volume}{110}}, \bibinfo{pages}{144444}
  (\bibinfo{year}{2024}).

\bibitem{YaoGoMokrousovMurakami2026}
\bibinfo{author}{Yao, D.}, \bibinfo{author}{Go, D.},
  \bibinfo{author}{Mokrousov, Y.} \& \bibinfo{author}{Murakami, S.}
\newblock \bibinfo{title}{Dynamical orbital angular momentum induced by
  circularly polarized phonons}.
\newblock \emph{\bibinfo{journal}{Physical Review Letters}}
  \textbf{\bibinfo{volume}{137}}, \bibinfo{pages}{066301}
  (\bibinfo{year}{2026}).

\bibitem{MrudulOppeneer2026}
\bibinfo{author}{Mrudul, M.~S.} \& \bibinfo{author}{Oppeneer, P.~M.}
\newblock \bibinfo{title}{Valley-polarized orbital and spin magnetism induced
  by femtosecond optical pulses in two-dimensional semiconductors}.
\newblock \emph{\bibinfo{journal}{arXiv preprint arXiv:2605.31263}}
  (\bibinfo{year}{2026}).

\bibitem{BurgosAtenciaAgarwalCulcer2024}
\bibinfo{author}{Burgos~Atencia, R.}, \bibinfo{author}{Agarwal, A.} \&
  \bibinfo{author}{Culcer, D.}
\newblock \bibinfo{title}{Orbital angular momentum of {Bloch} electrons:
  equilibrium formulation, magneto-electric phenomena, and the orbital {Hall}
  effect}.
\newblock \emph{\bibinfo{journal}{Advances in Physics: X}}
  \textbf{\bibinfo{volume}{9}}, \bibinfo{pages}{2371972}
  (\bibinfo{year}{2024}).

\bibitem{CysneSouzaRappoport2026}
\bibinfo{author}{Cysne, T.~P.}, \bibinfo{author}{Souza, I.} \&
  \bibinfo{author}{Rappoport, T.~G.}
\newblock \bibinfo{title}{Orbital {Hall} effect from orbital magnetic moments
  of {Bloch} states: The role of a new correction term}.
\newblock \emph{\bibinfo{journal}{Physical Review Research}}
  \textbf{\bibinfo{volume}{8}}, \bibinfo{pages}{033086} (\bibinfo{year}{2026}).

\bibitem{LeeBaekSastgesMokrousovLeeGo2026}
\bibinfo{author}{Lee, H.} \emph{et~al.}
\newblock \bibinfo{title}{Anatomy of the modern theory of orbital magnetism
  from first principles: Term-by-term analysis in the gauge-covariant
  formalism}.
\newblock \emph{\bibinfo{journal}{Physical Review B}}
  \textbf{\bibinfo{volume}{113}}, \bibinfo{pages}{214449}
  (\bibinfo{year}{2026}).

\bibitem{SastgesBaekLeeLeeMokrousovGo2026}
\bibinfo{author}{Sastges, M.} \emph{et~al.}
\newblock \bibinfo{title}{Modern approach to orbital {Hall} effect based on
  {Wannier} picture of solids}.
\newblock \emph{\bibinfo{journal}{arXiv preprint arXiv:2604.08280}}
  (\bibinfo{year}{2026}).

\bibitem{Blount1962}
\bibinfo{author}{Blount, E.~I.}
\newblock \bibinfo{title}{ in \textit{Formalisms of band theory}} (eds
  \bibinfo{editor}{Seitz, F.} \& \bibinfo{editor}{Turnbull, D.})
  \emph{\bibinfo{booktitle}{Solid State Physics}}, Vol.~\bibinfo{volume}{13}
  \bibinfo{pages}{305--373} (\bibinfo{publisher}{Academic Press},
  \bibinfo{address}{New York}, \bibinfo{year}{1962}).

\bibitem{Adams1953CrystalMomentum}
\bibinfo{author}{Adams, E.~N.}
\newblock \bibinfo{title}{The crystal momentum as a quantum mechanical
  operator}.
\newblock \emph{\bibinfo{journal}{The Journal of Chemical Physics}}
  \textbf{\bibinfo{volume}{21}}, \bibinfo{pages}{2013--2017}
  (\bibinfo{year}{1953}).

\bibitem{Fuchs2010TopologicalBerry}
\bibinfo{author}{Fuchs, J.~N.}, \bibinfo{author}{Pi{\'e}chon, F.},
  \bibinfo{author}{Goerbig, M.~O.} \& \bibinfo{author}{Montambaux, G.}
\newblock \bibinfo{title}{Topological {Berry} phase and semiclassical
  quantization of cyclotron orbits for two dimensional electrons in coupled
  band models}.
\newblock \emph{\bibinfo{journal}{The European Physical Journal B}}
  \textbf{\bibinfo{volume}{77}}, \bibinfo{pages}{351--362}
  (\bibinfo{year}{2010}).

\bibitem{Yao2008ValleyOptics}
\bibinfo{author}{Yao, W.}, \bibinfo{author}{Xiao, D.} \& \bibinfo{author}{Niu,
  Q.}
\newblock \bibinfo{title}{Valley-dependent optoelectronics from inversion
  symmetry breaking}.
\newblock \emph{\bibinfo{journal}{Physical Review B}}
  \textbf{\bibinfo{volume}{77}}, \bibinfo{pages}{235406}
  (\bibinfo{year}{2008}).

\bibitem{Bhowal2021OrbitalHall}
\bibinfo{author}{Bhowal, S.} \& \bibinfo{author}{Vignale, G.}
\newblock \bibinfo{title}{Orbital {Hall} effect as an alternative to valley
  {Hall} effect in gapped graphene}.
\newblock \emph{\bibinfo{journal}{Physical Review B}}
  \textbf{\bibinfo{volume}{103}}, \bibinfo{pages}{195309}
  (\bibinfo{year}{2021}).
\newblock \bibinfo{note}{Erratum: Phys. Rev. B 109, 239901 (2024)}.

\bibitem{Xiao2007Valley}
\bibinfo{author}{Xiao, D.}, \bibinfo{author}{Yao, W.} \& \bibinfo{author}{Niu,
  Q.}
\newblock \bibinfo{title}{Valley-contrasting physics in {Graphene}: Magnetic
  moment and topological transport}.
\newblock \emph{\bibinfo{journal}{Physical Review Letters}}
  \textbf{\bibinfo{volume}{99}}, \bibinfo{pages}{236809}
  (\bibinfo{year}{2007}).

\bibitem{MotlaghApalkov2021}
\bibinfo{author}{Motlagh, S. A.~O.} \& \bibinfo{author}{Apalkov, V.}
\newblock \bibinfo{title}{Anomalous ultrafast all-optical {Hall} effect in
  gapped graphene}.
\newblock \emph{\bibinfo{journal}{Nanophotonics}}
  \textbf{\bibinfo{volume}{10}}, \bibinfo{pages}{3677--3685}
  (\bibinfo{year}{2021}).

\bibitem{Faisal2022Graphene}
\bibinfo{author}{Faisal, F. H.~M.}
\newblock \bibinfo{title}{Intense laser pulse interaction with {Graphene} and
  {Graphene} ribbons}.
\newblock \emph{\bibinfo{journal}{Frontiers in Chemistry}}
  \textbf{\bibinfo{volume}{10}}, \bibinfo{pages}{859405}
  (\bibinfo{year}{2022}).

\bibitem{BuschZiolkowskiMertigHenk2023}
\bibinfo{author}{Busch, O.}, \bibinfo{author}{Ziolkowski, F.},
  \bibinfo{author}{Mertig, I.} \& \bibinfo{author}{Henk, J.}
\newblock \bibinfo{title}{Ultrafast dynamics of orbital angular momentum of
  electrons induced by femtosecond laser pulses: Generation and transfer across
  interfaces}.
\newblock \emph{\bibinfo{journal}{Physical Review B}}
  \textbf{\bibinfo{volume}{108}}, \bibinfo{pages}{104408}
  (\bibinfo{year}{2023}).

\bibitem{HeFrauenheimLi2024}
\bibinfo{author}{He, J.}, \bibinfo{author}{Frauenheim, T.} \&
  \bibinfo{author}{Li, S.}
\newblock \bibinfo{title}{Ultrafast chiral precession of spin and orbital
  angular momentum induced by circularly polarized laser pulse in elementary
  ferromagnets}.
\newblock \emph{\bibinfo{journal}{The Journal of Physical Chemistry Letters}}
  \textbf{\bibinfo{volume}{15}}, \bibinfo{pages}{2493--2498}
  (\bibinfo{year}{2024}).

\bibitem{DarmawanGrunerPentcheva2026}
\bibinfo{author}{Darmawan, A.}, \bibinfo{author}{Gruner, M.~E.} \&
  \bibinfo{author}{Pentcheva, R.}
\newblock \bibinfo{title}{Disentangling the dynamics of transient spin and
  orbital magnetization in {SrTiO$_3$} via the inverse {Faraday} effect from
  {RT-TDDFT}}.
\newblock \emph{\bibinfo{journal}{arXiv preprint arXiv:2602.20693}}
  (\bibinfo{year}{2026}).

\bibitem{SharmaGillKrishnaHarrisLeeDewhurstShallcross2026}
\bibinfo{author}{Sharma, S.} \emph{et~al.}
\newblock \bibinfo{title}{Giant magnetic moment increase by ultrafast laser
  light}.
\newblock \emph{\bibinfo{journal}{Nature Communications}}
  \textbf{\bibinfo{volume}{17}}, \bibinfo{pages}{5069} (\bibinfo{year}{2026}).

\end{thebibliography}

\begin{thebibliography}{10}
\expandafter\ifx\csname url\endcsname\relax
  \def\url#1{\burl{#1}}\fi
\expandafter\ifx\csname urlprefix\endcsname\relax\def\urlprefix{URL }\fi
\providecommand{\bibinfo}[2]{#2}
\providecommand{\eprint}[2][]{\url{#2}}
\providecommand{\doi}[1]{\url{https://doi.org/#1}}
\bibcommenthead

\bibitem{SI:Adams1953CrystalMomentum}
\bibinfo{author}{Adams, E.~N.}
\newblock \bibinfo{title}{The crystal momentum as a quantum mechanical
  operator}.
\newblock \emph{\bibinfo{journal}{The Journal of Chemical Physics}}
  \textbf{\bibinfo{volume}{21}}, \bibinfo{pages}{2013--2017}
  (\bibinfo{year}{1953}).

\bibitem{SI:AdamsBlount1959}
\bibinfo{author}{Adams, E.~N.} \& \bibinfo{author}{Blount, E.~I.}
\newblock \bibinfo{title}{Energy bands in the presence of an external force
  field---ii: Anomalous velocities}.
\newblock \emph{\bibinfo{journal}{Journal of Physics and Chemistry of Solids}}
  \textbf{\bibinfo{volume}{10}}, \bibinfo{pages}{286--303}
  (\bibinfo{year}{1959}).

\bibitem{SI:Blount1962}
\bibinfo{author}{Blount, E.~I.}
\newblock \bibinfo{title}{ in \textit{Formalisms of band theory}} (eds
  \bibinfo{editor}{Seitz, F.} \& \bibinfo{editor}{Turnbull, D.})
  \emph{\bibinfo{booktitle}{Solid State Physics}}, Vol.~\bibinfo{volume}{13}
  \bibinfo{pages}{305--373} (\bibinfo{publisher}{Academic Press},
  \bibinfo{address}{New York}, \bibinfo{year}{1962}).

\bibitem{SI:AversaSipe1995}
\bibinfo{author}{Aversa, C.} \& \bibinfo{author}{Sipe, J.~E.}
\newblock \bibinfo{title}{Nonlinear optical susceptibilities of semiconductors:
  Results with a length-gauge analysis}.
\newblock \emph{\bibinfo{journal}{Physical Review B}}
  \textbf{\bibinfo{volume}{52}}, \bibinfo{pages}{14636--14645}
  (\bibinfo{year}{1995}).

\bibitem{SI:Ventura2017}
\bibinfo{author}{Ventura, G.~B.}, \bibinfo{author}{Passos, D.~J.},
  \bibinfo{author}{Lopes~dos Santos, J. M.~B.}, \bibinfo{author}{Viana
  Parente~Lopes, J.~M.} \& \bibinfo{author}{Peres, N. M.~R.}
\newblock \bibinfo{title}{Gauge covariances and nonlinear optical responses}.
\newblock \emph{\bibinfo{journal}{Physical Review B}}
  \textbf{\bibinfo{volume}{96}}, \bibinfo{pages}{035431}
  (\bibinfo{year}{2017}).

\bibitem{SI:Parker2019}
\bibinfo{author}{Parker, D.~E.}, \bibinfo{author}{Morimoto, T.},
  \bibinfo{author}{Orenstein, J.} \& \bibinfo{author}{Moore, J.~E.}
\newblock \bibinfo{title}{Diagrammatic approach to nonlinear optical response
  with application to weyl semimetals}.
\newblock \emph{\bibinfo{journal}{Physical Review B}}
  \textbf{\bibinfo{volume}{99}}, \bibinfo{pages}{045121}
  (\bibinfo{year}{2019}).

\bibitem{SI:BonbienManchon2022}
\bibinfo{author}{Bonbien, V.} \& \bibinfo{author}{Manchon, A.}
\newblock \bibinfo{title}{Theory of perturbatively nonlinear quantum transport
  {II}: Hilbert space truncation, gauge invariance, and second order transport
  in a spatially uniform, time-varying electric field}.
\newblock \emph{\bibinfo{journal}{arXiv preprint arXiv:2202.05378}}
  (\bibinfo{year}{2022}).

\bibitem{SI:Desmarais2023}
\bibinfo{author}{Desmarais, J.~K.}, \bibinfo{author}{Kirtman, B.} \&
  \bibinfo{author}{R{\'e}rat, M.}
\newblock \bibinfo{title}{First-principles calculation of the optical rotatory
  power of periodic systems: Modern theory with modern functionals}.
\newblock \emph{\bibinfo{journal}{Physical Review B}}
  \textbf{\bibinfo{volume}{107}}, \bibinfo{pages}{224430}
  (\bibinfo{year}{2023}).

\bibitem{SI:Ma2024ProperSpin}
\bibinfo{author}{Ma, H.}, \bibinfo{author}{Cullen, J.~H.},
  \bibinfo{author}{Monir, S.}, \bibinfo{author}{Rahman, R.} \&
  \bibinfo{author}{Culcer, D.}
\newblock \bibinfo{title}{Spin-{Hall} effect in topological materials:
  Evaluating the proper spin current in systems with arbitrary degeneracies}.
\newblock \emph{\bibinfo{journal}{npj Spintronics}}
  \textbf{\bibinfo{volume}{2}}, \bibinfo{pages}{55} (\bibinfo{year}{2024}).

\bibitem{SI:An2026PositionOperators}
\bibinfo{author}{An, D.}, \bibinfo{author}{Jeon, J.} \& \bibinfo{author}{Kim,
  S.~K.}
\newblock \bibinfo{title}{Consistent evaluation of operators involving the
  position operator in the {Bloch} representation: Application to the orbital
  moment}.
\newblock \emph{\bibinfo{journal}{arXiv preprint arXiv:2606.11679}}
  (\bibinfo{year}{2026}).

\bibitem{SI:ThonhauserCeresoliVanderbiltResta2005}
\bibinfo{author}{Thonhauser, T.}, \bibinfo{author}{Ceresoli, D.},
  \bibinfo{author}{Vanderbilt, D.} \& \bibinfo{author}{Resta, R.}
\newblock \bibinfo{title}{Orbital magnetization in periodic insulators}.
\newblock \emph{\bibinfo{journal}{Physical Review Letters}}
  \textbf{\bibinfo{volume}{95}}, \bibinfo{pages}{137205}
  (\bibinfo{year}{2005}).

\bibitem{SI:CeresoliThonhauserVanderbiltResta2006}
\bibinfo{author}{Ceresoli, D.}, \bibinfo{author}{Thonhauser, T.},
  \bibinfo{author}{Vanderbilt, D.} \& \bibinfo{author}{Resta, R.}
\newblock \bibinfo{title}{Orbital magnetization in crystalline solids:
  Multi-band insulators, {Chern} insulators, and metals}.
\newblock \emph{\bibinfo{journal}{Physical Review B}}
  \textbf{\bibinfo{volume}{74}}, \bibinfo{pages}{024408}
  (\bibinfo{year}{2006}).

\bibitem{SI:ShiVignaleXiaoNiu2007}
\bibinfo{author}{Shi, J.}, \bibinfo{author}{Vignale, G.},
  \bibinfo{author}{Xiao, D.} \& \bibinfo{author}{Niu, Q.}
\newblock \bibinfo{title}{Quantum theory of orbital magnetization and its
  generalization to interacting systems}.
\newblock \emph{\bibinfo{journal}{Physical Review Letters}}
  \textbf{\bibinfo{volume}{99}}, \bibinfo{pages}{197202}
  (\bibinfo{year}{2007}).

\bibitem{SI:Xiao2007Valley}
\bibinfo{author}{Xiao, D.}, \bibinfo{author}{Yao, W.} \& \bibinfo{author}{Niu,
  Q.}
\newblock \bibinfo{title}{Valley-contrasting physics in {Graphene}: Magnetic
  moment and topological transport}.
\newblock \emph{\bibinfo{journal}{Physical Review Letters}}
  \textbf{\bibinfo{volume}{99}}, \bibinfo{pages}{236809}
  (\bibinfo{year}{2007}).

\bibitem{SI:Xiao2012SpinValley}
\bibinfo{author}{Xiao, D.}, \bibinfo{author}{Liu, G.-B.},
  \bibinfo{author}{Feng, W.}, \bibinfo{author}{Xu, X.} \& \bibinfo{author}{Yao,
  W.}
\newblock \bibinfo{title}{Coupled spin and valley physics in monolayers of
  {MoS$_2$} and other group-{VI} dichalcogenides}.
\newblock \emph{\bibinfo{journal}{Physical Review Letters}}
  \textbf{\bibinfo{volume}{108}}, \bibinfo{pages}{196802}
  (\bibinfo{year}{2012}).

\bibitem{SI:Xu2014SpinPseudospin}
\bibinfo{author}{Xu, X.}, \bibinfo{author}{Yao, W.}, \bibinfo{author}{Xiao, D.}
  \& \bibinfo{author}{Heinz, T.~F.}
\newblock \bibinfo{title}{Spin and pseudospins in layered transition metal
  dichalcogenides}.
\newblock \emph{\bibinfo{journal}{Nature Physics}}
  \textbf{\bibinfo{volume}{10}}, \bibinfo{pages}{343--350}
  (\bibinfo{year}{2014}).

\bibitem{SI:Yao2008ValleyOptics}
\bibinfo{author}{Yao, W.}, \bibinfo{author}{Xiao, D.} \& \bibinfo{author}{Niu,
  Q.}
\newblock \bibinfo{title}{Valley-dependent optoelectronics from inversion
  symmetry breaking}.
\newblock \emph{\bibinfo{journal}{Physical Review B}}
  \textbf{\bibinfo{volume}{77}}, \bibinfo{pages}{235406}
  (\bibinfo{year}{2008}).

\bibitem{SI:Bhowal2021OrbitalHall}
\bibinfo{author}{Bhowal, S.} \& \bibinfo{author}{Vignale, G.}
\newblock \bibinfo{title}{Orbital {Hall} effect as an alternative to valley
  {Hall} effect in gapped graphene}.
\newblock \emph{\bibinfo{journal}{Physical Review B}}
  \textbf{\bibinfo{volume}{103}}, \bibinfo{pages}{195309}
  (\bibinfo{year}{2021}).
\newblock \bibinfo{note}{Erratum: Phys. Rev. B 109, 239901 (2024)}.

\bibitem{SI:Xiao2010BerryRMP}
\bibinfo{author}{Xiao, D.}, \bibinfo{author}{Chang, M.-C.} \&
  \bibinfo{author}{Niu, Q.}
\newblock \bibinfo{title}{Berry phase effects on electronic properties}.
\newblock \emph{\bibinfo{journal}{Reviews of Modern Physics}}
  \textbf{\bibinfo{volume}{82}}, \bibinfo{pages}{1959--2007}
  (\bibinfo{year}{2010}).

\end{thebibliography}
\end{document}